\documentclass[
    paper=A4,
    twoside,
    BCOR=15mm,
    DIV=14,
    headsepline,
    parskip=half,
    titlepage,
    openright,
    11pt,
    abstracton,
    ngerman,
    english
]{scrreprt}
\usepackage{babel}                  %
\usepackage[utf8]{inputenc}
\usepackage[T1]{fontenc}
\usepackage{xspace}					%

\usepackage{graphicx}				%
\usepackage[usenames,dvipsnames,svgnames,table]{xcolor} %
\usepackage{array}

\usepackage{amsmath}
\usepackage{amssymb}
\usepackage{mathtools}
\usepackage{units}                  %
\usepackage{textcomp}               %
\usepackage{gensymb}                %
\usepackage{floatflt}               %
\usepackage{subfigure}              %
\usepackage{csquotes}               %
\usepackage{esdiff}                 %
\usepackage{caption}                %
\usepackage{listings}               %

\usepackage[version=3]{tex_include/mhchem}
\newcommand{\HRule}{\rule{\linewidth}{0.5mm}}
\newcommand{\sups}[1]{\ensuremath{^{\text{#1}}}}    %

\newcommand{\plotwidth}{0.80\textwidth}             %
\newcommand{\pdfce}[1]{%
    \texorpdfstring{\ce{#1}}{#1}%
}
\newcommand{\binmix}[4]{%
    \ce{#1}~\unit[#2]{\%} -- \ce{#3}~\unit[#4]{\%}%
}
\newcommand{\termix}[6]{%
    \binmix{#1}{#2}{#3}{#4} -- \ce{#5}~\unit[#6]{\%}%
}
\newcommand{\isotope}[2][]{\text{\sups{#1}#2}}      %
\newcommand{\ETp}{\ensuremath{E\,\nicefrac{T}{p}}}  %
\DeclareMathOperator{\RMS}{RMS}                     %
\newcommand{\pdiff}[2]{\ensuremath{%
    \frac{\partial #1}{\partial #2}%
}}
\newcommand{\dd}{\,\mathrm{d}}                        %

\newcommand{\gasDBURL}{\url{http://web.physik.rwth-aachen.de/gasDB/}}

\usepackage[numbers]{natbib}
\usepackage[nottoc]{tocbibind}          %

\usepackage[
  pdftitle={Master's thesis},
  pdfsubject={Measurements and simulations of drift gas properties},
  pdfauthor={Lukas Koch},
  pdfkeywords={Drift velocity, Townsend coefficient, Argon, Ar, Methane, CH4, Carbon dioxide, CO2, CF4, iC4H10, H2}
]{hyperref}

\begin{document}

\begin{titlepage}
\begin{center}
\Large
~\\[1.5cm]

\HRule\\[0.4cm]
{ \huge \bfseries Measurements and simulations of\\
    drift gas properties \\[0.4cm] }

\HRule\\[1.5cm]

von\\[0.5cm]

Lukas \textsc{Koch}\\

\vfill

\textsc{Masterarbeit in Physik}\\[0.5cm]

vorgelegt der\\[0.5cm]

\textsc{\large Fakultät für Mathematik, Informatik und Naturwissenschaften}\\
der \textsc{\large RWTH Aachen}\\[0.5cm]

im September, 2013\\[0.5cm]

angefertigt im\\[0.5cm]

\textsc{III. Physikalischen Institut B}\\[0.5cm]

bei\\[0.5cm]

PD Dr.~Stefan \textsc{Roth}

\end{center}
\end{titlepage}

\thispagestyle{empty}

\begin{abstract}
For the successful design and operation of gas based particle detectors,
one needs a good understanding of the drift properties of the deployed
gas. This includes the drift velocity of electrons, their diffusion
and the gas amplification in different electric and magnetic fields.

This work presents simulations and precision
measurements of the drift velocity $v_{d}$ in low electric fields
($< \unitfrac[400]{V}{cm}$) for argon-based gas mixtures with up to two additives.
The additives used are \ce{CH4}, \ce{CO2}, \ce{CF4}, \ce{iC4H10} and \ce{H2}.

The simulations were done using Garfield++ and Magboltz.
The measurements were taken with T2K/ND280-type monitoring drift chambers, which allow $v_d$~measurements
in the range of about $\unitfrac[20]{\micro{m}}{ns}$ -- $\unitfrac[100]{\micro{m}}{ns}$.
Experiment and simulation are in good agreement and most remaining deviations
are below \unitfrac[1]{\micro{m}}{ns}.
To make the obtained data available for interested parties, a database was established at \gasDBURL.

Finally, a promising method for the measurement of the first Townsend coefficient $\alpha_{T}$ is presented.
\end{abstract}

\thispagestyle{empty}

\tableofcontents
\renewcommand{\listfigurename}{List of figures}
\listoffigures

\chapter{Introduction}

\begin{figure}
    \centering
    \includegraphics[width=0.99\textwidth]{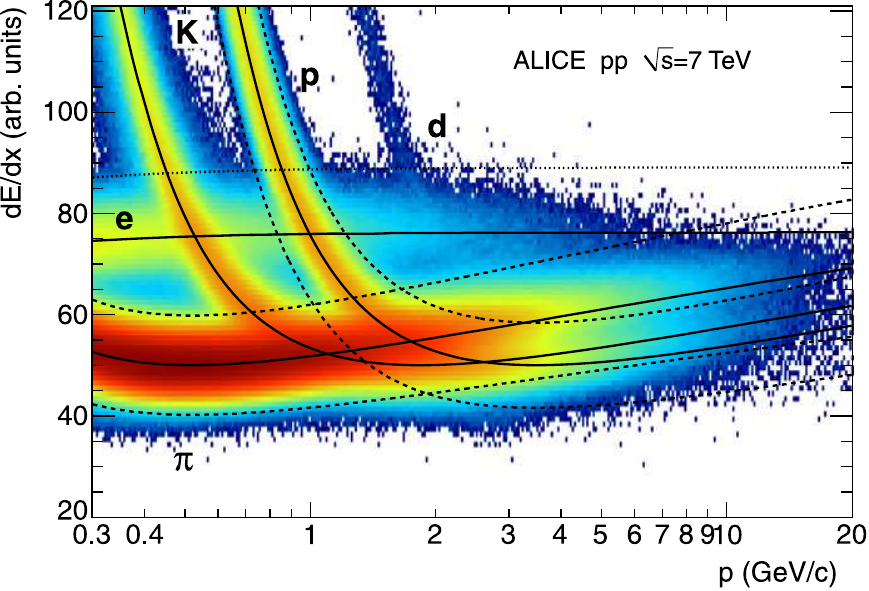}
    \caption[Particle identification by energy loss and momentum]
            {Particle identification by energy loss and momentum \cite{Aamodt2011442}}
    \label{fig:dEdxp}
\end{figure}

Gas based particle detectors have proven to be a reliable technology
and offer a few advantages over other detector technologies:

\begin{itemize}
\item They allow large sensitive volumes with very little material in the particles' path.
\item Utilising gas amplification, one can use less sensitive electronic amplification.
\item They can measure the energy loss of the particle.
Together with a momentum measurement (reconstructed track in a magnetic field), this allows particle
identification (see figure \ref{fig:dEdxp}).
\item In the case of TPCs\footnote{Time Projection Chamber},
they offer a fine track reconstruction with comparatively few read-out channels.
\end{itemize}

To make use of these advantages, one needs a good understanding of the properties of the
employed gas mixture. This includes everything from the first ionisation of the gas by a passing
high energy particle, to the signal generation when the created free electrons are amplified
and collected at the anode. This thesis will concentrate on the transportation of the electrons from the
place of ionisation to the amplification region, namely it will study the drift velocity. In addition,
this chapter will give a short overview over the whole process from ionisation to detection.

\section{Electron transport in gases}

The transport of electrons in a gas is a very complex process and this introduction can only cover
the most important derivations and results. For an in-depth discussion of the processes involved,
please refer to reference \cite{Blum2008}, as most of the following is taken from there.

\subsection{Primary and secondary ionisation}
\label{sec:ionisation}

\begin{figure}
    \centering
    \includegraphics[width=0.99\textwidth]{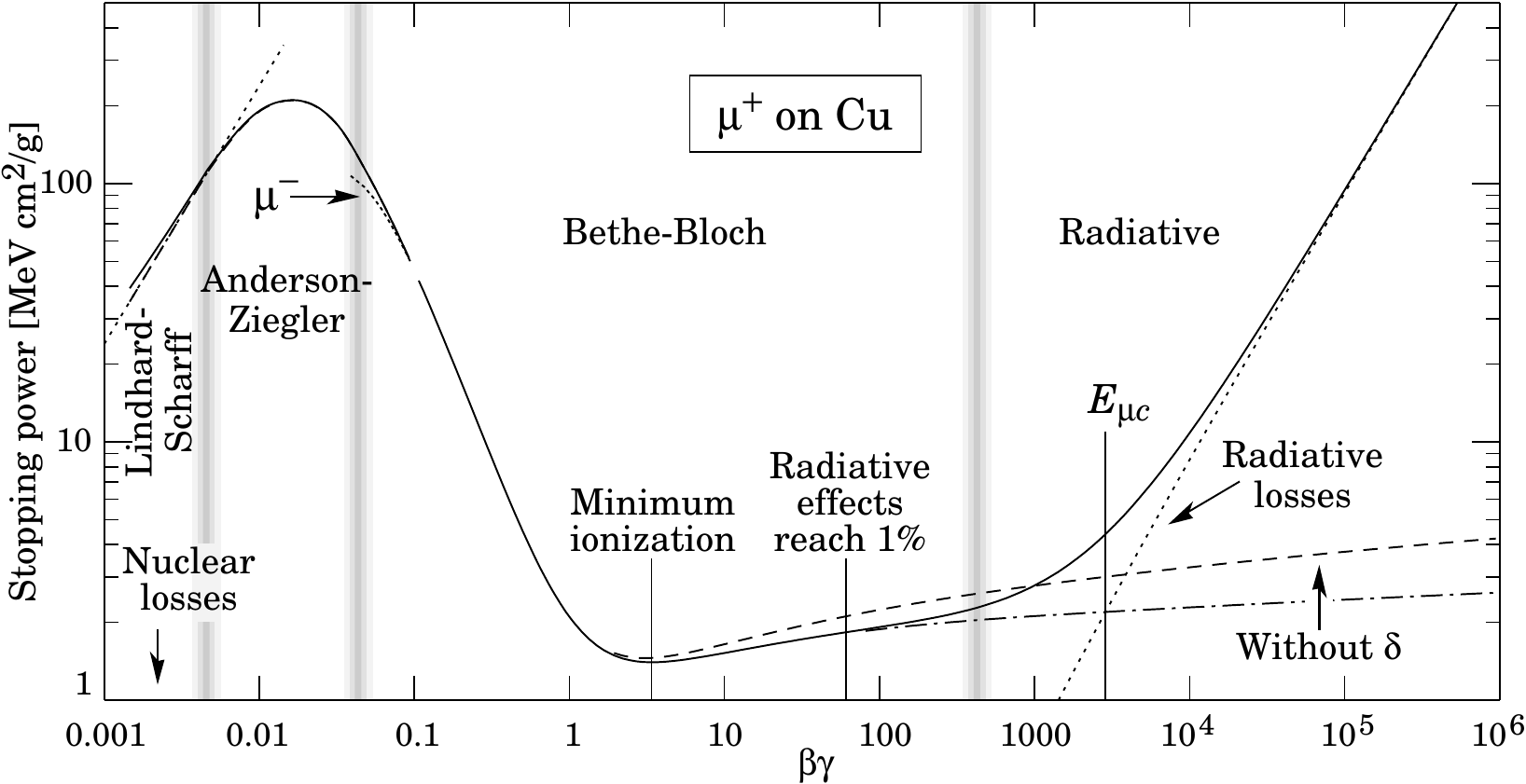}
    \caption[Energy loss of antimuons in copper]
            {Energy loss of antimuons in copper \cite{PDG2012}.
             The behaviour is very similar for other particles and materials.}
    \label{fig:stopping}
\end{figure}

When high energy charged particles traverse a gas, they lose energy due to interactions with the matter.
Depending on the particle's energy, different processes contribute to the energy loss (see figure~\ref{fig:stopping}).
This energy excites the gas molecules and if the incident particle's energy is high enough,
some of the deposited energy goes into ionising the molecules along its trajectory.
One distinguishes between primary and secondary ionisation. Primary ionisation is the direct ionisation
of a gas molecule $A$ by a passing charged particle $P^\pm$.
\[
    P^\pm + A \rightarrow P^\pm + A^+ + e^-
\]

Secondary ionisation occurs if a molecule is not ionised directly by the passing particle, but by a product
of the interaction with another molecule. This can either be a comparatively highly energetic ionisation electron, also called $\delta$-electron,
\begin{align*}
    P^\pm + A &\rightarrow P^\pm + A^+ + e^- \\
    e^- + B &\rightarrow e^- + B^+ + e^- ,
\end{align*}
or an excited state of a molecule,
\begin{align*}
    P^\pm + A &\rightarrow P^\pm + A^* \\
    A^* + B &\rightarrow A + B^+ + e^- .
\end{align*}
In the latter case, the ionisation energy of molecule $B$ must be lower than the energy of the excited state
$A^*$. Ionisation by excited gas molecules is known as Penning effect, if the energy transfer from $A^*$ to $B$
happens through a collision, or Jesse effect, if it happens via an emitted photon. Most ionisation electrons
along a particle track are created by secondary ionisation.

By measuring the separated charge along the particle track, one can reconstruct the energy loss of the
ionising particle. Unfortunately, only a certain fraction of the energy that a particle loses when traveling through
the gas does go into ionisation of the gas molecules. Therefore, one cannot simply use the ionisation energies of
the gases to derive the energy loss. For a correct calculation, one has to use the average energy $W$ that is needed to produce
one ionisation pair. If a particle traverses a distance $L$ in a gas mixture, it will, on average, lose the energy
$\langle\Delta E\rangle$ and produce $\langle N\rangle$ ion electron pairs. $W$ is then the ratio of energy loss
to the number of produced ionisation pairs:

\[
    W = \frac{\langle \Delta E \rangle}{\langle N\rangle} =\left\langle\diff{E}{x}\right\rangle \frac{L}{\langle N\rangle}
\]

This value depends on the type of ionising particle, its energy and, of course, the gas mixture that
is being ionised. The energy dependence vanishes for high energies, i.e. a few \unit{keV} for ionising
electrons and a few \unit{MeV} for ionising alpha particles. If one knows $W$ (through measurement or simulation),
one can use it to reconstruct the energy loss of the traversing particle by measuring the ionisation charges that
were separated along its track. This information can then be used to identify the particle. Typical values of $W$
range between \unit[20]{eV} and \unit[50]{eV} (see appendix \ref{chap:gasprop}).

\subsection{Electron drift in an electric field}

To prevent the ion-electron-pairs from recombining and to transport the electrons to an anode where the charge
can be measured, one applies an electric field to the gas. The electrons will then drift along the field lines%
\footnote{For now we assume that there is no magnetic field present.} towards the anode.

On their way the electrons collide with gas molecules, and due to the large mass difference between
molecules and electrons, they are scattered isotropically with an average instantaneous microscopic velocity $v_\text{inst}$.
This velocity does not directly contribute to the drift velocity $v_d$, due to the isotropic nature of the scattering.
Between collisions the electrons, and thus their mean position $\bar{z}$, are accelerated by the electric field
in the $z$-direction. This macroscopic movement is stopped
at the next collision, when the direction is randomised again. The time between two collisions $t$ is exponentially
distributed with the mean value $\tau$. The additional velocity that has been picked up during that time is the
macroscopic drift velocity $v_d$.
\begin{equation} \label{eq:vd}
    v_d = \frac{\langle \Delta\bar{z}\rangle}{\langle t\rangle} 
        = \frac{\langle\frac{1}{2}at^2\rangle}{\langle t\rangle}
        = \frac{\frac{1}{2}a\langle t^2\rangle}{\langle t\rangle}
        = \frac{\frac{1}{2}a 2\tau^2}{\tau}
        = \frac{eE}{m_e}\tau
\end{equation}

This additional velocity corresponds to an energy gain of the electron. In equilibrium, when the drift velocity and
the electron energy $\varepsilon$ are constant over time, this additional energy must (on average) be lost at the next
collision. We therefore introduce the average fractional energy loss per collision $\lambda$. The energy gained from
drifting a distance $\Delta z$ through the electric field must be equal to the energy loss due to collisions with the gas
molecules:
\begin{equation} \label{eq:lambda}
    eE\Delta z = \varepsilon \lambda \frac{\Delta z}{v_d\tau}
\end{equation}

The mean time between collisions can be expressed with the microscopic velocity $v_\text{inst}$, the number density of
gas molecules $n$ and the cross-section of the scattering process $\sigma$.
\begin{equation}
    \tau = \frac{1}{v_\text{inst}n\sigma} \label{eq:tau}
\end{equation}
The electron energy $\varepsilon$ is also a function of $v_\text{inst}$.
\begin{equation}
    \varepsilon = \frac{1}{2}m_ev_\text{inst}^2 \label{eq:epsilon}
\end{equation}

Combining equations \eqref{eq:vd} through \eqref{eq:epsilon} yields the following
expression:
\begin{equation*}
    v_d^2 = v_\text{inst}^2\frac{\lambda}{2}
          = \frac{e}{m_e\sigma}\sqrt{\frac{\lambda}{2}}\frac{E}{n}
\end{equation*}
These equations suggest a proportional relationship between $E$ and $v_d^2$, but $\sigma$ and $\lambda$
strongly depend on the electron energy $\varepsilon$ and thus on $v_\text{inst}$. One often combines these dependencies
into the electron mobility $\mu$ which is then a function of $\nicefrac{E}{n}$.
\[
    v_d = \sqrt{\frac{e}{m_e\sigma}\sqrt{\frac{\lambda}{2}}\frac{n}{E}}\,\frac{E}{n} = \mu(\nicefrac{E}{n})\frac{E}{n}
\]

Up to now we assumed that all
electrons behave like the average electron with a single distinct value of $v_\text{inst}$ and thus $\sigma$ and $\lambda$.
In reality these values are statistically distributed and this has to be considered in the calculations.
A calculation that takes the distribution of $v_\text{inst}$ into account, but assumes constant $\sigma$ and $\lambda$,
yields \cite[p.\,81]{Blum2008}
\begin{align}
    \langle v_\text{inst}^2\rangle &= 0.854\frac{e}{m_e\sigma}\sqrt{\frac{2}{\lambda}}\frac{E}{n}, \nonumber \\
    v_d^2 &= 0.855\frac{e}{m_e\sigma}\sqrt{\frac{\lambda}{2}}\frac{E}{n}. \label{eq:vd-comp}
\end{align}

It is noteworthy that the dependency on the electric field $E$ only appears in the fraction $\nicefrac{E}{n}$.
Hence, if the density of a drift gas changes (due to temperature or pressure changes), one can compensate for that by changing
the electric field accordingly.

\begin{figure}
    \centering
    \subfigure{
        \includegraphics[width=\plotwidth]{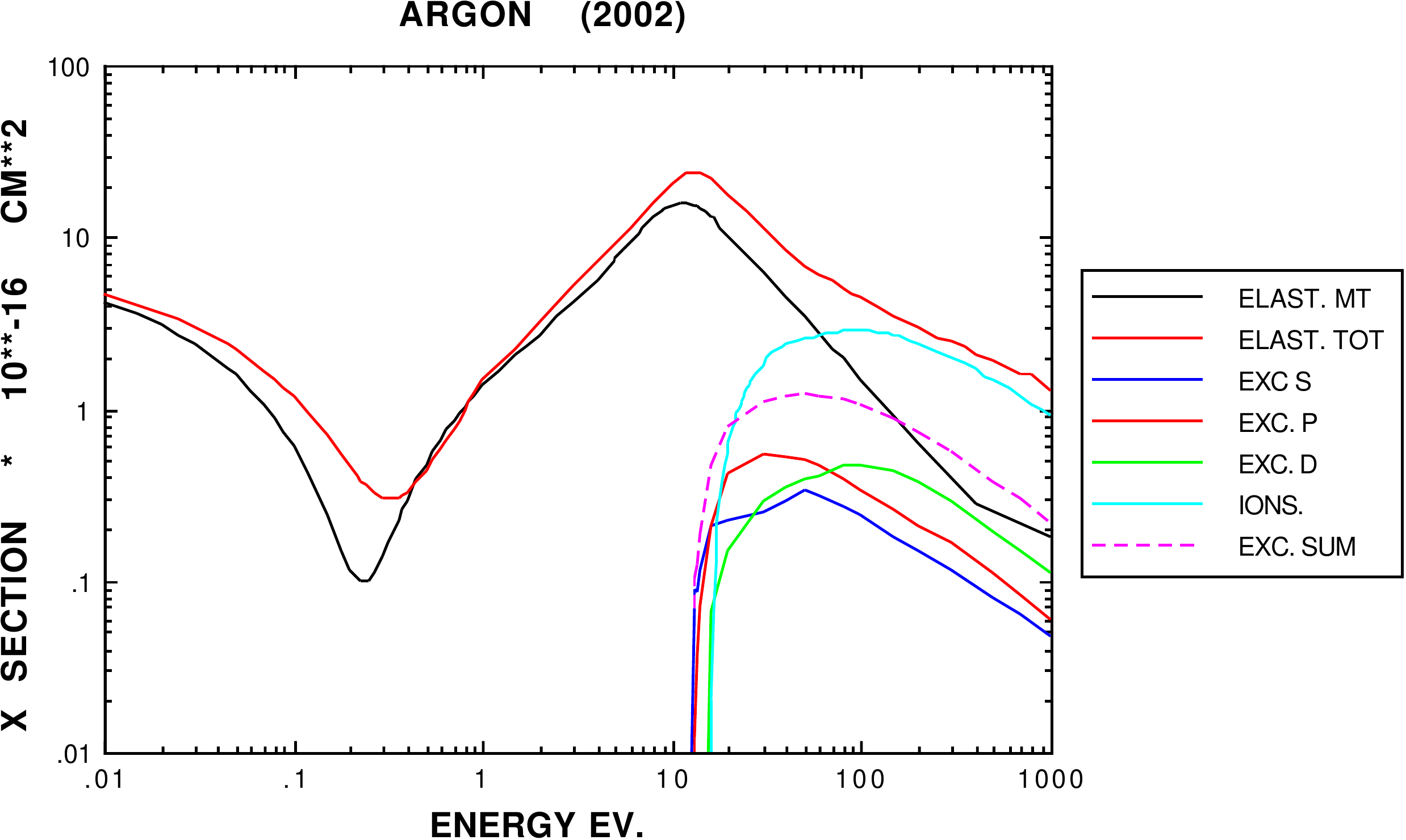}
    }
    \hfill\\[10pt]
    \subfigure{
        \includegraphics[width=\plotwidth]{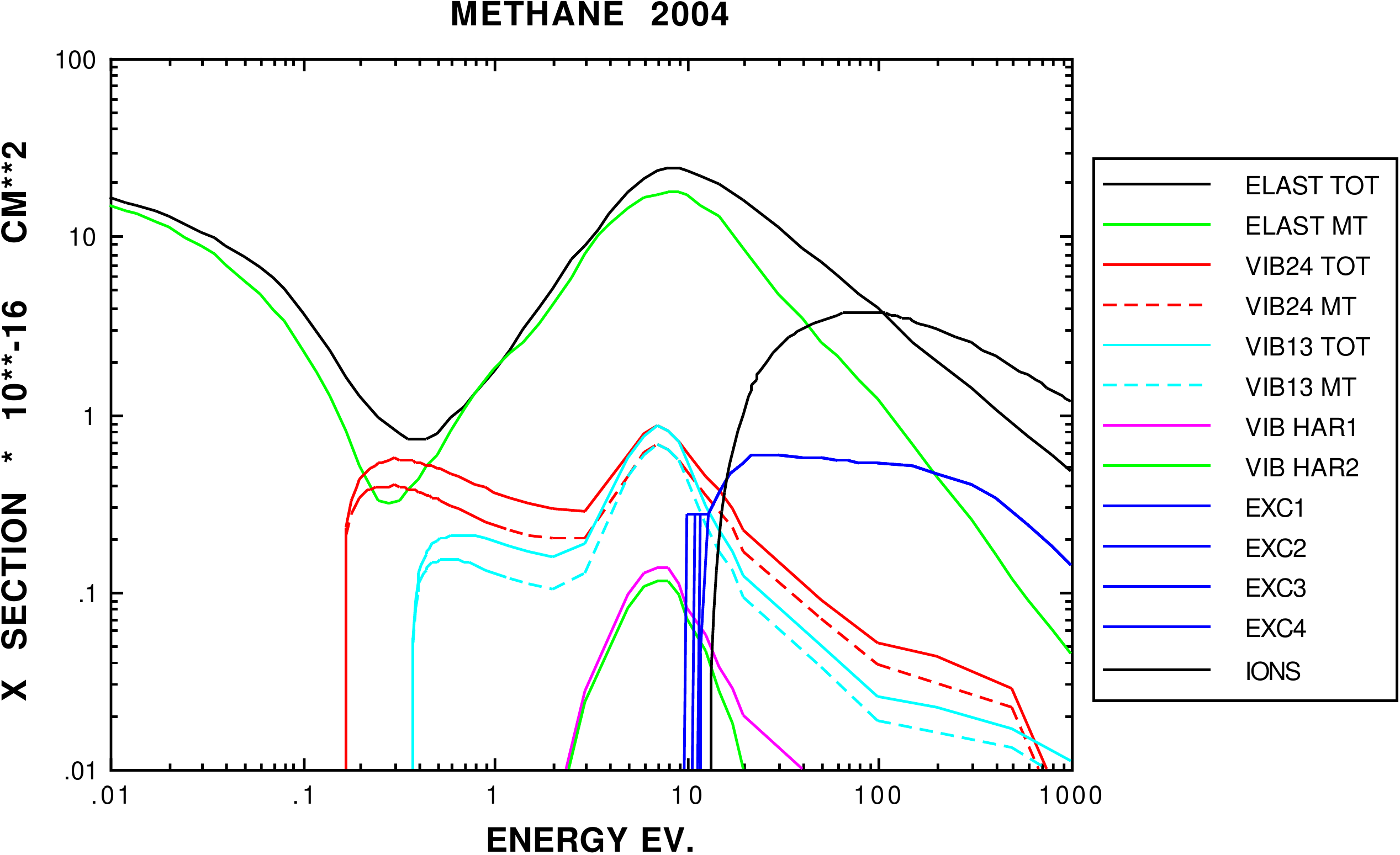}
    }
    \caption[Electron cross-sections for argon and methane]
            {Elastic and inelastic electron cross-sections for argon and methane in dependence of the electron energy \cite{MAGCROSS}}
    \label{fig:cross-sections}
\end{figure}

For an exact reproduction of the real behaviour of electrons in the field, one has to consider the (energy
dependent) cross-sections of all relevant elastic scattering, excitation and ionisation processes of the
different gases in the mix. This can only be done numerically. Figure~\ref{fig:cross-sections} shows example
cross-sections for argon and methane.

The prominent dip in the elastic scattering cross-section at around \unit[0.2]{eV} is called the Ramsauer minimum.
According to equation \eqref{eq:vd-comp}, a low $\sigma$ means a high drift velocity, but argon has no excitation states
in that energy region. Therefore, the electrons lose only very little energy in the elastic collisions and
will quickly be accelerated beyond the Ramsauer minimum to higher energies and slower drift velocity.

Methane, on the other hand, has considerable cross-sections for inelastic processes in that energy region.
So by adding methane to the gas mixture, one can lower the average electron energy towards the Ramsauer minimum,
thereby increasing the drift velocity. This is the general principle of drift gas mixtures, the details of which
depend on the excitation states and cross-sections of the deployed gases. Usually the drift velocity increases with
increasing electric field and then reaches a plateau or a maximum due to combined effects of the elastic and inelastic
scattering cross-sections.

\subsection{Diffusion}

Due to the random isotropic scattering of the electrons, the electron cloud traversing the drift volume will
diffuse. In the simplest case, we can assume the diffusion to be isotropic while the electron cloud as a whole
drifts through the gas with the velocity $v_d$. The electron number is conserved, so the current density $\vec{j}$   %
and number density $n_e$ have to satisfy the following equations:
\begin{align}
    \pdiff{n_e}{t} + \vec\nabla \cdot \vec{j} &= 0 \label{eq:diff1}\\
    \vec{j} + D \vec\nabla n_e  &= 0 \label{eq:diff2}
\end{align}
Here $D$ is the diffusion constant. Combining \eqref{eq:diff1} and \eqref{eq:diff2} we get
\[
    \pdiff{n_e}{t} - D \Delta n_e = 0.
\]
For an initially point-like electron cloud that is created at $x=y=z=t=0$, this equation is satisfied by the Gaussian distribution
\[
    n_e = \left(\frac{1}{\sqrt{4\pi Dt}}\right)^3\exp\left(\frac{-r^2}{4Dt}\right),
\]
where $r^2 = x^2 + y^2 + (z-v_dt)^2$ is the distance from centre of the drifting electron cloud.
The mean squared deviation from the centre (in any direction) is thus
\[
    \sigma_{r}^2 = 2Dt.
\]

Further calculations reveal that the diffusion constant depends on the microscopic electron velocity $v_\text{inst}$ and the
mean free drift time $\tau$ \cite[p.\,68]{Blum2008}:
\[
    D = \frac{v_\text{inst}^2\tau}{3} \overset{\text{\eqref{eq:tau}}}{=} \frac{v_\text{inst}}{3n\sigma}
\]
Please note that $n$ is the number density of gas molecules, while $n_e$ denotes the number density of the
drifting electrons. This simple relationship again assumes a singular value of $v_\text{inst}$ and more complicated
calculations have to be performed in order to take the $v_\text{inst}$~distribution into account. Also, due to the differential
movement of the electron cloud with respect to the gas, the diffusion is not exactly isotropic. Electrons
moving parallel to the direction of drift have a higher $v_\text{inst}$ with respect to the gas than electrons
moving antiparallel to $v_d$.

Sometimes it is more useful to treat the diffusion as dependent from the drift length and not the time,
e.g. when choosing a gas mixture for a detector of given length. Since the electron cloud is drifting at a constant
velocity $v_d$, the relation is simple:
\[
    \sigma_r = \sqrt{2Dt} = \underbrace{\sqrt{\frac{2D}{v_d}}}_{:=\,d} \sqrt{\Delta z} = d \sqrt{\Delta z}
\]
The unit of the new diffusion constant $d$ is $\sqrt{\unit{cm}}$.

\subsection{Influence of a magnetic field}

If one also applies a magnetic field to the drift volume, the rotational symmetry around the $E$-field vector
is broken\footnote{With the exception of $\vec{E} \parallel \vec{B}$, of course.} and the drift direction is not
parallel to the $E$-field anymore. Mathematically we take this into account by replacing the scalar electron
mobility $\mu$ from \eqref{eq:vd-comp} with a tensor $\boldsymbol{\mu}$ and treating the drift velocity and E-field
as the vector quantities they are.
\[
    \vec{v}_d = \boldsymbol{\mu}(\nicefrac{E}{n}, \nicefrac{B}{n}, \varphi) \frac{\vec{E}}{n}
\]
Here $\varphi$ denotes the angle between E- and B-field.

The diffusion gets treated similarly and we replace the
scalar diffusion constant with a tensor. The mean squared deviation of the cloud in in the direction $\vec{e}$
is then
\[
    \sigma_e^2 = 2\vec{e}\boldsymbol{D}(\nicefrac{E}{n}, \nicefrac{B}{n}, \varphi)\vec{e}t.
\]

\subsection{Gas amplification at high electric fields}
\label{sec:amp}

Once the electrons reach the anode, they create an electric signal. But without amplification, the charge created by
the ionising particle would not be sufficient for a reliable detection. One can use gas amplification to boost the signal into
regions that are more easily detectable by electronic readouts.

In high electric fields, the electron energy $\varepsilon$ will be sufficient to ionise gas molecules, thus creating more
free electrons. Those electrons will then ionise even more gas molecules and so on. This exponential amplification process
is called an electron cascade.

Mathematically the amplification is described by the so called Townsend coefficient $\alpha_T$. It describes the
number of ionisations by a single electron per unit length.
The ionisation cross-section is energy dependent and thus a function of $\nicefrac{E}{n}$. Additionally the ionisation rate
is proportional to the number density of gas molecules, so the functional dependence of $\alpha_T$ can be written as
\[
    \alpha_T(\nicefrac{E}{n}, n) = \left\langle \frac{N}{\Delta z} \right\rangle = f(\nicefrac{E}{n})\,n,
\]
where $N$ is the number of ionisations along the drift length $\Delta z$.
The gain factor $G$ can then be calculated from the integrated Townsend coefficient along the drift path.
\[
    G = \left\langle \frac{N_e}{N_{e,0}} \right\rangle = \exp\left( \int \alpha_T(\nicefrac{E(z)}{n}, n) \dd z \right)
\]
Here $N_{e,0}$ and $N_e$ are the total number of free electrons before and after the amplification.

In principle, it is possible to calculate $\alpha_T$ just like $v_d$ if one knows the cross-sections for the
scattering and ionisation processes. But the ionisation rate calculated this way only covers direct ionisation by the electrons
and, as described in section \ref{sec:ionisation}, a large fraction of the separated charges is produced by indirect secondary
ionisation. The Penning and Jesse effects, at the moment, cannot be calculated and have to be measured. Mathematically
they are modelled by a probability $P$ that an excited gas molecule with an excitation energy over the lowest ionisation
energy in the gas mix will produce a secondary ionisation.
\[
    P = \left\langle \frac{N_{e,\text{sec}}}{N^*_{E>E_\text{min}}} \right\rangle
\]
Here $N^*_{E>E_\text{min}}$ is the number of excited molecules with an excitation energy above the lowest ionisation potential
in the gas mixture and $N_{e,\text{sec}}$ is the number of free electrons produced by the Penning and Jesse effects.

Gas amplification starts in electric fields of the order of \unitfrac[10]{kV}{cm}. To create such high fields without
having to create unmanageable voltages, one uses small structures such as thin wires or micro patterns.

\section{Examples of gas based particle detectors}

\subsection{Drift tubes}

\begin{figure}
    \centering
    \includegraphics[width=0.99\textwidth]{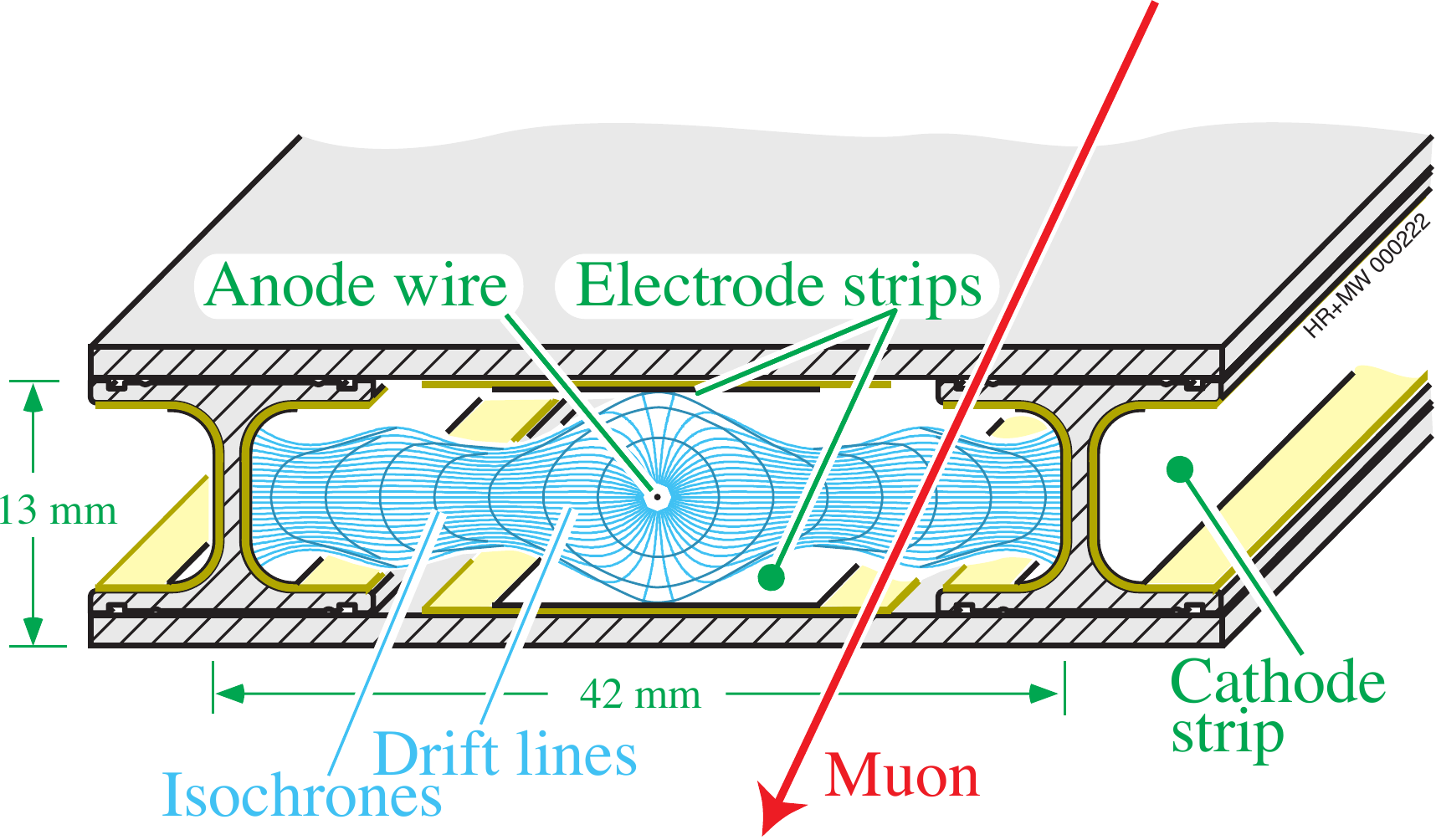}
    \caption[A drift tube of the CMS muon system]
            {A drift tube of the CMS muon system \cite{REIT2010}}
    \label{fig:drift-tube}
\end{figure}

Drift tubes are probably the simplest type of gas-based particle detectors that use the drift velocity
of the electrons to reconstruct the ionising particles location. They consist of an anode wire
inside a gaseous volume that is encased by a cathode. If one knows the time of the ionisation and the
drift velocities in the volume, one can reconstruct the distance of the track from the wire.
Drift tubes don't have to be cylindrical, as is shown in figure~\ref{fig:drift-tube}, which shows a drift tube
of the muon system of the CMS\footnote{Compact Muon Solenoid} detector at the LHC\footnote{Large Hadron Collider}.

\subsection{Time projection chambers}

\begin{figure}
    \centering
    \includegraphics[width=0.99\textwidth]{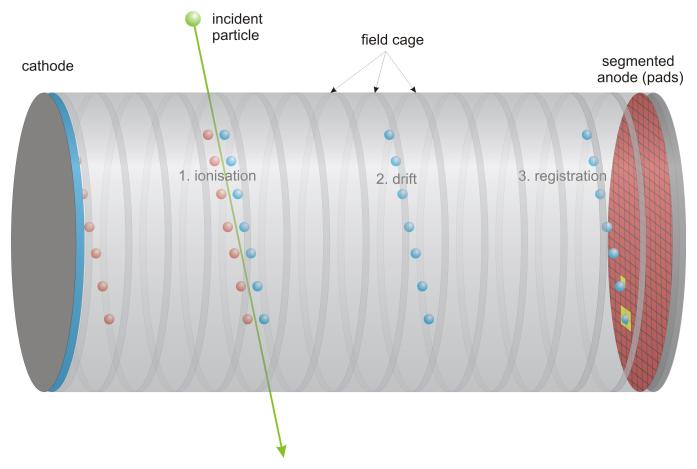}
    \caption[Principle of operation of time projection chambers]
            {Principle of operation of time projection chambers \cite{SCHAEFER2013}}
    \label{fig:TPC}
\end{figure}

Time projection chambers consist of a large gaseous volume in which one creates a homogeneous electric field
(see figure \ref{fig:TPC}). Ionisation tracks are transported to the anode, where a two dimensional detector records
the position as well as the time of the signal. Since the electric field in the chamber is homogeneous,
the drift velocity is also constant and one can easily reconstruct the full three-dimensional coordinates of a track.

If one also applies a magnetic field to the gas volume, one can use the track curvature to measure the particle's
momentum. Together with the $\nicefrac{\dd E}{\dd x}$ information from the amount of ionisation, this can be used to
identify the ionising particle.

\begin{figure}
    \centering
    \subfigure[Exploded view of the ND280 detector and position of the TPCs]{
        \includegraphics[width=\plotwidth]{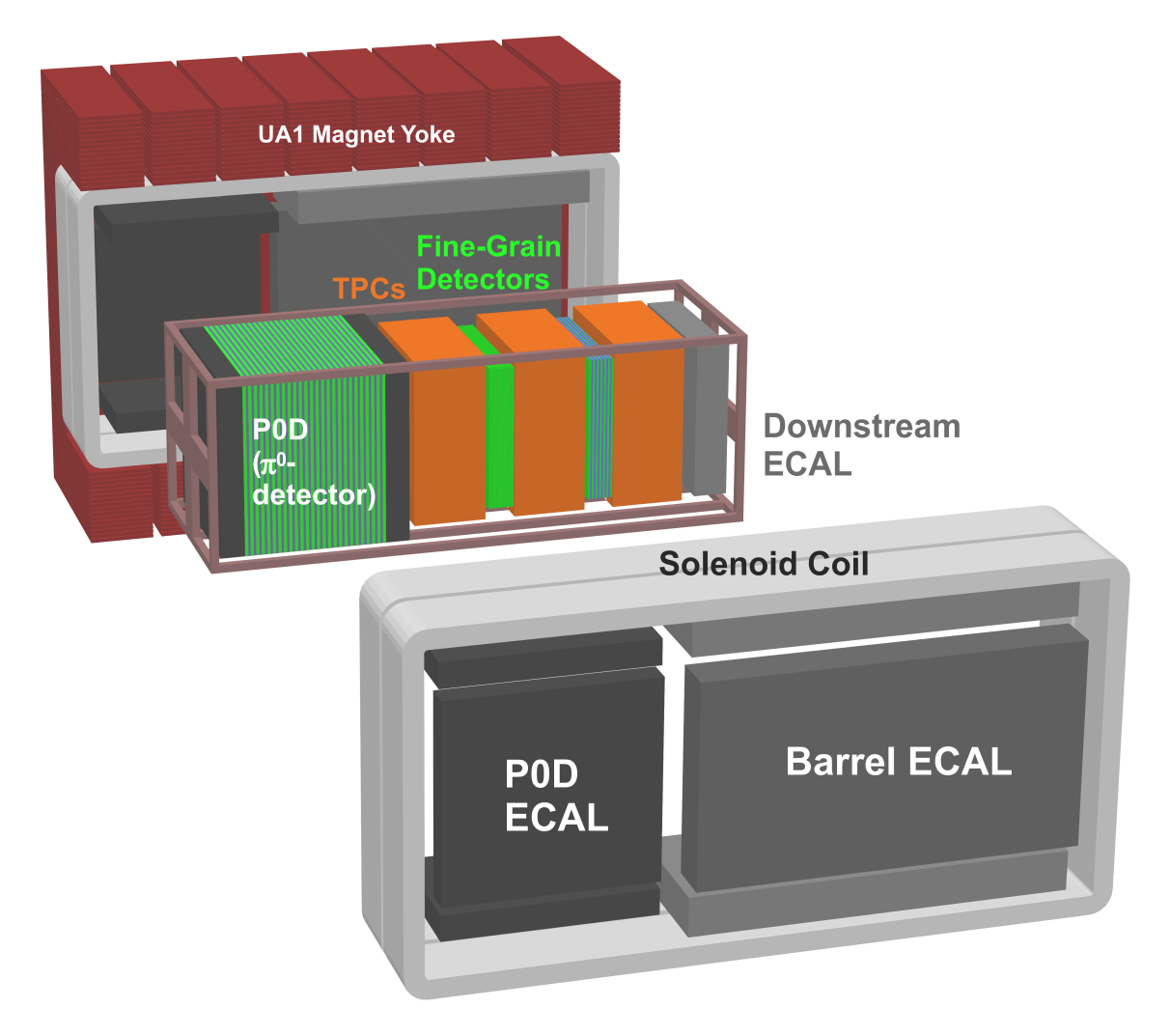}
    }
    \hfill
    \subfigure[One of the TPCs during assembly]{
        \includegraphics[width=\plotwidth]{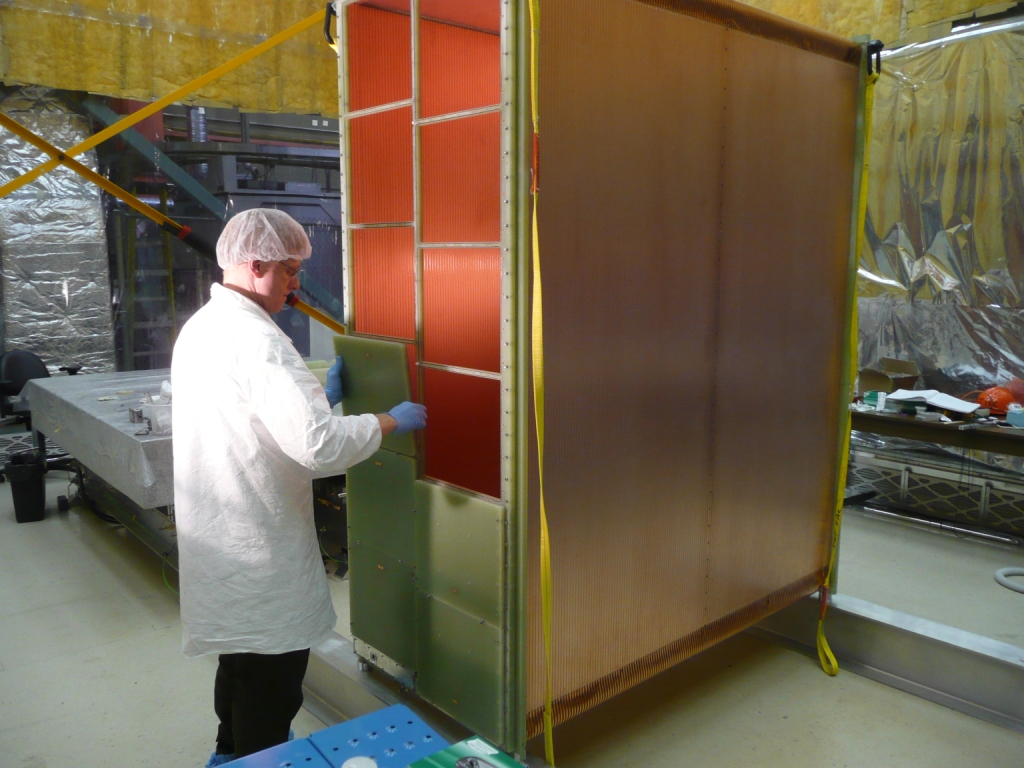}
    }
    \caption[The T2K/ND280 TPCs]
            {The T2K/ND280 TPCs \cite{T2KORG}}
    \label{fig:ND280}
\end{figure}

One experiment that uses these capabilities is the ND280\footnote{Near Detector \unit[280]{m}} near detector
of the long baseline neutrino experiment T2K\footnote{Tokai To Kamioka}.
Figure \ref{fig:ND280} shows the whole detector and one of the three TPCs. Together the TPCs account for a
sensitive volume of $\sim \unit[8.5]{m^3}$, which is read out with $\sim 124\,000$ channels. This provides a
space point resolution of about \unit[0.7]{mm} \cite{Abgrall201125}. If one were to instrument the whole volume instead
of just the anode sides (as would be necessary for a silicon pixel detector), one would need orders of magnitude
more read-out channels.

\chapter{Setup}

\section{The drift chambers}

\begin{figure}
    \centering
    \subfigure{
        \includegraphics[width=0.47\textwidth]{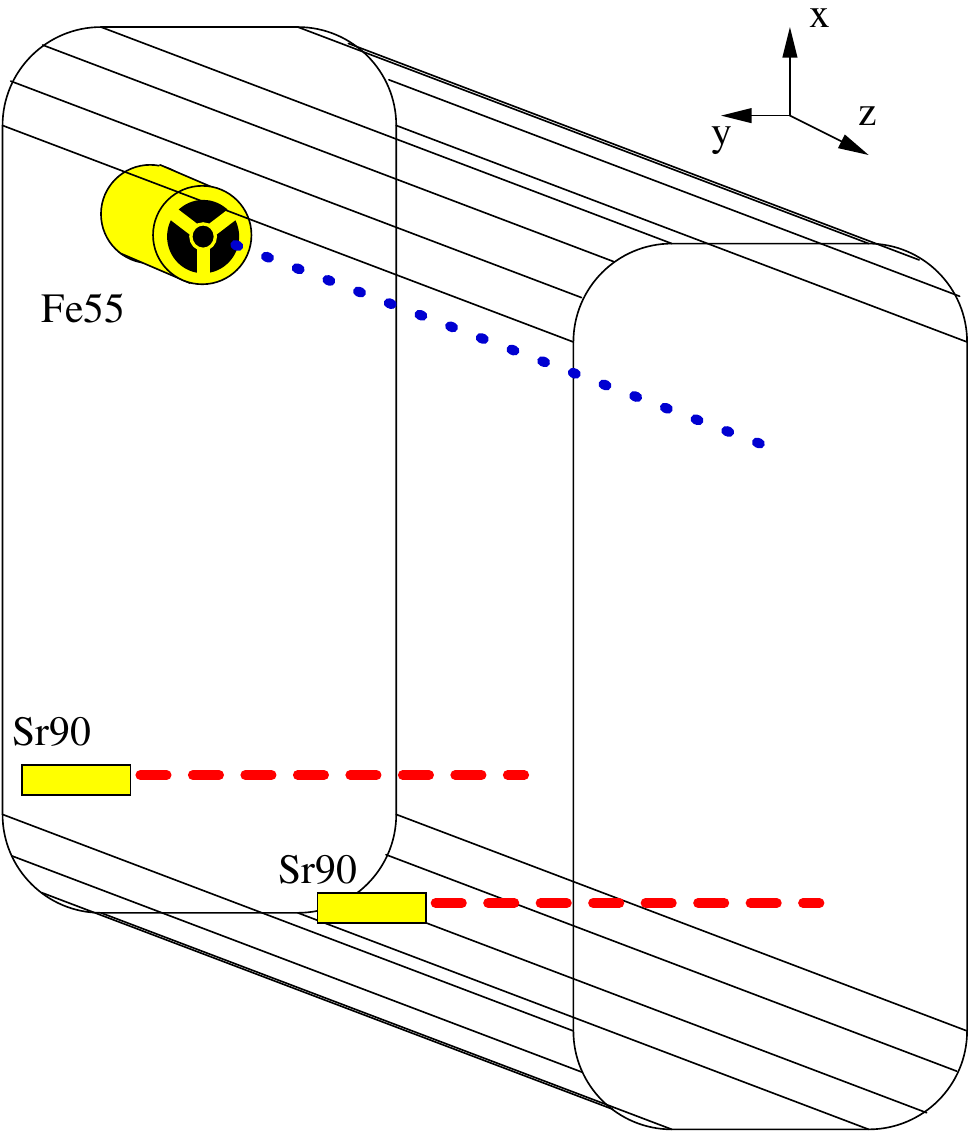}
    }
    \hfill
    \subfigure{
        \includegraphics[width=0.47\textwidth]{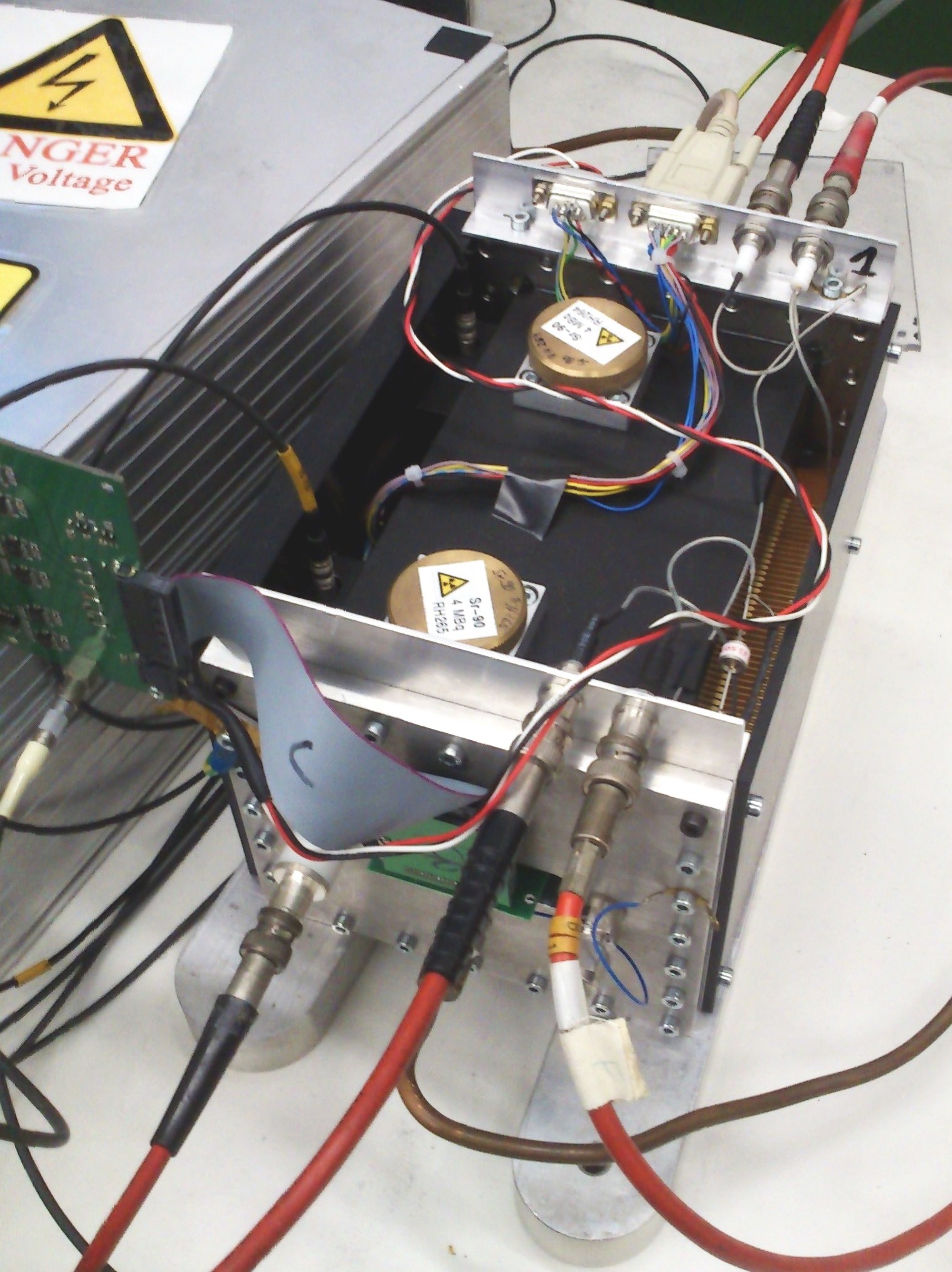}
    }
    \caption[Schematic view and picture of the used drift chambers]
            {Schematic view (left) \cite{TERHORST2008} and picture (right) of one of the monitoring chambers.
             One can see the \isotope[90]{Sr} sources mounted on top of the chamber.
             An electric field is applied in the $z$-direction inside
             the chamber, perpendicular to the trajectory of the ionising electrons.}
    \label{fig:Chamber}
\end{figure}

We measure the electron drift velocity using T2K/ND280-type monitoring
chambers \cite{TERHORST2008,STEINMANN2010,WROBEL2011}. They are specifically
designed for the measurement of the electron drift velocity and feature a gas
volume with a field cage of \unit[14.8]{cm} length. One of the chambers can be seen
in figure~\ref{fig:Chamber}. For a $v_{d}$ measurement, electrons of $\sim\unit[0.5]{MeV}$ are emitted
into the chamber by a \isotope[90]{Sr}~source. These electrons traverse the chamber
and ionise the gas along their trajectory. At the other end of the chamber
the electrons hit a scintillator and trigger the measurement. An electric field
along the $z$-axis transports the ionisation electrons towards a
MicroMeGaS\footnote{Micro Mesh Gaseous Structure} at the end of the chamber, where the
signal is amplified and recorded so one can determine the drift time of the electrons.

\begin{figure}
    \centering
    \subfigure{
        \includegraphics[width=0.47\textwidth]{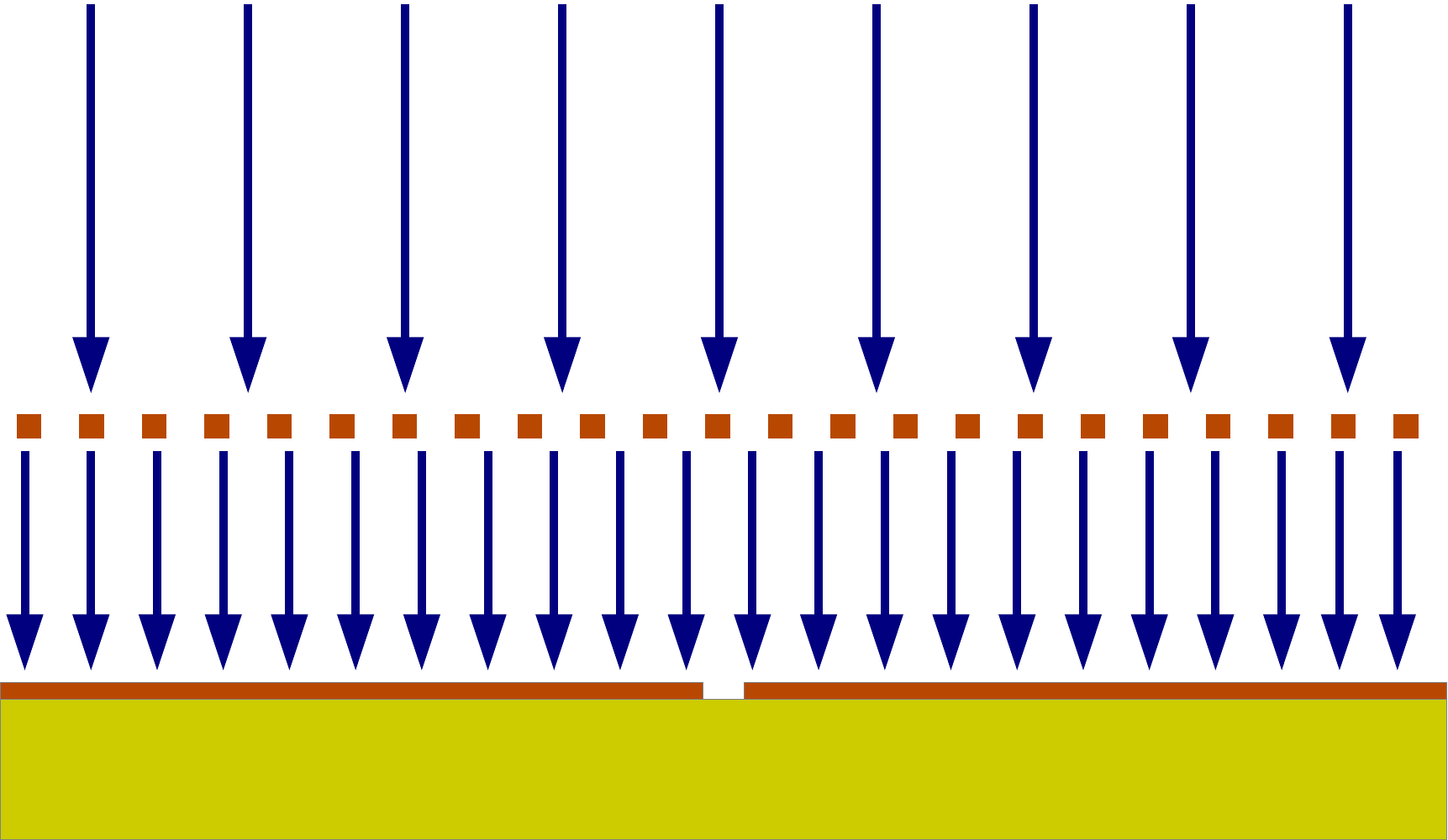}
    }
    \hfill
    \subfigure{
        \includegraphics[width=0.47\textwidth]{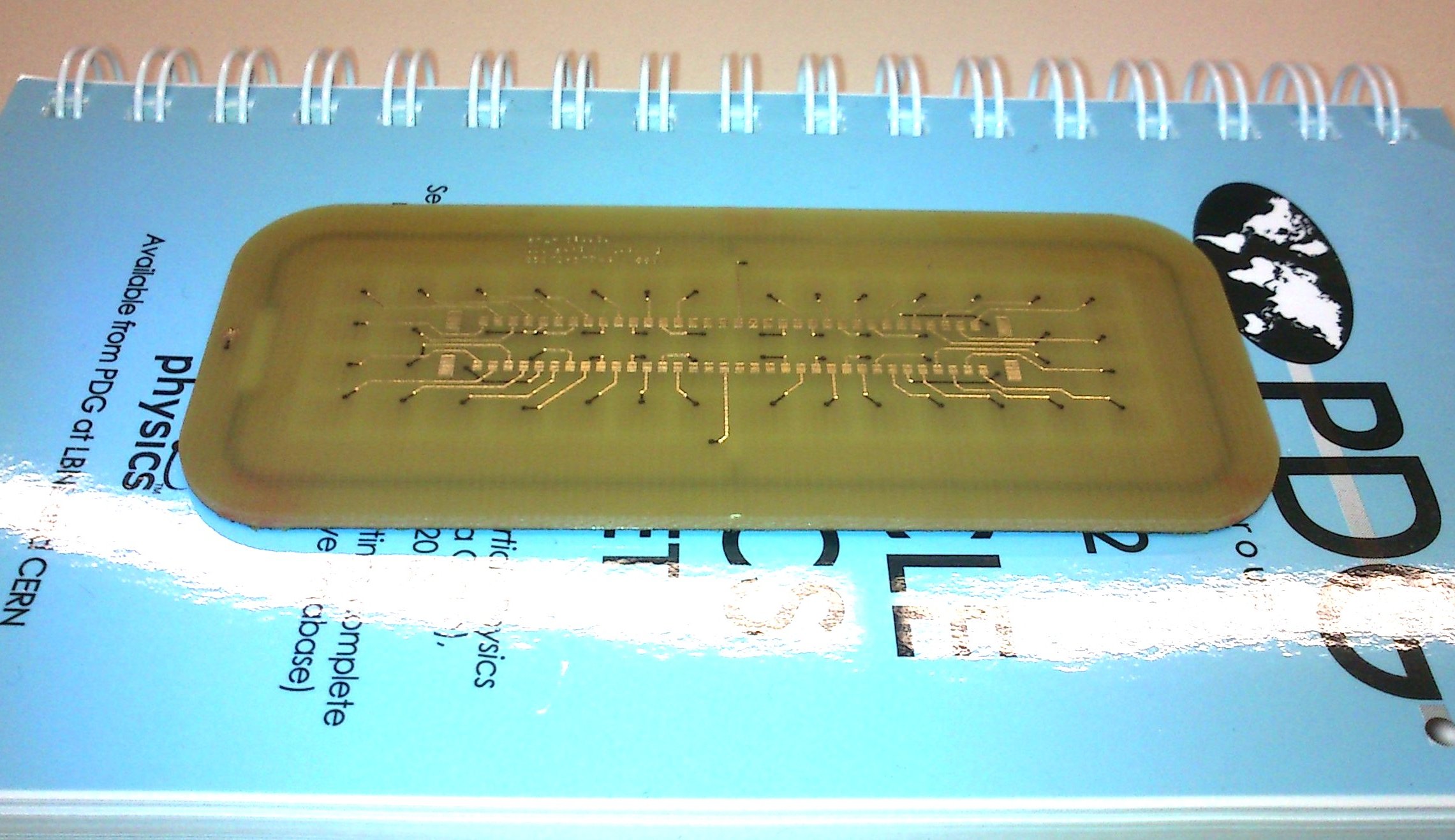}
    }
    \caption[Schematic view and picture of a MicroMeGaS]
            {Schematic view (left) and picture (right) of a MicroMeGaS as used in this study.
             The arrows symbolise the applied electric field.
             The picture shows the back side of the PCB with the connectors facing the camera.
             One can see the anode pads shining through the PCB.}
    \label{fig:MicroMeGaS}
\end{figure}

MicroMeGaS consist of a printed circuit board (PCB) with anode pads
and a fine wire mesh, suspended $\sim\unit[100]{\micro{m}}$ above
those pads (figure~\ref{fig:MicroMeGaS}). If one applies a voltage
of about \unit[300]{V} -- \unit[500]{V} between mesh and pads, the
electric field will be strong enough to achieve gas amplification
of the electrons drifting into the structure.

There are two \isotope[90]{Sr} sources at different $z$-positions to minimise systematic errors due to
effects of the amplification region.
The drift time at both $z$-positions is measured and the drift velocity
\[ v_d = \frac{\Delta z}{\Delta t} = \frac{z_1 - z_2}{t_1 - t_2} \]
is then calculated only from the drift path between the two \isotope[90]{Sr} positions. This way, systematic offsets on
the time measurement (electronic delays, field inhomogeneities near the walls of the chamber, etc.)
cancel out, since they are the same for both measurements.

The ND280 monitoring chambers have a $\Delta z$ of \unit[120.4]{mm}, which in combination with the used
DAQ\footnote{Data acquisition} setup allows $v_d$ measurements between about \unitfrac[20]{\micro{m}}{ns} and \unitfrac[100]{\micro{m}}{ns}.
The chambers also feature a slot for the insertion of an \isotope[55]{Fe}
gamma source for the measurement of the gain at the MicroMeGaS, but this is not used in this study.

\section{DAQ}

\begin{figure}
    \centering
    \includegraphics[width=0.99\textwidth]{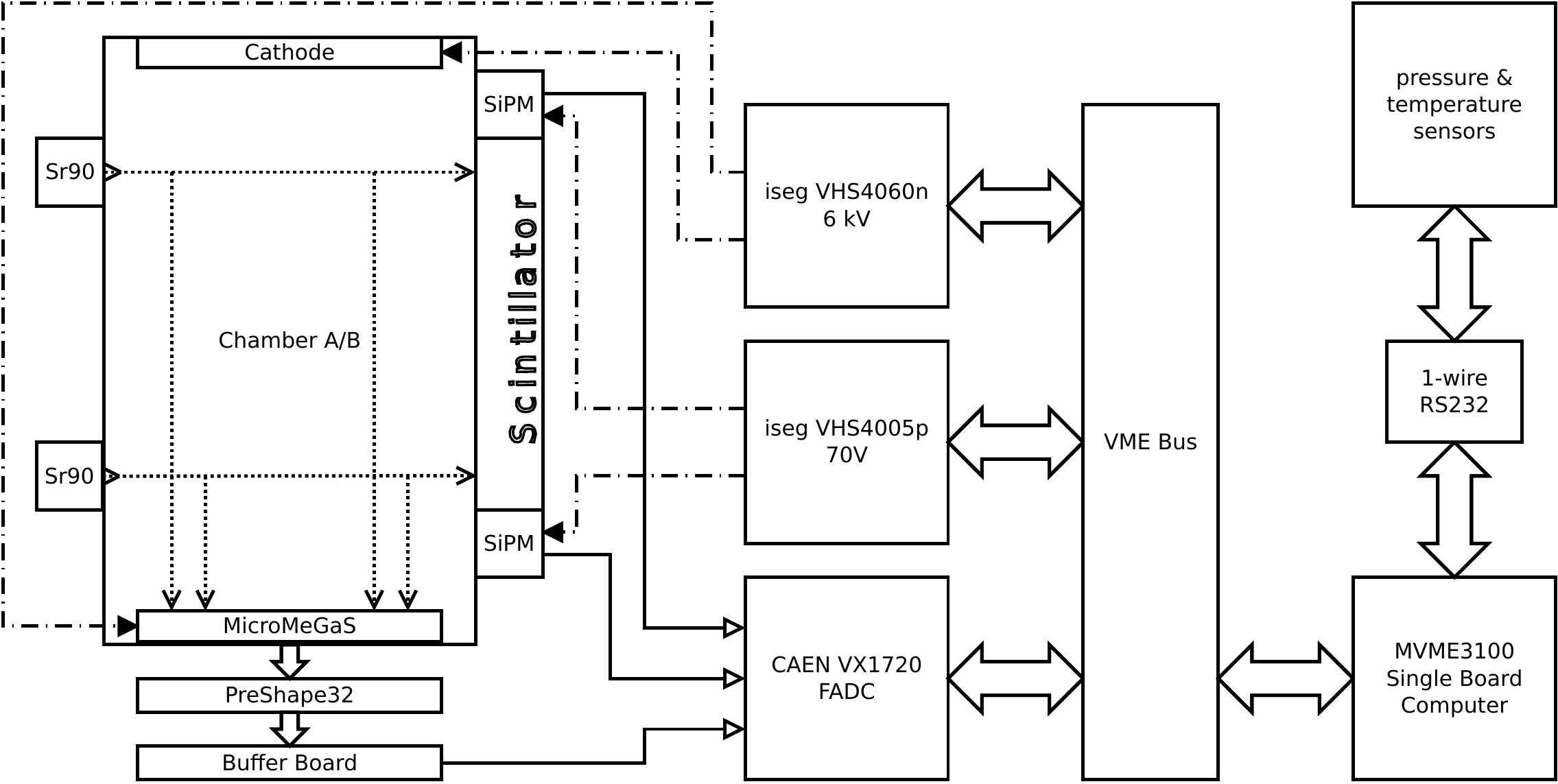}
    \caption{DAQ setup}
    \label{fig:DAQ}
\end{figure}

The basic DAQ setup can be seen in figure~\ref{fig:DAQ}. It is built around an MVME3100 
single board computer and a CAEN VX1720 Flash ADC\footnote{Analog to Digital Converter}.
The ADC uses a sampling rate of \unit[250]{MHz} and is triggered by the coincidence of
the signals of two SiPMs\footnote{Silicon Photo Multiplier} connected to a scintillating fibre.
Using the coincidence signal of two SiPMs rather than just one, suppresses the noise
always present at a single SiPM.

\begin{figure}
    \centering
    \includegraphics[width=0.99\textwidth]{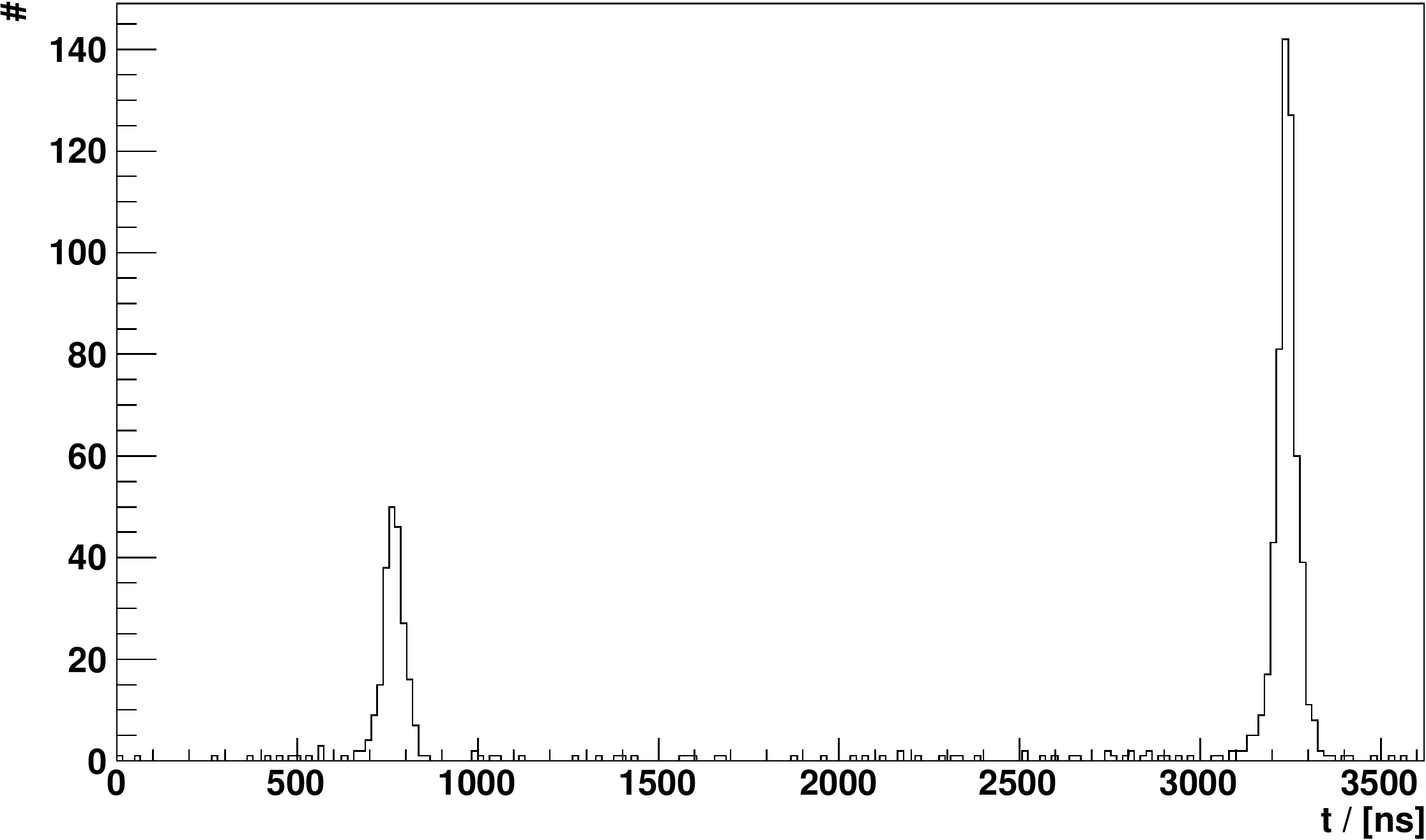}
    \caption[Histogram of measured drift times]
            {Histogram of measured drift times.
             The two peaks correspond to the two $z$-positions of the tracks.
             The different peak heights can be explained by a difference of activity of the two \isotope[90]{Sr}~sources
             or a slight misalignment of source, slits and scintillator for one of the sources.}
    \label{fig:t_hist}
\end{figure}

This setup cannot distinguish between the two $z$-positions. So to measure the drift velocity,
a lot of time measurements are taken and the values are put into a histogram like the one shown
in figure~\ref{fig:t_hist}. The two different $z$-positions show themselves as two peaks
in the $t$~distribution. A fit to these peaks allows the determination of the mean
values of $t_1$ and $t_2$ and thus $\Delta t$. About 2000 time measurements make up
one $v_d$~measurement. This takes about one minute, during which we can assume the 
temperature and the pressure of the drift gas to be stable.

\begin{figure}
    \centering
    \includegraphics[width=0.99\textwidth]{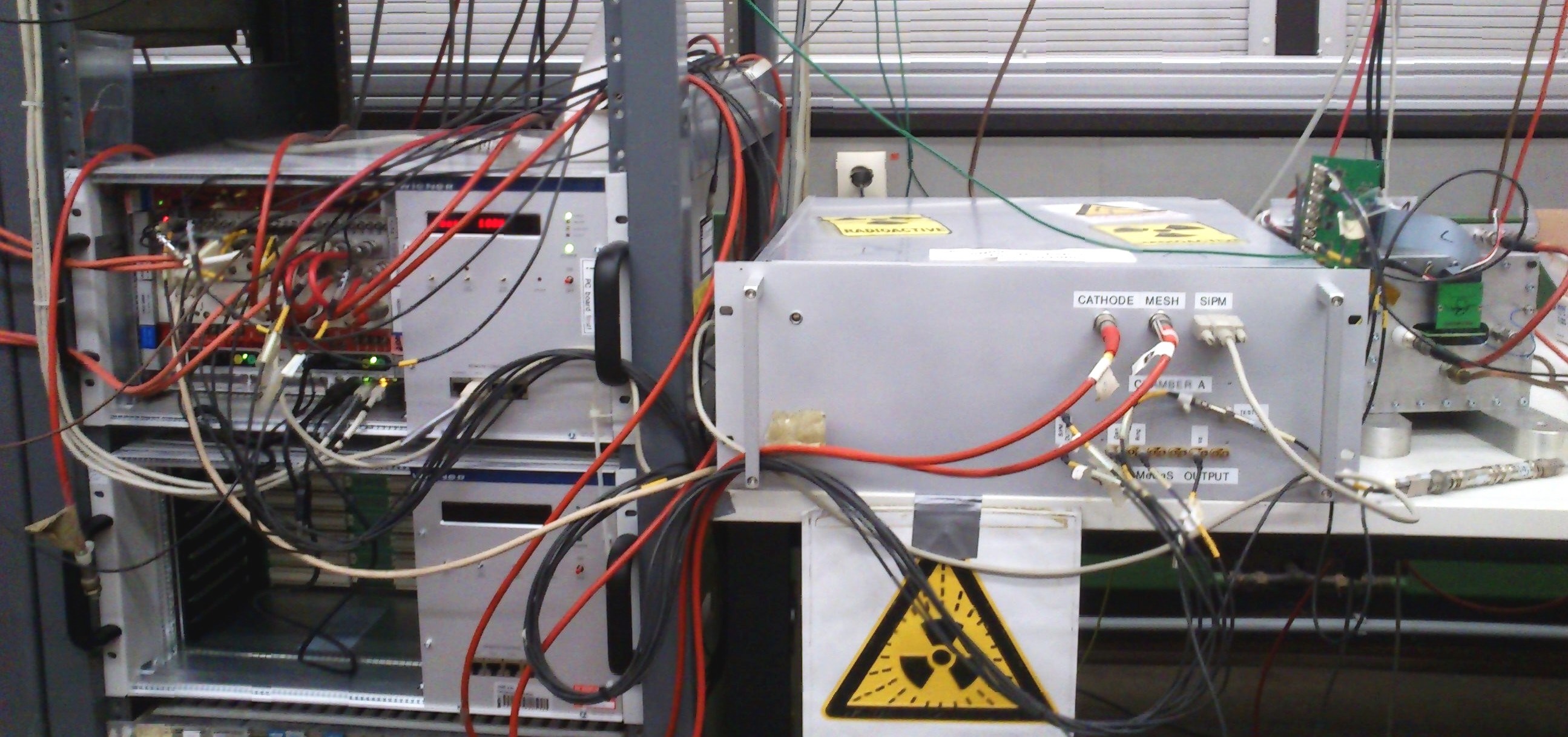}
    \caption{Picture of the experimental setup}
    \label{fig:setup}
\end{figure}

Pressure and temperature are measured at multiple points in and around the chambers.
All sensors are connected to a 1-wire network, as those measurements are not time critical.
A programmable \unit[6]{kV} power supply creates the drift and amplification fields.
A second power supply feeds the SiPMs with about $\unit[70]{V}$. A picture of the whole
DAQ setup can be seen in figure~\ref{fig:setup}.

\section{Gas mixing}

The gases analyzed in this work were mixed using the UGMA\footnote{Universelle Gas-Misch-Anlage}
gas mixing station of \foreignlanguage{ngerman}{III.~Physikalisches Institut~B} at RWTH Aachen University. It was operated in partial
pressure mode, which allows the mixing of up to two gas additives to one base gas, with volume
fraction errors below \unit[0.1]{vol.-\%} \cite{STEINMANN2013a}.

Gas mixtures are commonly quoted as volume fractions $\eta_V$, e.g. \enquote{\binmix{Ar}{95}{CH4}{5}}.
For ideal gases, these values unambiguously describe the mixtures since for ideal gas mixtures
volume fractions, partial pressure fractions and mole fractions are equal at all temperatures and pressures:
\[
    \eta_V = \eta_p = \eta_n = const,\quad\forall\,T,p
\]

For real gases this does not hold anymore. For a given mixture, e.g. in a pressure vessel,
only the mole fraction stays constant over all temperatures and pressures, since we do not add or
remove any molecules from the mix. The volume and pressure fraction of the constituents, on the other
hand, become dependent of the temperature and pressure in the vessel:
\begin{gather*}
    \eta_n = const,\quad\forall\,T,p \\
    \eta_V = \eta_V(T, p) \\
    \eta_p = \eta_p(T, p)
\end{gather*}

This means that the given volume fraction of the mix is actually only valid for certain values of $T$
and $p$. If one uses the gas mixture in other conditions, the mole fraction will stay the same,
but the volume fractions will be different.
In this work, we will quote gas mixtures as volume fractions at \unit[1013.25]{mbar} and \unit[0]{\celsius},
in accordance to DIN~1343~\cite{DIN1343}, which regulates the standard volumes used by gas distributors in Germany.

The UGMA is able to regulate both the gas flow and the pressure inside the drift chambers.
We chose a supply flow of $\unitfrac[5]{l}{h}$ and a pressure of \unit[1000]{mbar}. The flow ensures about five volume
exchanges per hour and the slight overpressure is intended to keep the contamination with atmosphere to a minimum.
This works very well and we achieve a water contamination of less than \unit[5]{ppm} and oxygen levels below
\unit[1]{ppm}.

In cases where the atmospheric pressure surpasses \unit[1000]{mbar}, the chambers are still operated at
slight overpressure, since the gas flow creates a pressure gradient in the pipes leading from the chambers
to the UGMA. This pressure drop is in the order of a few mbar and keeps the chamber pressure above the atmospheric
pressure even if the pressure regulation in the UGMA is fully opened (as is the case when the nominal pressure
is below atmospheric pressure).
This is considered in the calculations, as the pressure measurements used for the $v_d$~analysis happen directly at the chambers.

As a security feature, a maximum differential pressure of \unit[20]{mbar} was established. When the ambient pressure
falls below \unit[980]{mbar}, the nominal chamber pressure is lowered accordingly to prevent damaging the
chambers.

\section{Simulation software}

There exists no comprehensive analytical formula to predict the electron drift velocity in
arbitrary gas mixtures. To get a theoretical prediction of the drift properties, we therefore
use the numerical simulation program Magboltz\footnote{Version 9.0.1, \url{http://consult.cern.ch/writeup/magboltz/}}
embedded in the framework of Garfield++\footnote{Revision 309, \url{http://garfieldpp.web.cern.ch/garfieldpp/}}.

Garfield++ and Magboltz use mole fractions to describe the simulated gas mixtures.
Since we use volume fractions to describe the gas mixtures, those have to be converted before the simulation.
To do so, the molecular masses and densities at standard conditions (as given in appendix~\ref{chap:gasprop}) are used.
\[
    \eta_{n,i} = \frac{ \eta_{V,i} \cdot \frac{ \rho_{0,i} }
                                        { M_{mol,i} }
                   }
                   { \sum\limits_{j} \eta_{V,j} \cdot \frac{ \rho_{0,j} }
                                                        { M_{mol,j} }
                   }
\]

The difference between mole and volume fractions is small for noble gases and small molecules, but can
become important for large-molecule carbon hydrates. As an example, the molar fractions
of \termix{Ar}{93}{CH4}{5}{CO2}{2} are
\begin{align*}
    \eta_{n,\ce{Ar}} &= \unit[92.98]{\%}, \\
    \eta_{n,\ce{CH4}} &= \unit[5.01]{\%}, \\
    \eta_{n,\ce{CO2}} &= \unit[2.01]{\%},
\end{align*}
while the fractions of \termix{Ar}{95}{CF4}{3}{iC4H10}{2} are
\begin{align*}
    \eta_{n,\ce{Ar}} &= \unit[94.92]{\%}, \\
    \eta_{n,\ce{CF4}} &= \unit[3.01]{\%}, \\
    \eta_{n,\ce{iC4H10}} &= \unit[2.07]{\%}
\end{align*}
and show a much larger deviation from the volume fractions due to the isobutane admixture.

Magboltz can simulate the drift velocity and the three-dimensional diffusion along the track
as well as the Townsend coefficient (excluding the Penning and Jesse effects)
for arbitrary $E$- and $B$-fields. It is written in Fortran, but Garfield++ provides an easy to use
C++~API. This API was used to build a program, called Magsim, that runs on our Condor%
\footnote{\url{http://research.cs.wisc.edu/htcondor/}} computing cluster. It allows the automated and fast
simulation of gas mixtures with up to 5 gas additives.

During the simulations we noticed that Magboltz overestimates the statistical errors on
its data points. This has been fixed in the newest versions of Magboltz, which could not yet be
included into Garfield++ and Magsim.
But since this bug has no influence on the position of the data points themselves, it is irrelevant for
the plots presented in this work.

\section{Calibration and systematic errors}

The drift velocity $v_d = \nicefrac{\Delta z}{\Delta t}$ does not only depend on $E$, but also on
the gas density $n$ and thus the pressure $p$ and the temperature $T$.
Therefore, a handle on the systematic errors of all these variables is needed.

\subsection{Drift length}

The drift length $\Delta z$ is the distance between the two \isotope[90]{Sr} electron tracks. The position
of these tracks is given by slits in the casing of the chambers and the $z$-distance between those slits
(centre to centre) is nominally \unit[120.4]{mm}. The main uncertainty of the drift length stems from the width
of the slits, which is \unit[1.2]{mm}.
\[
    \frac{\sigma_{\Delta z}}{\Delta z} = \frac{\nicefrac{2\,\cdot\,\unit[0.6]{mm}}{\sqrt{12}}}{\unit[120.4]{mm}} \lesssim \unit[3]{\perthousand}
\]

\subsection{Timing}

The drift time $\Delta t$ is the difference between the drift time measurements of the two
\isotope[90]{Sr} electron tracks. To check the accuracy of those time measurements, a digital signal generator
was connected to the inputs of the Flash ADC. The DAQ setup was then fed with
pulses of defined intervals and the output of the setup was compared with the settings of the signal
generator. This way the whole hardware and software chain that produces our final time measurement could be tested.
\[
    \frac{\sigma_{\Delta t}}{\Delta t} \lesssim \unit[2]{\perthousand}
\]

\subsection{Electric field}

The electric field is created by a voltage $U_f$ which is applied over an electric field cage
of length $l_f$. The electric field and its error are thus:
\begin{align*}
    E &= \frac{U_f}{l_f} \\
    \frac{\sigma_E}{E} &= \sqrt{ \left( \frac{\sigma_{U_f}}{U_f} \right)^2 +
                                \left( \frac{\sigma_{l_f}}{l_f} \right)^2 }
\end{align*}

\begin{figure}
    \centering
    \includegraphics[width=0.99\textwidth]{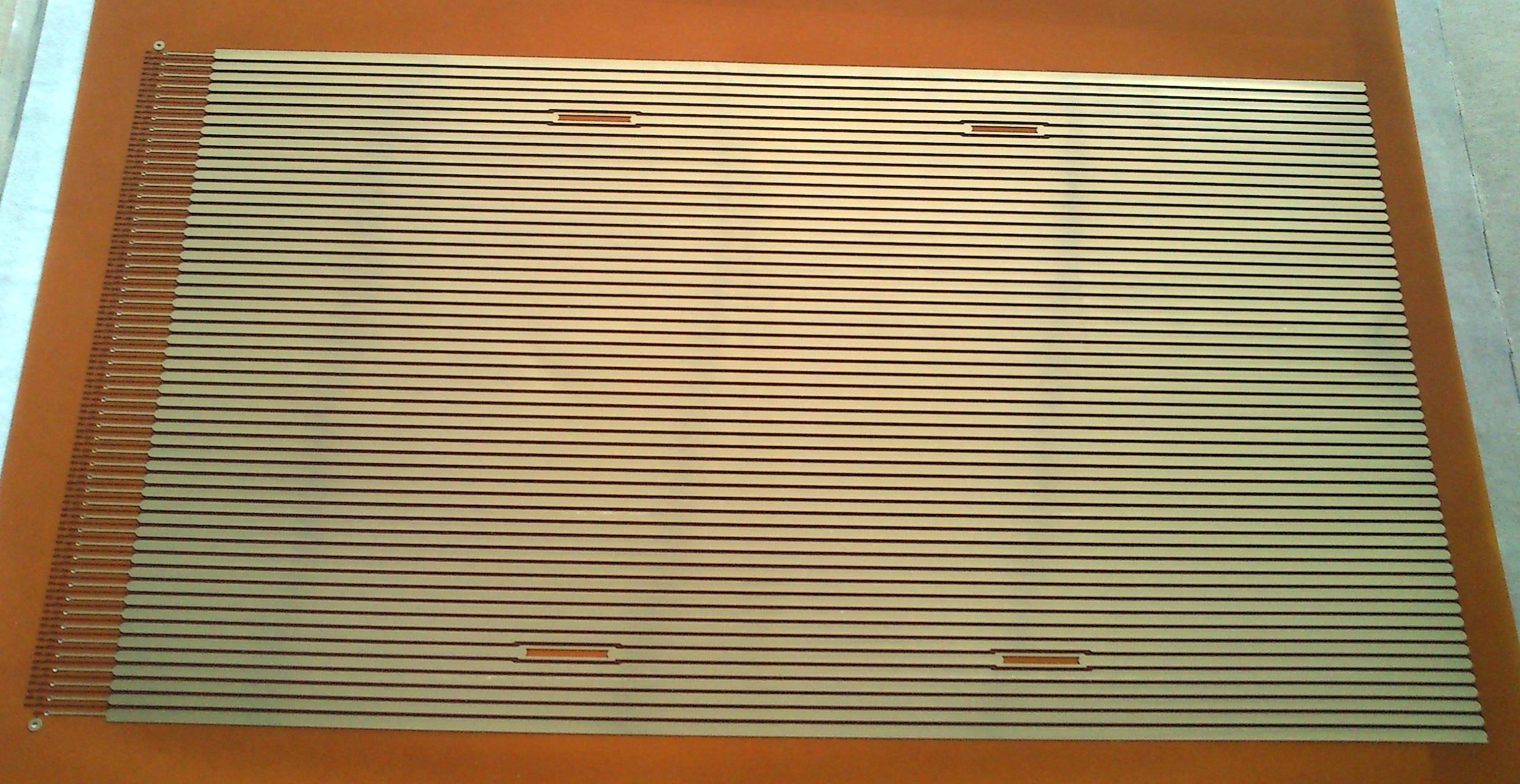}
    \caption[Field strips of unassembled field cage]
            {Field strips of unassembled field cage}
    \label{fig:fieldstrips}
\end{figure}

Since the actual field cage dimensions cannot be measured in its assembled state, the distance between
the centres of the first and the last field strip of an unassembled field cage were measured
(figure \ref{fig:fieldstrips}).
\begin{align*}
    l_f &= \left(148.0 \pm 0.5 \right) \unit{mm} \\
    \frac{\sigma_{l_f}}{l_f} &\lesssim \unit[3]{\perthousand} \\
\end{align*}

The accuracy of the voltage measurements was estimated by measuring
the applied voltages with a Keithley source meter and comparing these values to the
setpoints and measurements of the programmable power supply. The values were found to be in very good agreement with each other.
\begin{align*}
    \frac{\sigma_{U_f}}{U_f} &\lesssim \unit[0.5]{\perthousand}
\end{align*}

\subsection{Temperature}

\begin{figure}
    \centering
    \includegraphics[width=\plotwidth]{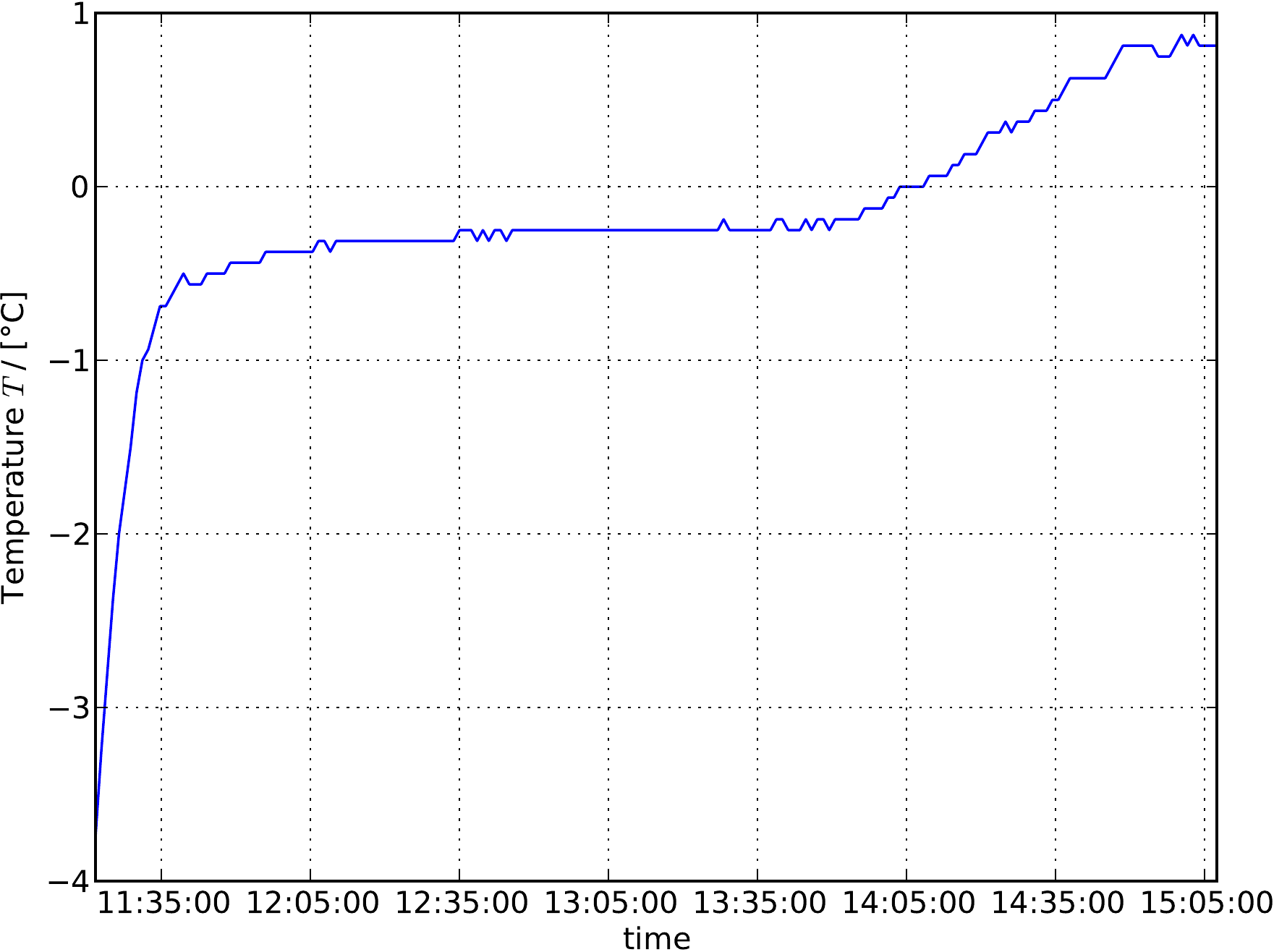}
    \caption{Temperature curve of a DS18S20 temperature sensor in melting ice}
    \label{fig:T-cal}
\end{figure}

The temperature logging uses DS18S20 digital temperature sensors in the gas flow right before and after the drift chambers,
as well as outside the chambers for ambient measurements.
These sensors have a guaranteed precision of \unit[0.5]{K} \cite{DS18S20}. To verify this, one of the sensors was frozen
in water and the temperature curve during the melting process recorded. Due to latent heat, one expects
a plateau at $\unit[0]{\celsius}$. Figure \ref{fig:T-cal} shows such a plateau at around $\unit[-0.2]{\celsius}$, which is well within
the specified limits. Since all our measurements are done at room temperature, a temperature of $\unit[25]{\celsius}$
or $\unit[298]{K}$ can be used for a simple estimate of the relative error of the temperature measurements.
\[
    \frac{\sigma_T}{T} = \frac{\unit[0.5]{K}}{\unit[298]{K}} \lesssim \unit[2]{\perthousand}
\]

\subsection{Pressure}

\begin{figure}
    \centering
    \subfigure[{\unit[1006.5]{mbar}}]{
        \includegraphics[width=0.47\textwidth]{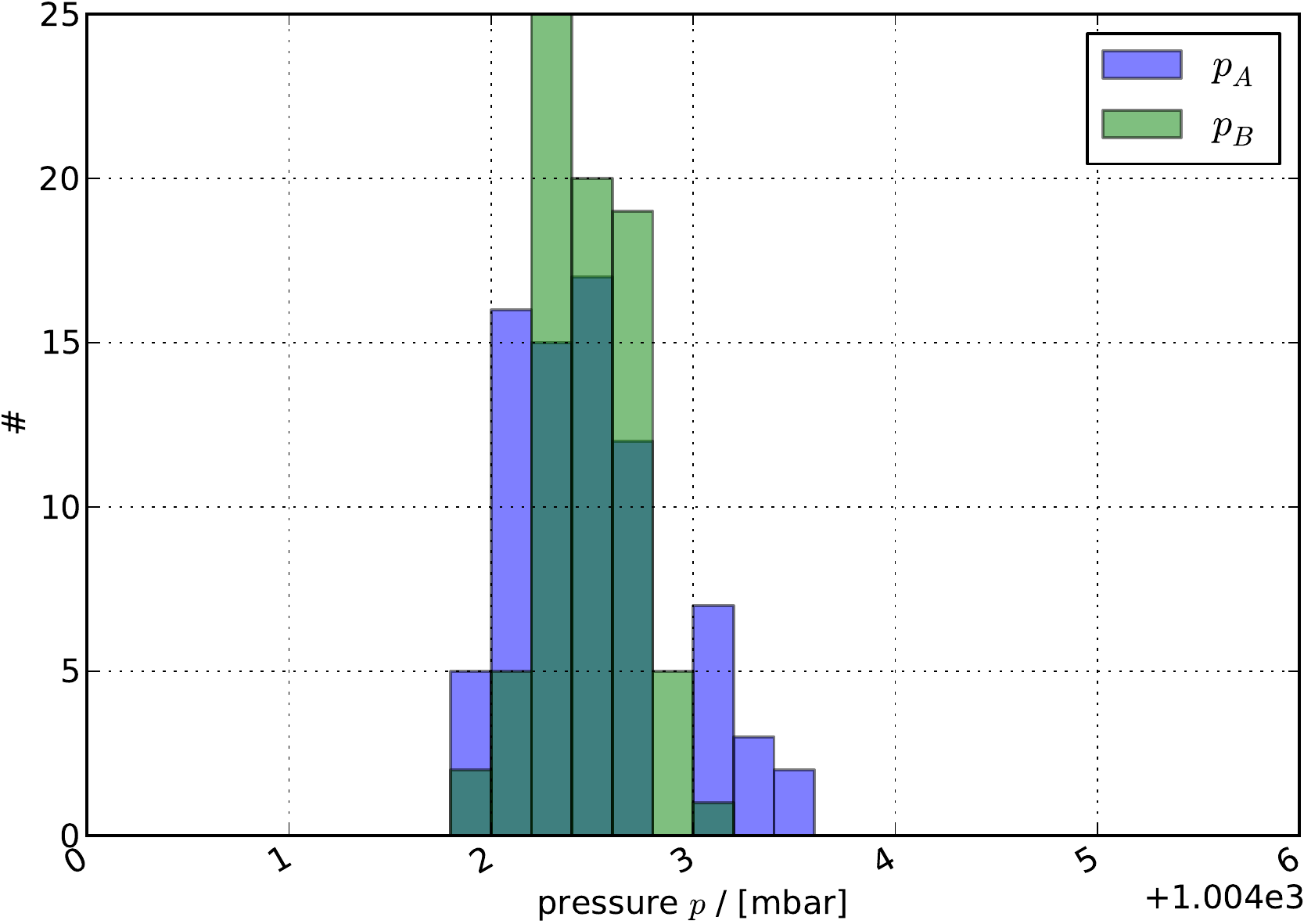}
        \label{fig:p-cal-1006}
    }
    \hfill
    \subfigure[{\unit[975.55]{mbar}}]{
        \includegraphics[width=0.47\textwidth]{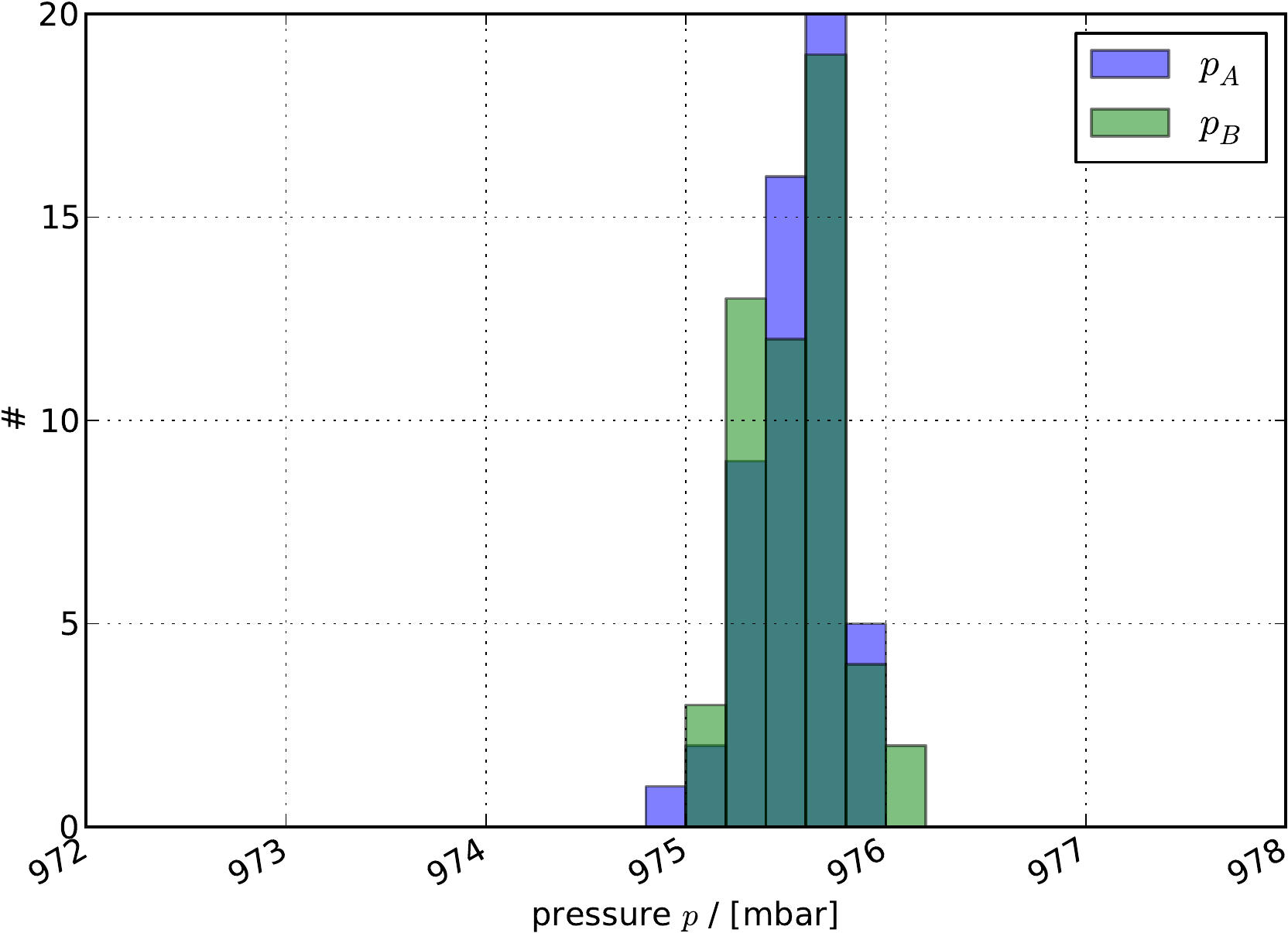}
        \label{fig:p-cal-975}
    }
    \caption[Pressure distribution at calibration points]
            {Pressure distribution at calibration points}
    \label{fig:p-cal}
\end{figure}

The pressure in the chambers is measured with pressure transmitters manufactured by WIKA, which are read out by
DS2450 ADCs. These sensors show a linear relation between pressure and signal.
For the calibration of the sensors, they were put on atmospheric pressure and their signals were compared
with the measurements of a calibrated Baratron manometer next door in the same building.
This was done at two different days with ambient pressures of \unit[1006.5]{mbar} and \unit[975.55]{mbar}
respectively. From these two data points a linear calibration for the pressure sensors was calculated.
The pressure distribution at the calibration points after the calibration can be seen in figure
\ref{fig:p-cal}.

\begin{figure}
    \centering
    \includegraphics[width=\plotwidth]{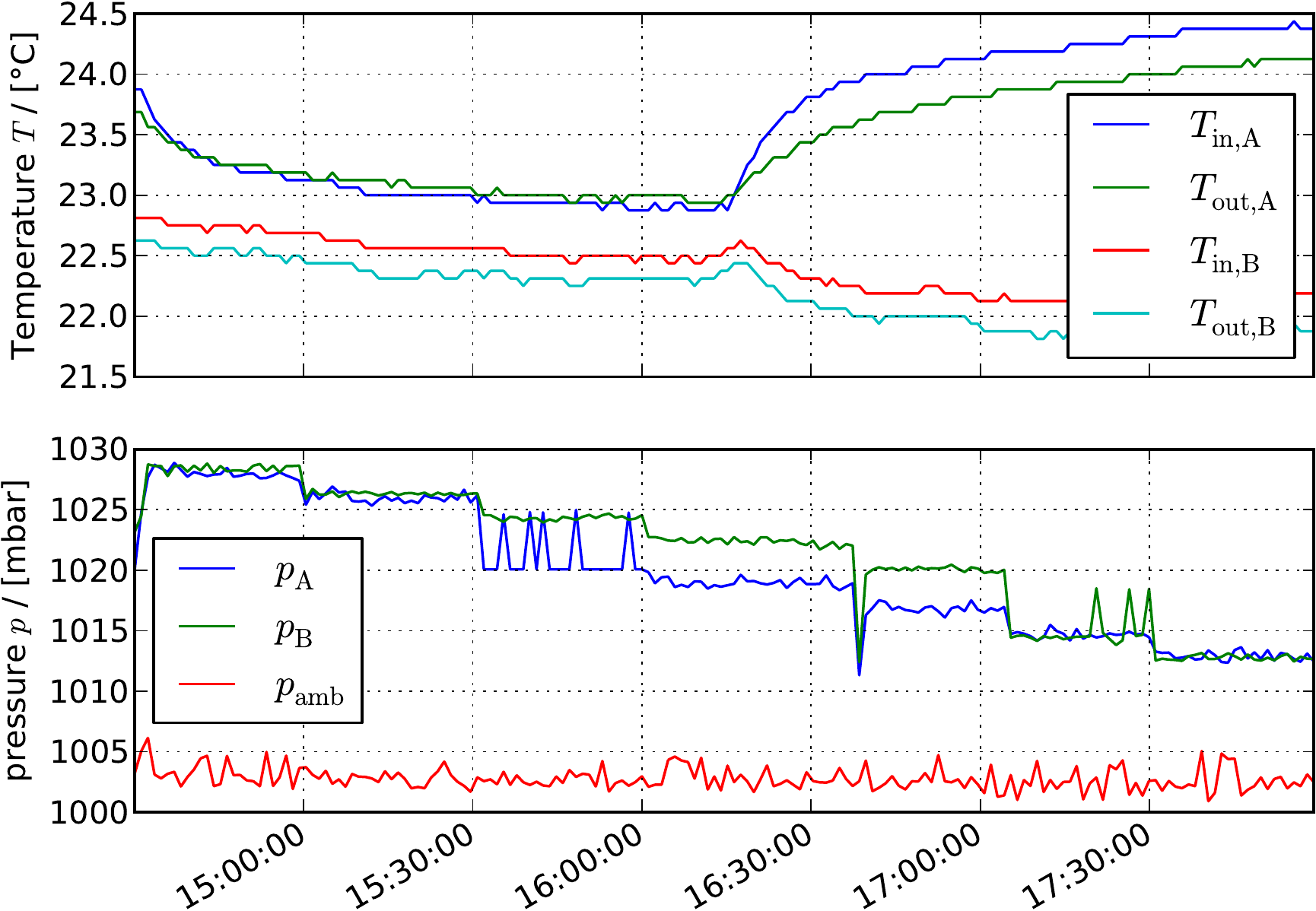}
    \caption[Jump in pressure signals due to missing codes in the ADCs]
            {Jump in pressure signals due to missing codes in the ADCs.
             The Temperatures were measured at the in- and outlets of the two chambers.}
    \label{fig:p-jump}
\end{figure}

\begin{figure}
    \centering
    \includegraphics[width=0.75\textwidth]{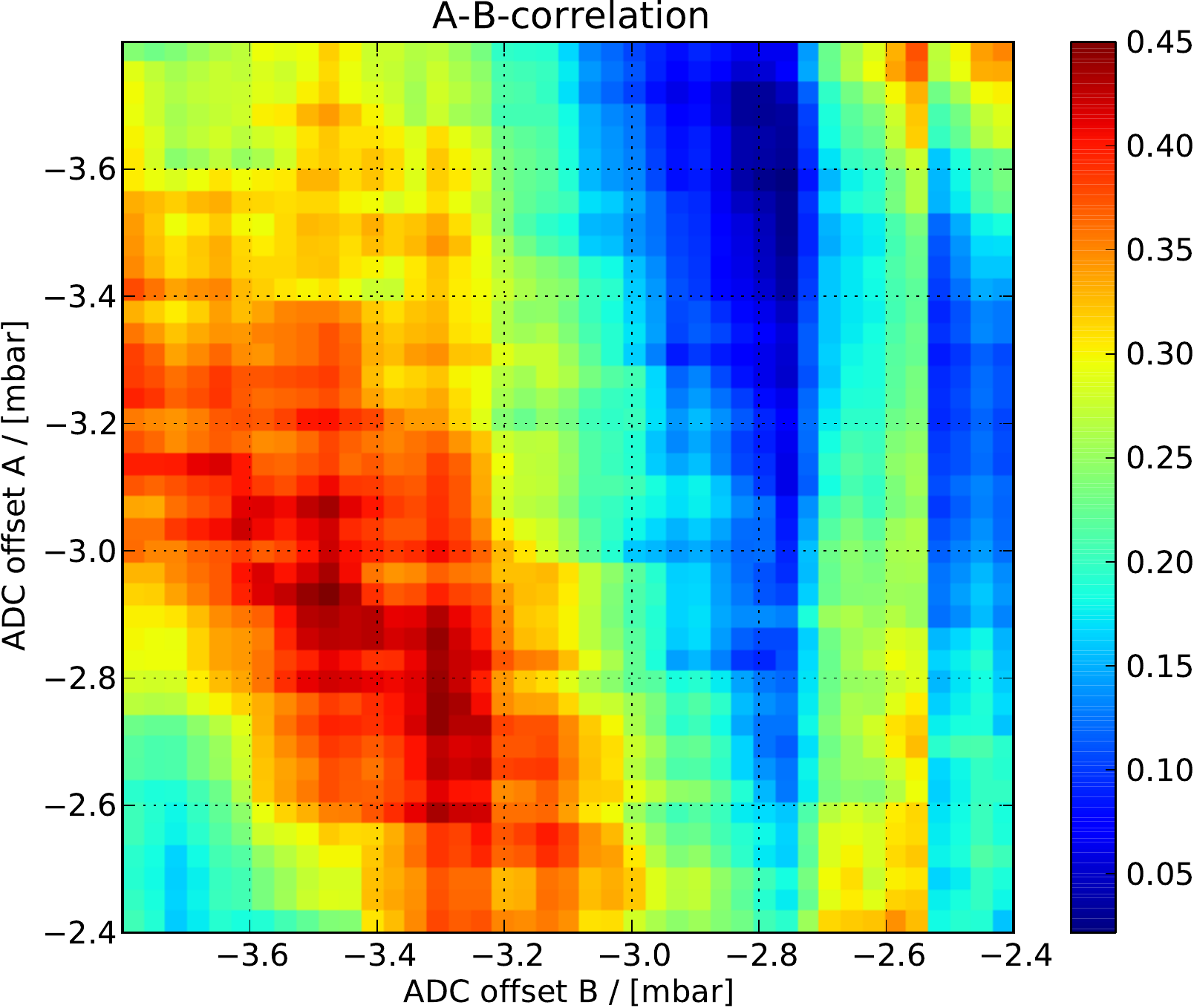}
    \caption{Correlation between pressure signals for different offset corrections}
    \label{fig:p-cor}
\end{figure}

\begin{figure}
    \centering
    \includegraphics[width=\plotwidth]{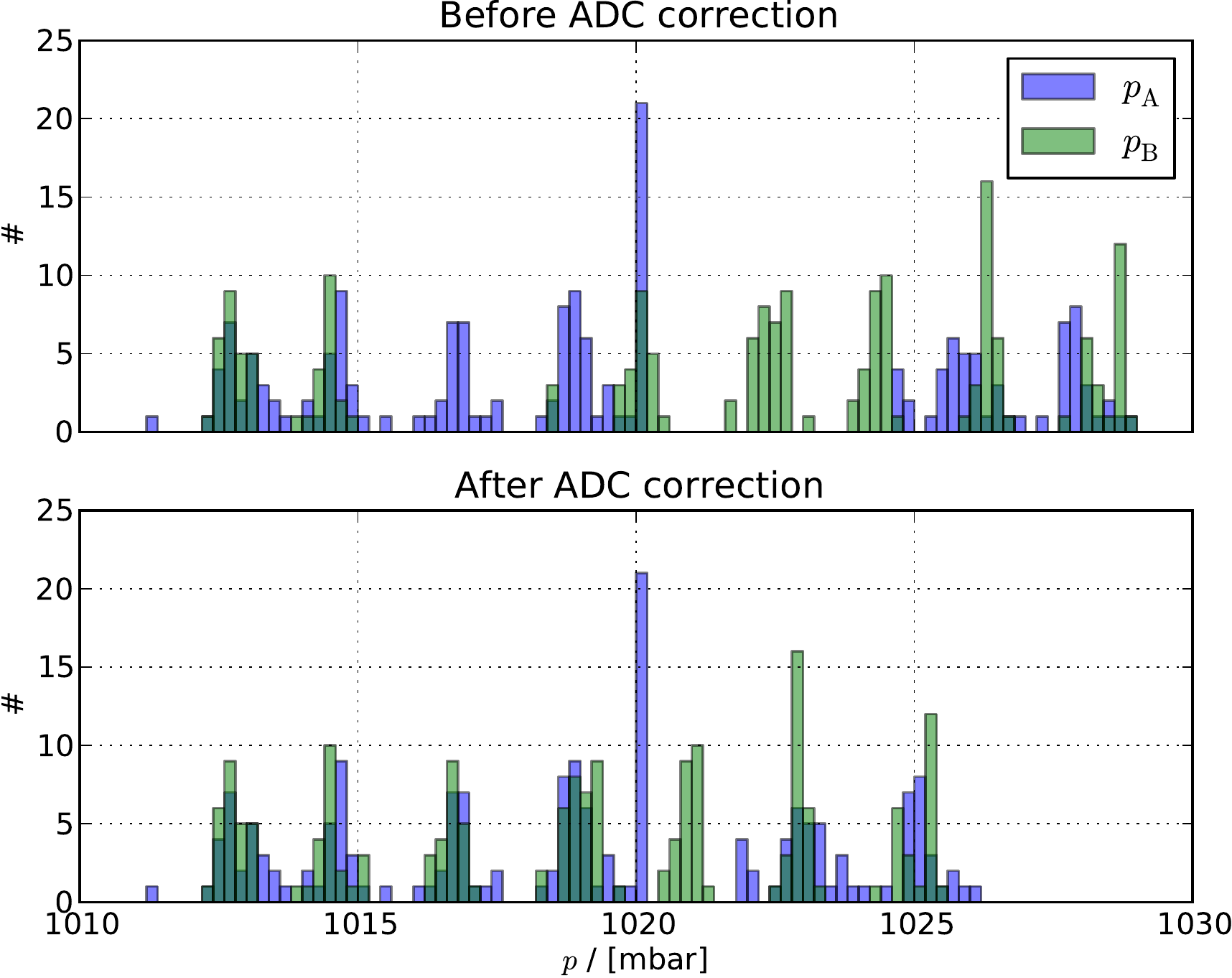}
    \caption{Histogram of pressures before and after ADC correction}
    \label{fig:p-hist}
\end{figure}

Due to missing codes in the used ADCs, the pressure sensors show jumps in their signals at \unit[1022]{mbar}
and \unit[1016]{mbar} respectively (figure \ref{fig:p-jump}). One can compensate for this behaviour by
shifting all pressure values above the jumps (as the calibration happened below) down.
Since these jumps occur at different pressures for the two sensors, the proper values for the offsets can
be deduced by employing cross correlation between the signals. Figure \ref{fig:p-cor} shows
the correlation between the signals for different offsets. We chose an offset of $\unit[-2.8]{mbar}$
for sensor~A and $\unit[-3.4]{mbar}$ for sensor~B. Figure \ref{fig:p-hist} shows the histograms of
the measurements in figure \ref{fig:p-jump} before and after the ADC correction. The discontinuity has
been reduced considerably, but it is still visible. For most measurements presented in this work,
this is not a problem though, since the pressure in the chambers is regulated and set to a value
below the pressure where the discontinuity appears.

Altogether the systematics of the pressure measurements are expected to be not larger than \unit[1]{mbar}.
The nominal pressure in the chambers is set to \unit[1000]{mbar}, so the estimated relative error is
\[
    \frac{\sigma_p}{p} \lesssim \frac{\unit[1]{mbar}}{\unit[1000]{mbar}} = \unit[1]{\perthousand}
\]


\clearpage
\section{Putting it all together}

Combining the errors of all the variables yields the systematic uncertainties
\begin{gather*}
    \left( \frac{\sigma_{v_d}}{v_d} \right)_{sys} = \sqrt{ \left(\frac{\sigma_{\Delta z}}{\Delta z}\right)^2
                                                     + \left(\frac{\sigma_{\Delta t}}{\Delta t}\right)^2 }
                                                   \lesssim \unit[4]{\perthousand}, \\
    \shortintertext{and}
    \left( \frac{\sigma_{\ETp}}{\ETp} \right)_{sys} = \sqrt{ \left(\frac{\sigma_{U_f}}{U_f}\right)^2
                                                            + \left(\frac{\sigma_{l_f}}{l_f}\right)^2
                                                            + \left(\frac{\sigma_{T}}{T}\right)^2
                                                            + \left(\frac{\sigma_{p}}{p}\right)^2 }
                                                     \lesssim \unit[4]{\perthousand}.
\end{gather*}
These errors proved to be reasonable as we used two chambers to cross check for chamber
specific systematics. No differences outside the expected errors were found.

\chapter{Results}

The drift velocity of electrons in a gas mixture depends not only on the electric field
strength $E$, but also on the gas molecule density $n$. So if one wanted to characterise
a gas mixture in its entirety, one would have to do so at a wide range of environments.
Conversely, if one wants to measure the drift velocity at a certain $E$-field, one would have
to regulate the temperature and pressure of the gas precisely, which can be hard to achieve.

Fortunately the dependence on $E$ and $n$ only appears in the form of $\nicefrac{E}{n}$.
So instead of interpreting $v_d$ as a function of $E$ with additional parameters $T$ and $p$,
it can be seen as a function of $\nicefrac{E}{n}$, or as a function of $\ETp$ if we assume
an ideal gas. Since we have no direct way to measure $n$, we use $\ETp$ as argument
for the drift velocity functions:
\[
    v_d = v_d(\ETp)
\]

All data presented in this work can also be accessed via the gasDB at \gasDBURL.
Every $v_d$ data point actually consists of multiple measurements
at the same electric field that have been combined into one mean point $\overline{v_d}$ and the statistical
error $\sigma_{v_d,\text{stat}}$ is just the error of that mean.
The same is of course true for $\ETp$.

\begin{figure}
    \centering
    \includegraphics[width=\plotwidth]{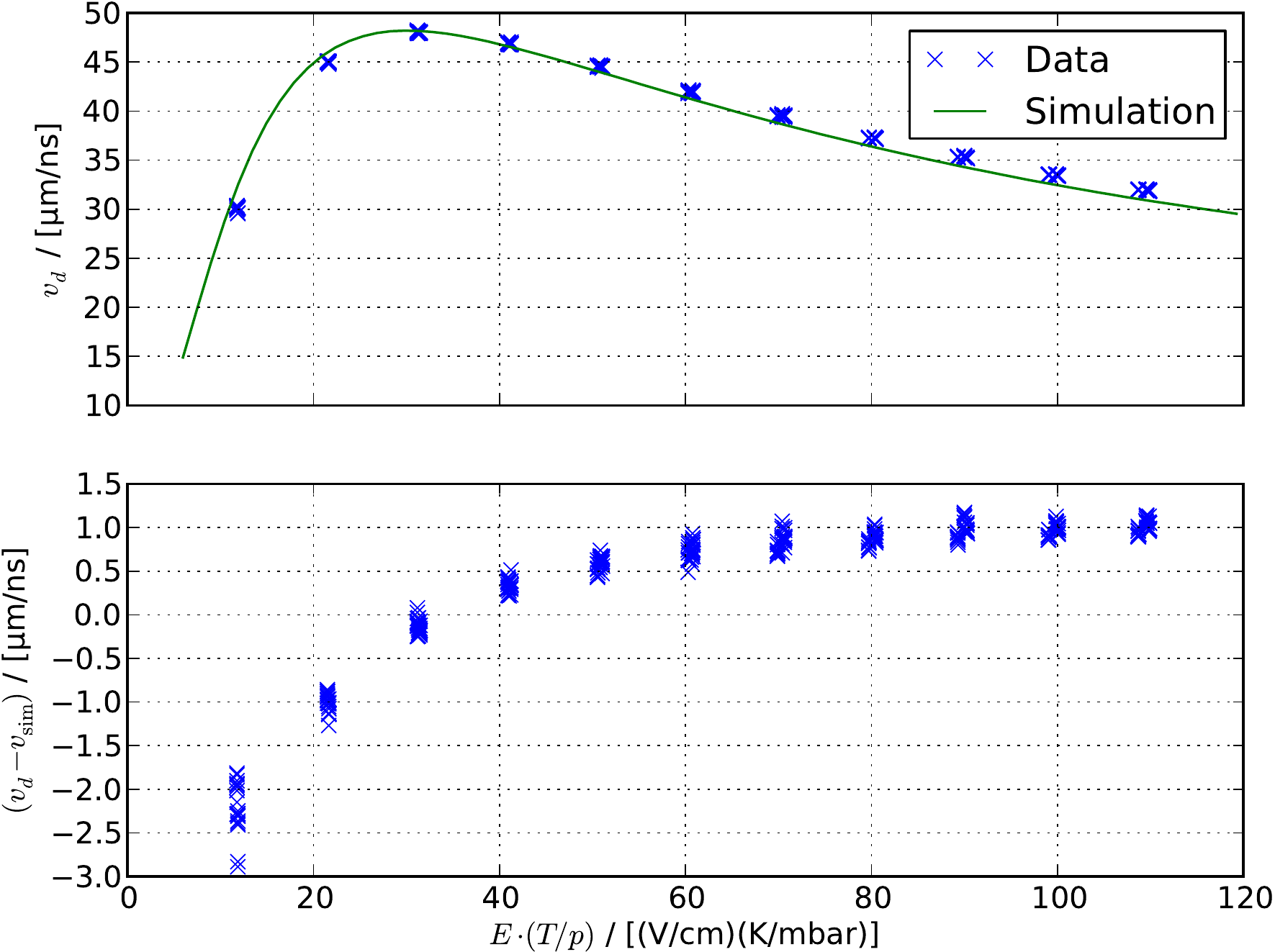}
    \caption{Example  of raw $v_d$ measurements with \binmix{Ar}{93}{CH4}{7}}
    \label{fig:vd-example}
\end{figure}

\begin{figure}
    \centering
    \includegraphics[width=\plotwidth]{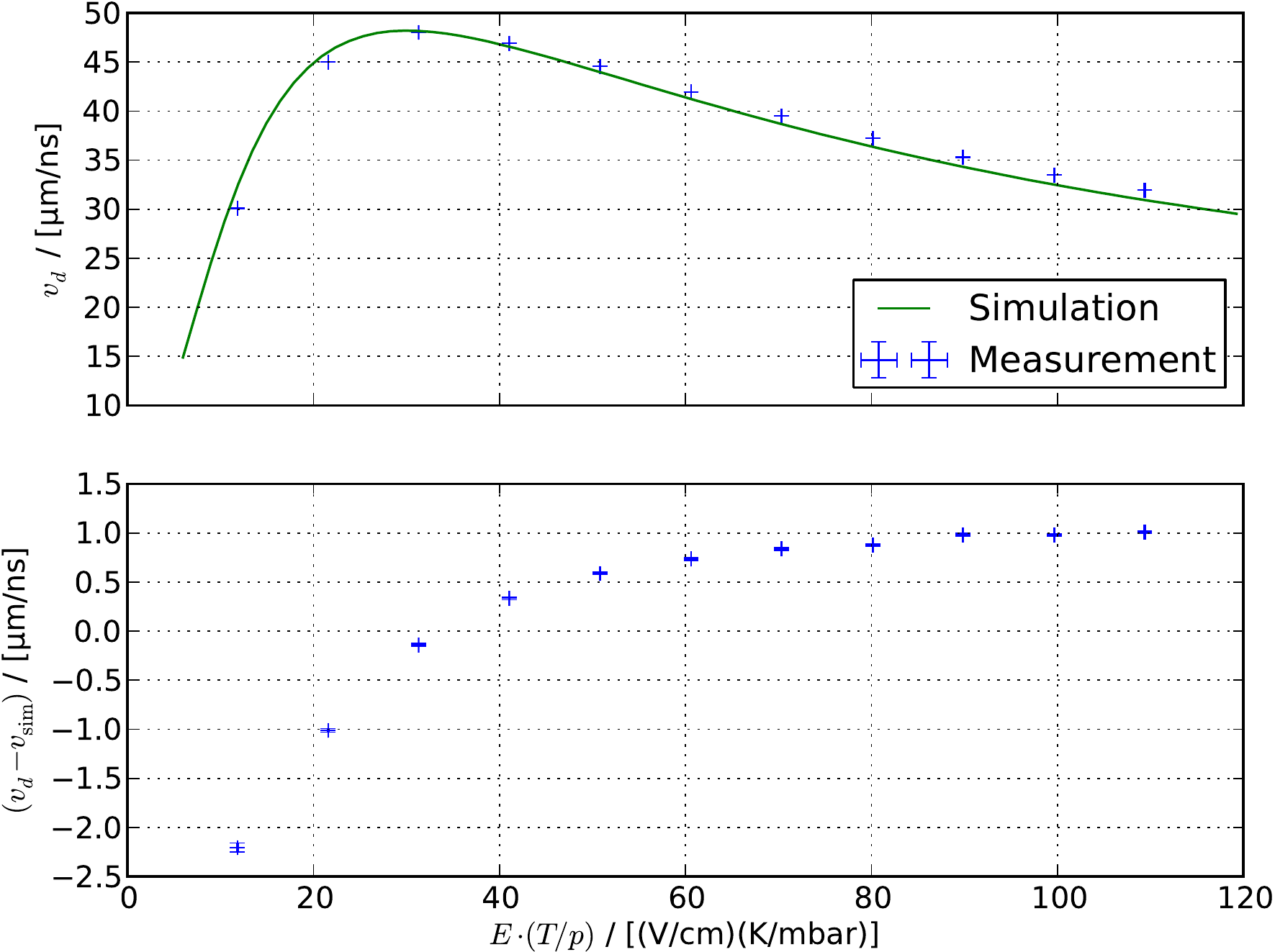}
    \caption[Example \enquote{condensed} $v_d$ measurements with \binmix{Ar}{93}{CH4}{7}]
            {Example \enquote{condensed} $v_d$ measurements with \binmix{Ar}{93}{CH4}{7}.
             Please note that the statistical errors are too small to be seen for most points in this plot.}
    \label{fig:vd-prof-example}
\end{figure}

An example of a set of measured drift velocities at different electric fields can be seen
in figure~\ref{fig:vd-example}. Each cluster of points is reduced to one single data point
in figure~\ref{fig:vd-prof-example}. Clusters with an $\RMS(v_d)$ of \unitfrac[2]{\micro{m}}{ns}
and above are ignored, as such high RMS~values are usually the result of a $v_d$ above or
below the accessible $v_d$~range of the experimental setup (though this was not necessary in this example).
Also, all points outside a $3 \cdot \RMS(v_d)$ range of the cluster are removed and the mean and RMS are recalculated.
This automatically removes implausible outliers and is only done once per cluster. For most
data points this has little to no effect on the final mean and RMS value.

From now on, unless explicitly stated otherwise, $\overline{v_d}$ will be written simply as
$v_d$ and $\overline{\ETp}$ as $\ETp$.
All $v_d$ plots in this work include the statistical error on $v_d$ as well as $\ETp$, but
since they are in the order of \unitfrac[0.01]{\micro{m}}{ns} and \unitfrac[\unitfrac[0.01]{V}{cm}]{K}{mbar}
respectively, they often cannot be seen clearly.
In general, the statistical errors are in the order of \unit[1]{\perthousand} or better.

\section{\pdfce{Ar}-\pdfce{CH4}-\pdfce{CO2} mixtures}

\ce{Ar}-\ce{CH4} mixtures are classical drift gases that have been used, and continue to be used, in many gas based detectors.
They are often named PX, where X is the percentage of \ce{CH4} in the mix, e.g. P5 for
\binmix{Ar}{95}{CH4}{5} or P20 for \binmix{Ar}{80}{CH4}{20}.

\begin{figure}
    \centering
    \includegraphics[width=\plotwidth]{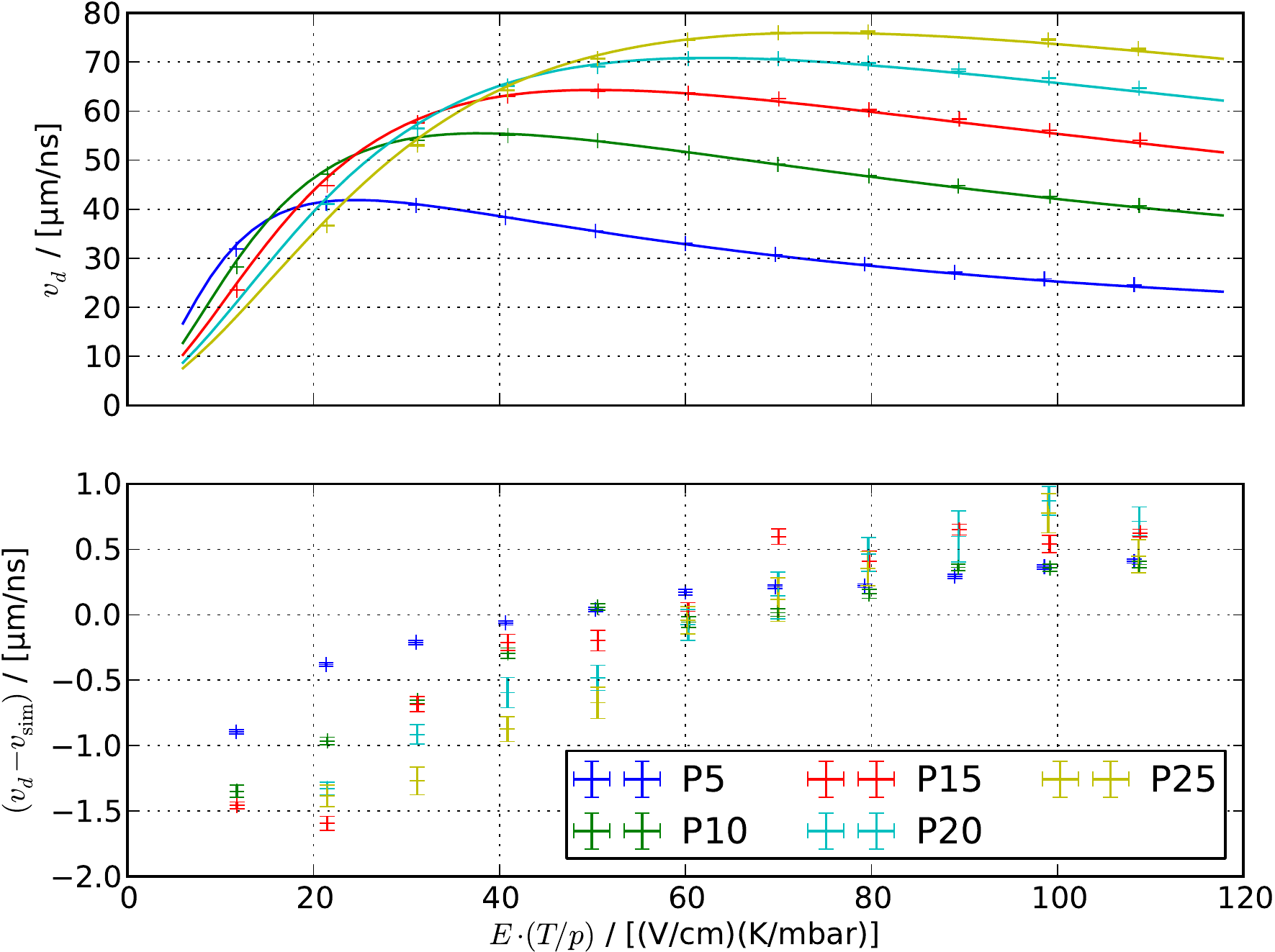}
    \caption[$v_d$ measurements and simulations of P5, P10, P15, P20 and P25]
            {$v_d$ measurements and simulations of P5, P10, P15, P20 and P25}
    \label{fig:P5-25-comb}
\end{figure}

Figure~\ref{fig:P5-25-comb} shows our measurements for P5, P10, P15, P20 and P25. They all show similar
deviations from the simulated drift velocities. These deviations cannot be attributed to gas
mixing errors alone. Varying the \ce{CH4} fraction for the simulation does not affect the drift
velocities at low  electric fields as much as it does at high electric fields. And it is not possible
to tweak the simulated gas fraction in a way as to create a perfect fit to the data. The deviations are
therefore either an inaccuracy in the Magboltz simulations, effects of impurities of the mixed raw gases
or not understood systematics in our experimental setup, or a combination of all three.

\begin{figure}
    \centering
    \includegraphics[width=\plotwidth]{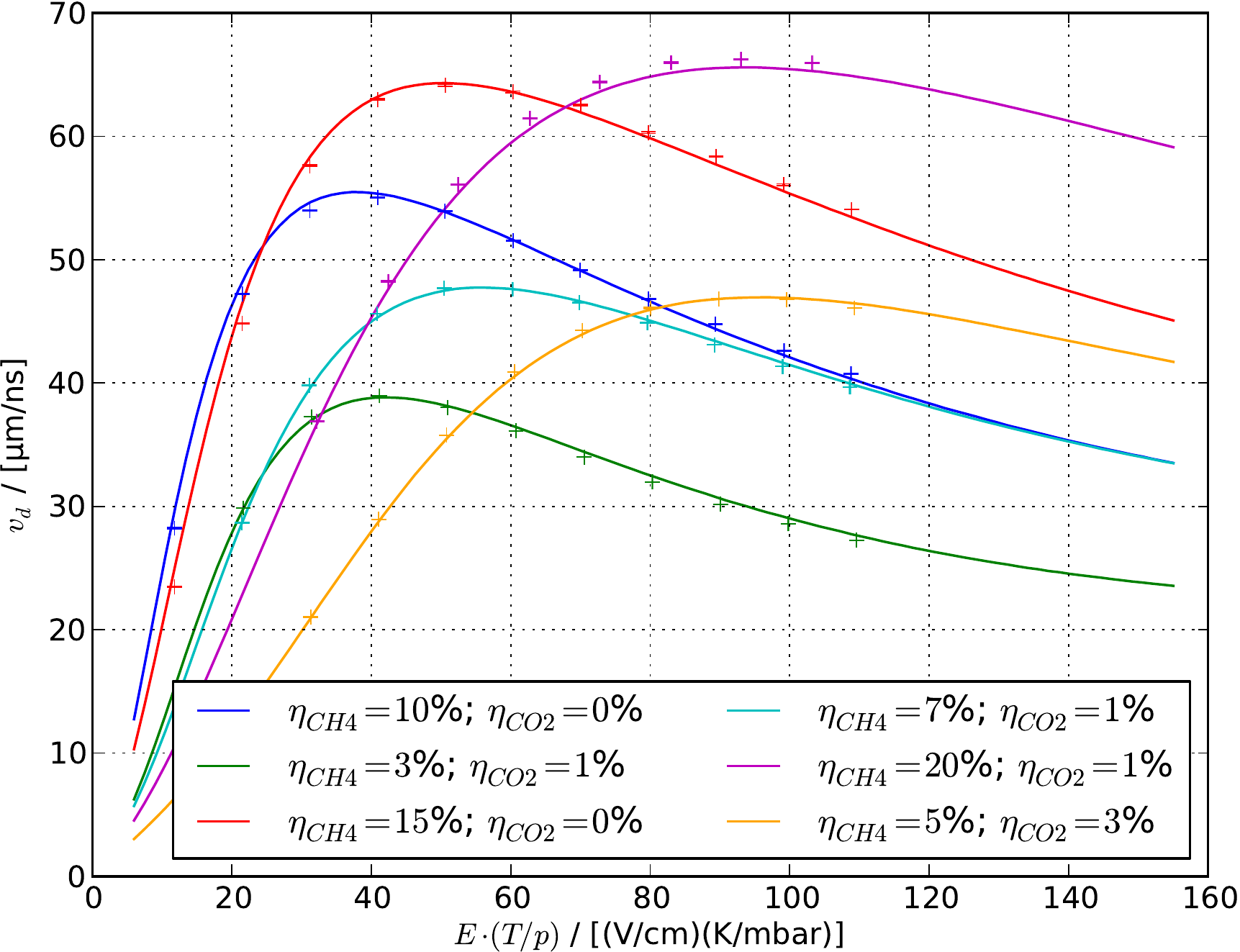}
    \caption{$v_d$ curve shape examples}
    \label{fig:vd-shape}
\end{figure}

Drawing multiple $v_d$~curves into one plot is feasible as long as one only varies one gas fraction.
Adding a third gas (in this case \ce{CO2}) and varying its fraction as well would render such
a plot cluttered and unreadable. We therefore chose a different approach to visualise the impact
of the different additives. Most $v_d$~curves show the same behaviour: The drift velocity increases
with an increasing $\ETp$ until it reaches a maximum. It then falls off again, and the curvature
around the maximum is higher if the maximum appears at low $\ETp$ values (see figure \ref{fig:vd-shape}).

Knowing the position of the maximum thus already gives a rather good impression of the features
of the curve. Furthermore, ideally one would like to operate a detector at the maximum of the curve
since there the dependence of $\ETp$ is minimal and fluctuations in temperature, pressure and electric
field have the smallest effect on the drift velocity and thus on the reconstruction of the original particle track.
Since every $v_d$~curve is reduced to this single working point, the effects of two
gas additives can be comfortably visualised in a 2D~plot.

\begin{figure}
    \centering
    \includegraphics[width=\plotwidth]{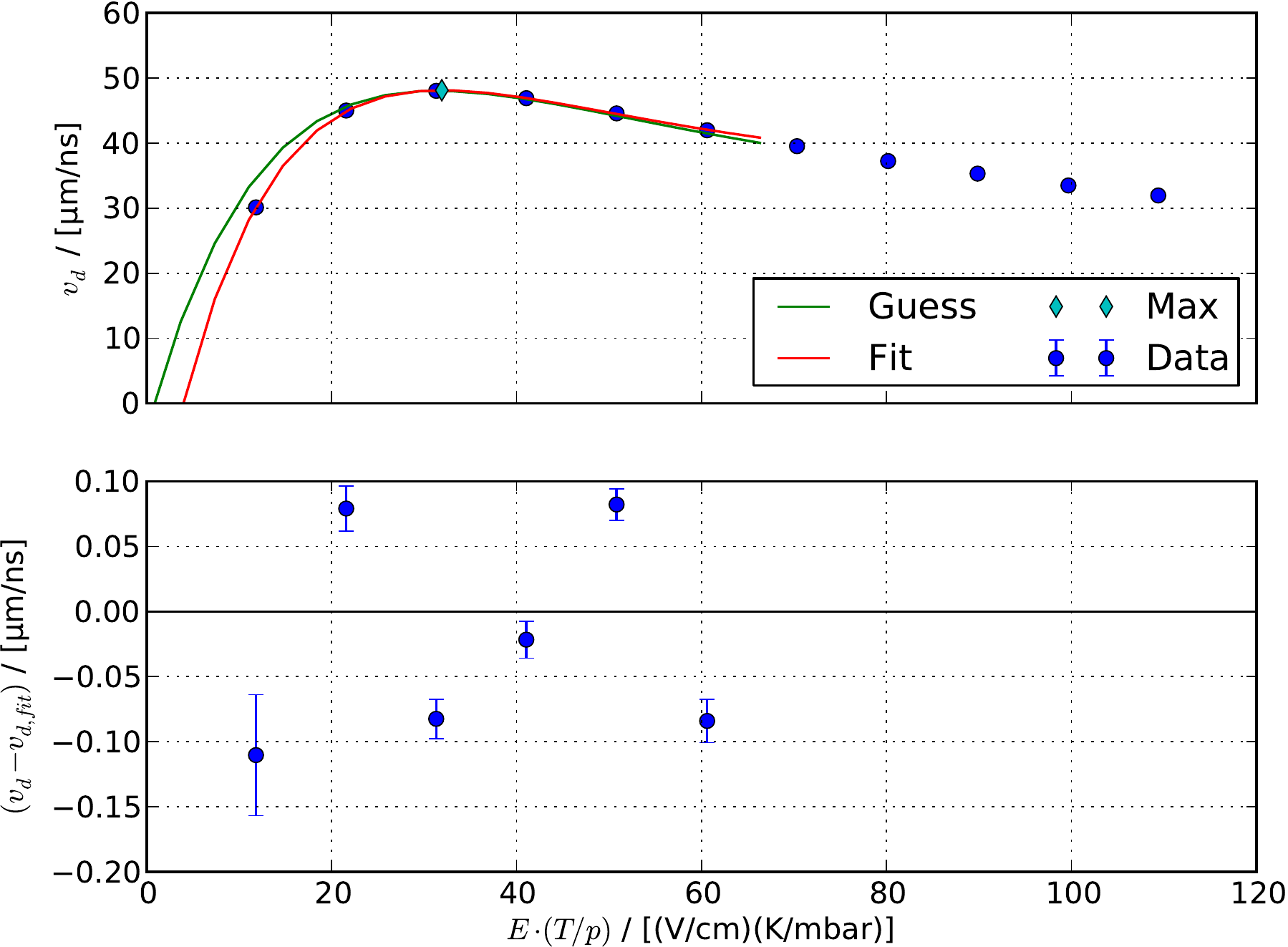}
    \caption[Example working point fit for \binmix{Ar}{93}{CH4}{7}]
            {Example working point fit for \binmix{Ar}{93}{CH4}{7}.
             The curve labeled as \enquote{guess} shows the seed parameters for the fit algorithm.}
    \label{fig:WP-fit}
\end{figure}

Both the measured and simulated working points are determined by fitting a function to the simulated or measured
data points and calculating the position of the maximum from the best fit parameters.
\begin{align*}
    f_{\text{fit}}(\ETp) &= (a + b\cdot\ETp) \exp(-d\cdot\ETp) + c \\
    \ETp|_\text{WP} &= \frac{ad-b}{-bd} \\
    v_d|_\text{WP} &= f_{\text{fit}}(\ETp|_\text{WP})
\end{align*}
Figure \ref{fig:WP-fit} shows an example of such a fit. Although there is no theoretical
motivation for the chosen function, it fits the data around the $v_d$ maxima very well.
This holds true even for extremely wide maxima, as will be shown in section \ref{sec:AMH}.

Due to limitations of the experimental setup,
we can only measure $\ETp$ values up to about \unitfrac[\unitfrac[110]{V}{cm}]{K}{mbar} and determine
working points up to \unitfrac[\unitfrac[100]{V}{cm}]{K}{mbar}. The position of the simulated working
points is only limited by the chosen simulated electric field range.

\begin{figure}
    \centering
    \includegraphics[width=\plotwidth]{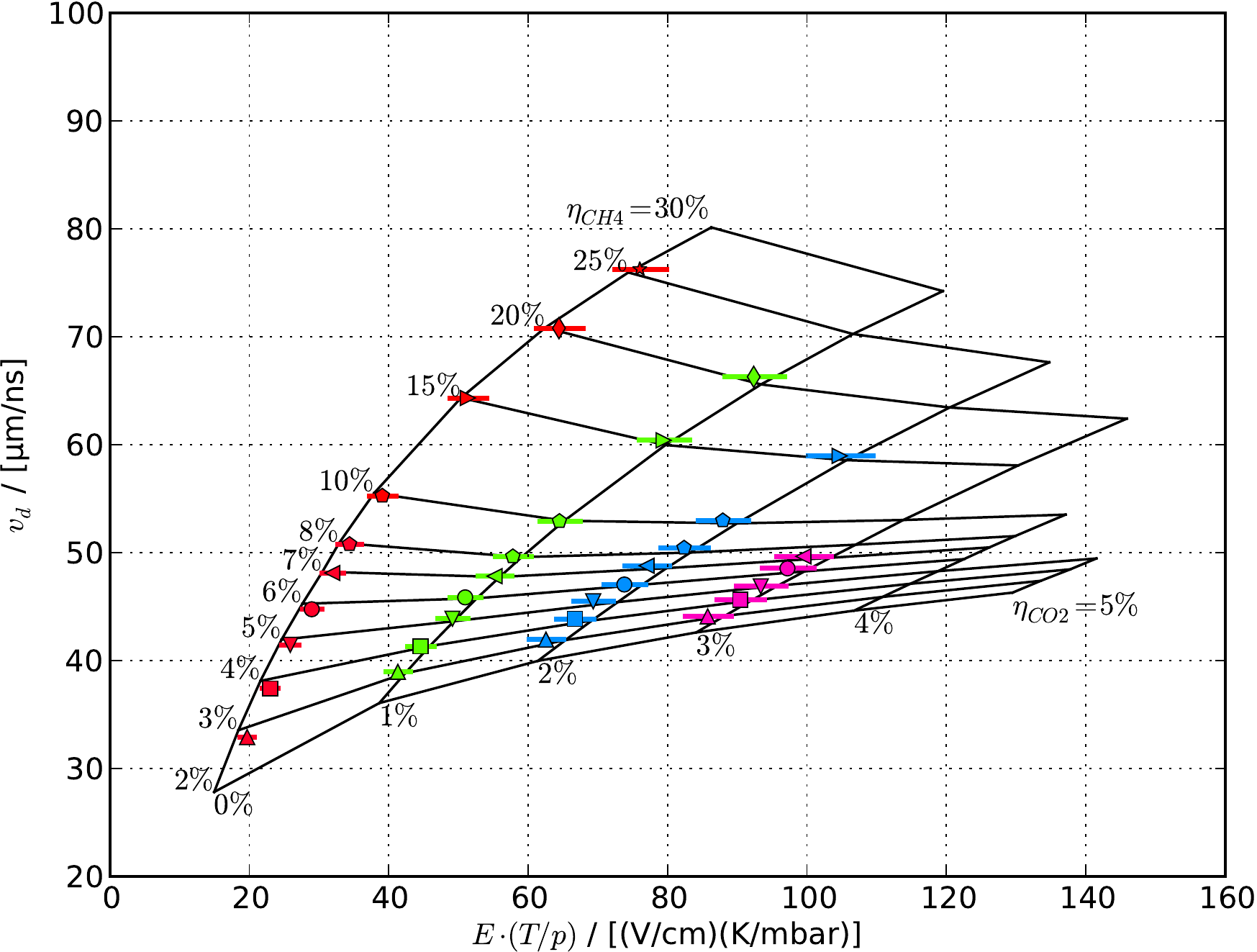}
    \caption[Working points for \ce{Ar}-\ce{CH4}-\ce{CO2} mixtures]
            {Working points for \ce{Ar}-\ce{CH4}-\ce{CO2} mixtures.
             The horizontal lines mark the $\ETp$~range where $\nicefrac{\Delta v_d}{v_d|_\text{WP}} < \unit[1]{\perthousand}$.}
    \label{fig:Ar-CH4-CO2-WP}
\end{figure}

The effect of varying \ce{CH4} and \ce{CO2} fractions on the position of the working point is shown in
figure \ref{fig:Ar-CH4-CO2-WP}. The intersections of the solid lines were determined from simulated data,
while the coloured marks show measured data. Points with the same marker shape have the same \ce{CH4} fraction,
while same colour indicates a shared \ce{CO2} fraction. Additionally to the position of the maximum, the horizontal
lines show the width of the peak, defined as the $\ETp$~range where
\[
    \frac{\Delta v_d}{v_d|_\text{WP}} = \frac{v_d|_\text{WP} - v_d}{v_d|_\text{WP}} < \unit[1]{\perthousand}.
\]

\begin{figure}
    \centering
    \includegraphics[width=\plotwidth]{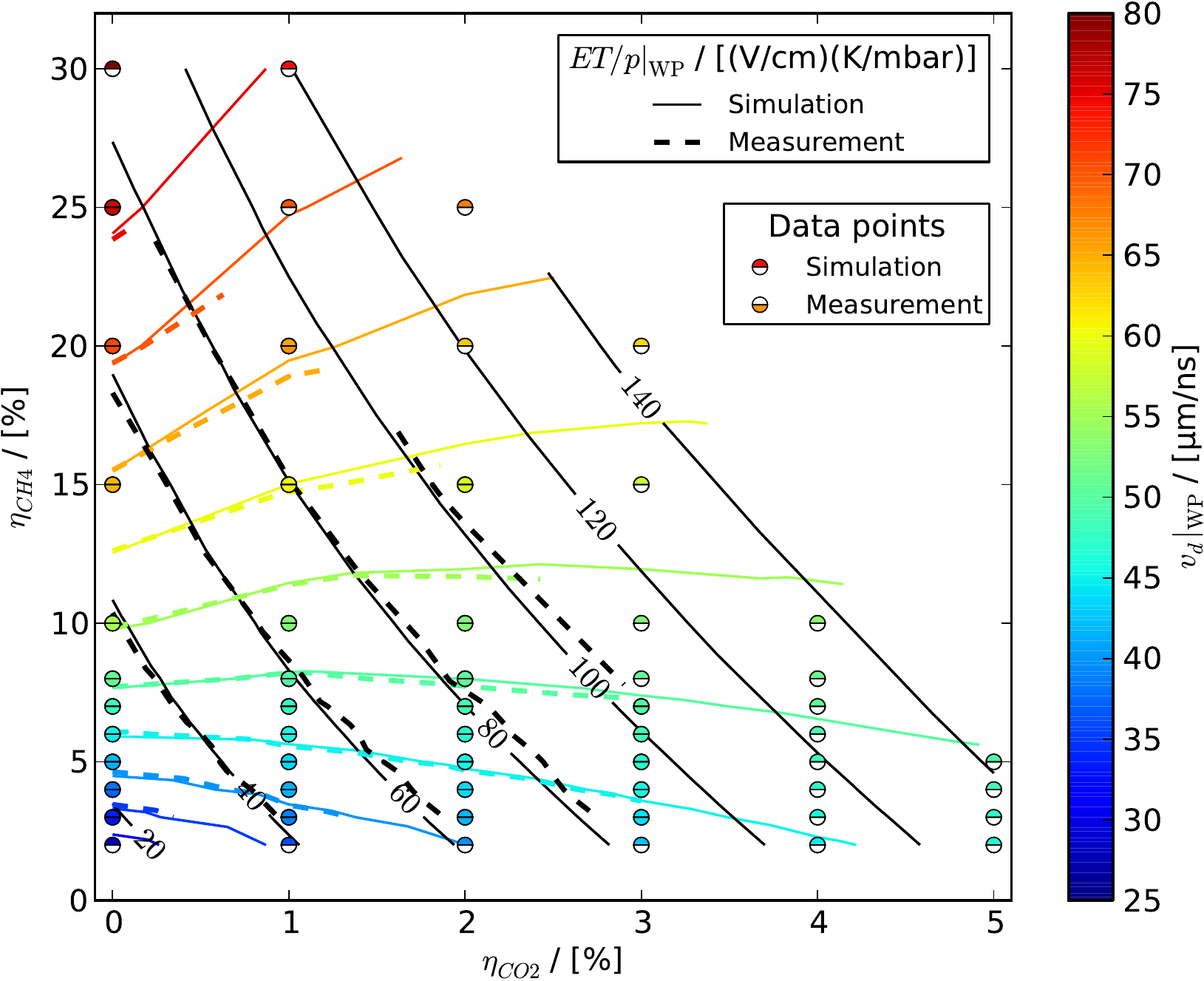}
    \caption{Working points for \ce{Ar}-\ce{CH4}-\ce{CO2} mixtures in dependence of $\eta$}
    \label{fig:Ar-CH4-CO2-WP-inv}
\end{figure}

One can also plot the values of $v_d|_\text{WP}$ and $\ETp|_\text{WP}$ directly in dependence of the gas fractions
as seen in figure \ref{fig:Ar-CH4-CO2-WP-inv}.
These contour plots were created with exactly the same data as the spider web plot in figure \ref{fig:Ar-CH4-CO2-WP}
and the space in-between the data points is interpolated. The solid lines are interpolated from the simulated data
and the dashed lines show the interpolation of the measured data. The black lines show mixtures with a constant
$\ETp|_\text{WP}$ and the coloured lines represent a constant $v_d|_\text{WP}$. The colours of the data points also
indicate the $v_d|_\text{WP}$ value of those mixtures.

\section{\pdfce{Ar}-\pdfce{CF4}-\pdfce{iC4H10} mixtures}

\begin{figure}
    \centering
    \includegraphics[width=\plotwidth]{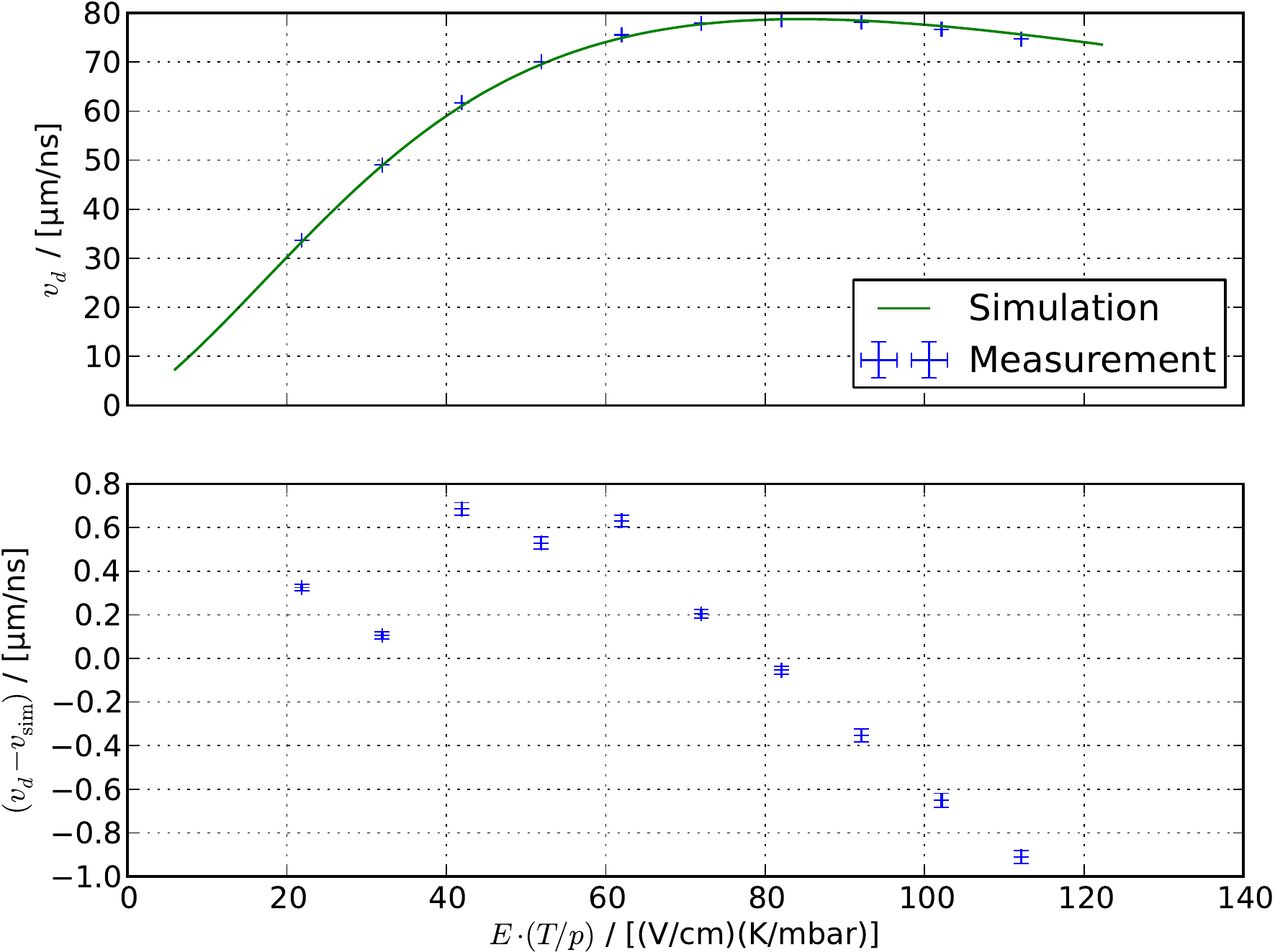}
    \caption{$v_d$ curve of \termix{Ar}{95}{CF4}{3}{iC4H10}{2}, also known as T2K-gas}
    \label{fig:T2K-gas}
\end{figure}

\termix{Ar}{95}{CF4}{3}{iC4H10}{2} is also known as T2K-gas since it is being used in the ND280 near detector
of the T2K neutrino experiment. Its $v_d$~curve can be seen in figure \ref{fig:T2K-gas}.
\ce{Ar}-\ce{CF4}-\ce{iC4H10} mixtures' $v_d$~curves have a shape that is very similar to those of
\ce{Ar}-\ce{CH4}-\ce{CO2} mixtures. One can therefore use the same visualization technique to show the effects
of varying volume fractions of the additives.

\begin{figure}
    \centering
    \includegraphics[width=\plotwidth]{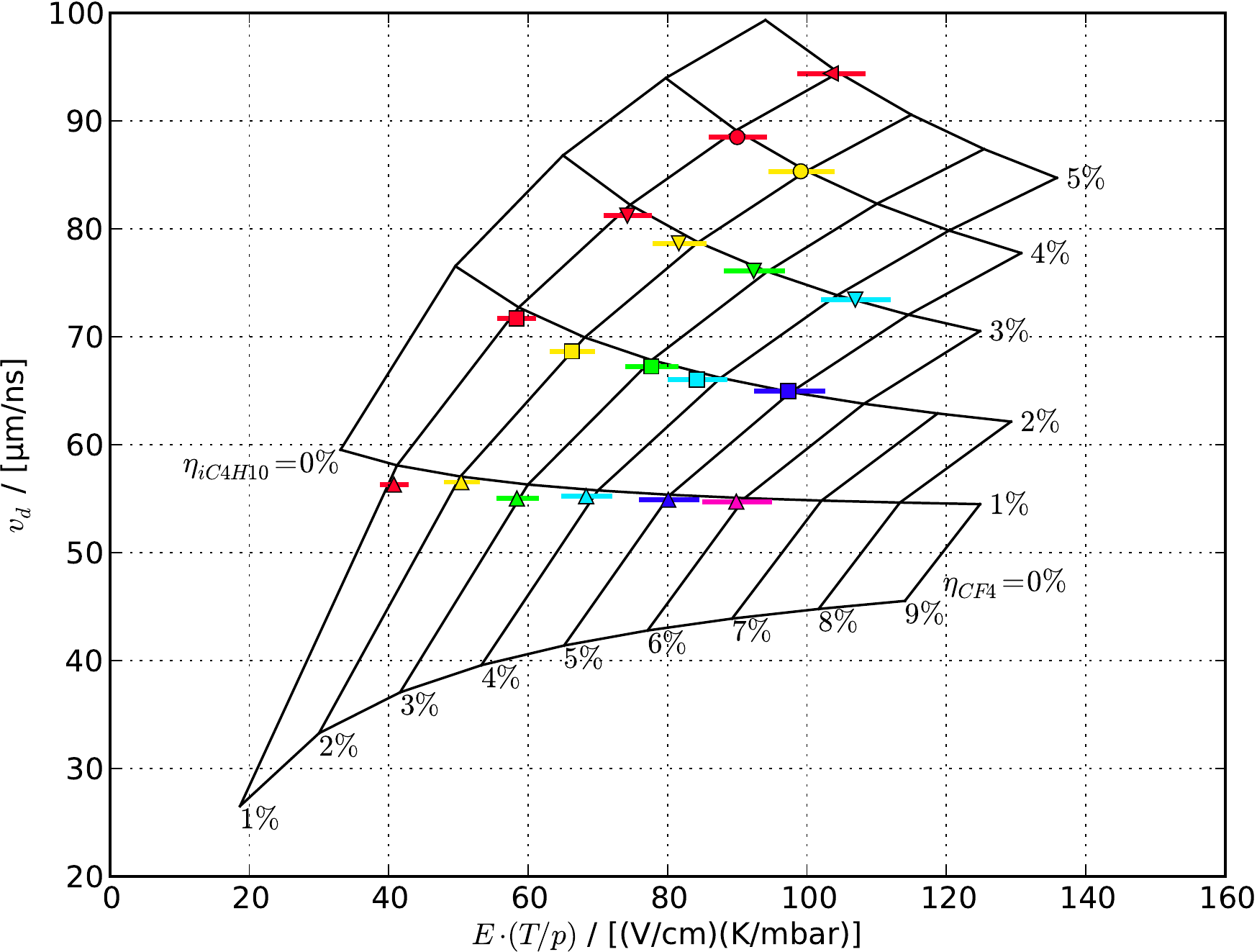}
    \caption[Working points for \ce{Ar}-\ce{CF4}-\ce{iC4H10} mixtures]
            {Working points for \ce{Ar}-\ce{CF4}-\ce{iC4H10} mixtures.
             The horizontal lines mark the $\ETp$~range where $\nicefrac{\Delta v_d}{v_d|_\text{WP}} < \unit[1]{\perthousand}$.}
    \label{fig:Ar-CF4-iC4H10-WP}
\end{figure}

\begin{figure}
    \centering
    \includegraphics[width=\plotwidth]{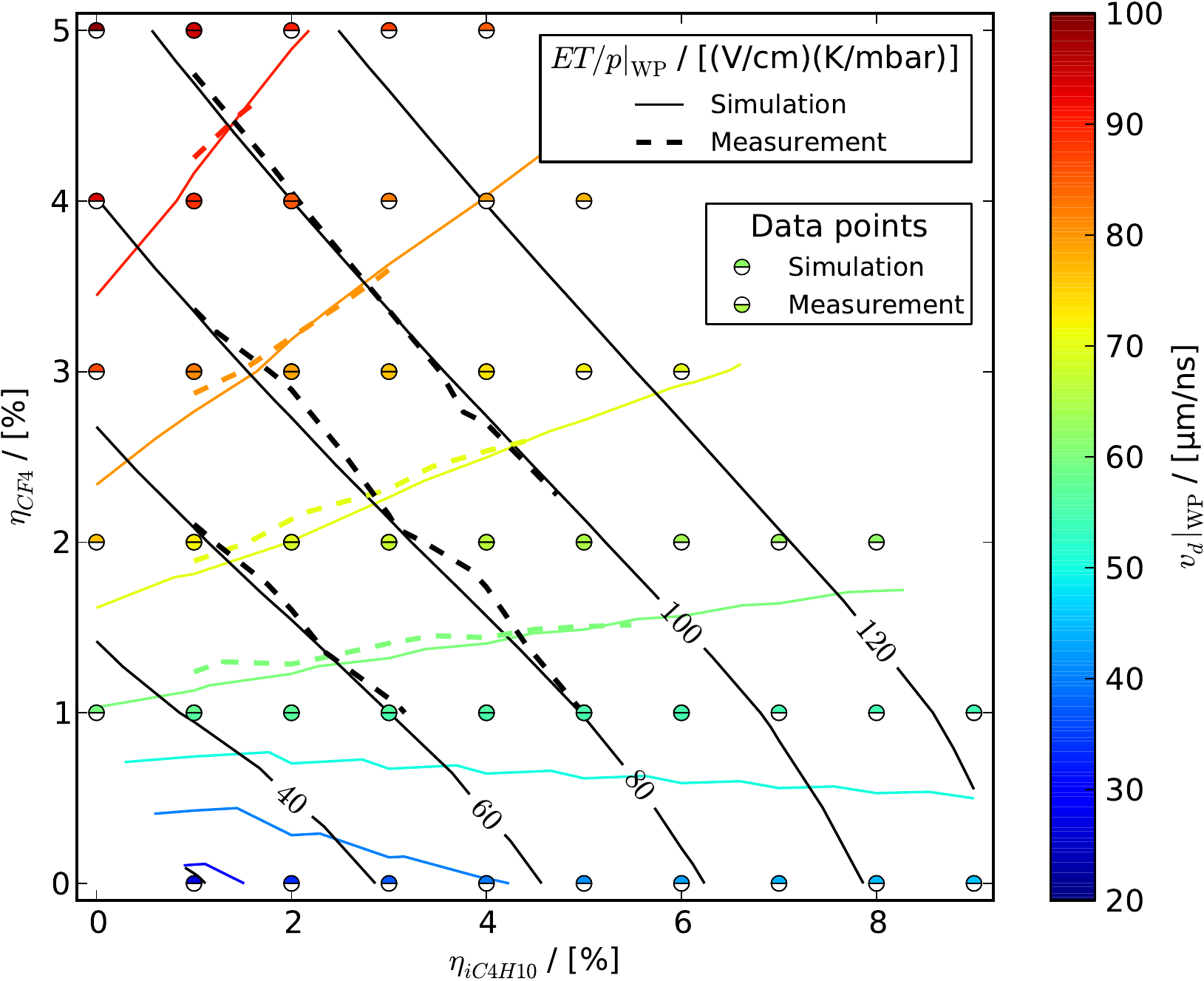}
    \caption{Working points for \ce{Ar}-\ce{CF4}-\ce{iC4H10} mixtures in dependence of $\eta$}
    \label{fig:Ar-CF4-iC4H10-WP-inv}
\end{figure}

This is shown in figures \ref{fig:Ar-CF4-iC4H10-WP} and \ref{fig:Ar-CF4-iC4H10-WP-inv}.
\ce{CF4} and \ce{CH4} have a very similar influence on the working
point position, but one needs far less \ce{CF4} to achieve the same effect as with \ce{CH4}.
It is the other way around for \ce{iC4H10} and \ce{CO2}: \ce{iC4H10} has a similar, but smaller influence on the
drift velocity compared to \ce{CO2}.

\section{\pdfce{Ar}-\pdfce{CH4}-\pdfce{H2} mixtures}
\label{sec:AMH}

\begin{figure}
    \centering
    \includegraphics[width=\plotwidth]{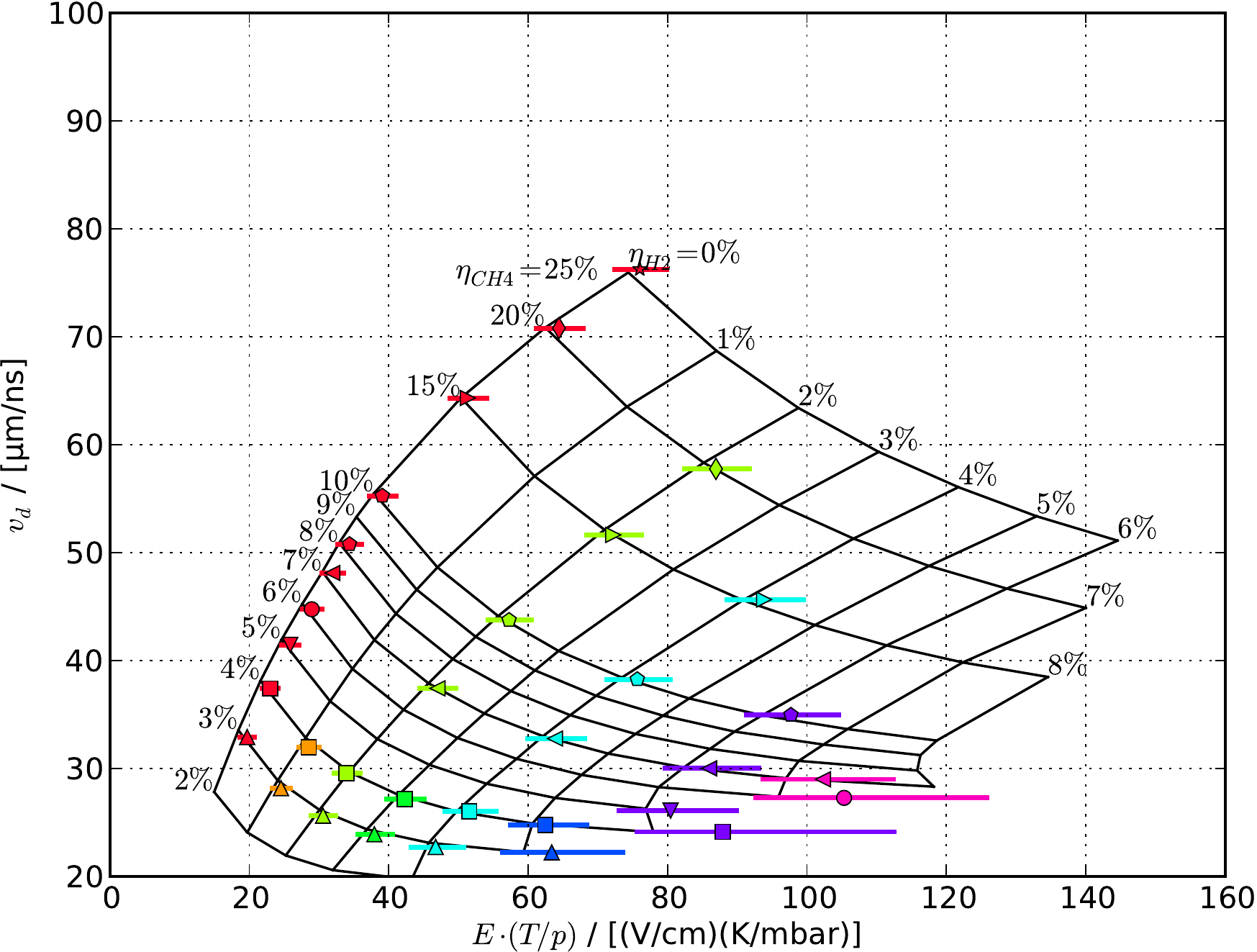}
    \caption[Working points for \ce{Ar}-\ce{CH4}-\ce{H2} mixtures]
            {Working points for \ce{Ar}-\ce{CH4}-\ce{H2} mixtures.
             The horizontal lines mark the $\ETp$~range where $\nicefrac{\Delta v_d}{v_d|_\text{WP}} < \unit[1]{\perthousand}$.}
    \label{fig:Ar-CH4-H2-WP}
\end{figure}

\begin{figure}
    \centering
    \includegraphics[width=\plotwidth]{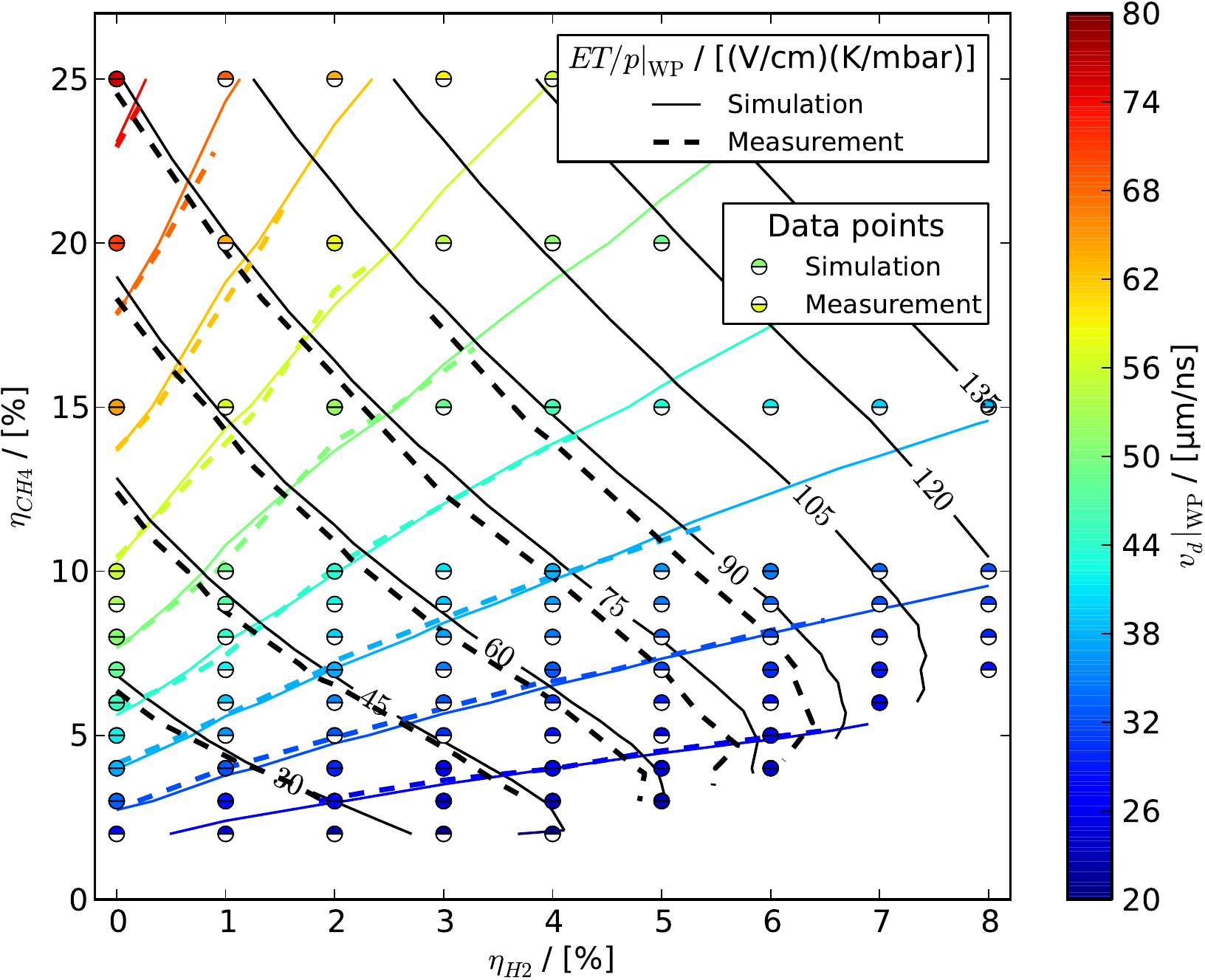}
    \caption{Working points for \ce{Ar}-\ce{CH4}-\ce{H2} mixtures in dependence of $\eta$}
    \label{fig:Ar-CH4-H2-WP-inv}
\end{figure}

The influence of \ce{H2} is quite different from \ce{CO2} and \ce{iC4H10}. All three pull the working point
to higher $\ETp$ values. But while \ce{CO2} and \ce{iC4H10} pull it towards $v_d$ values of about \unitfrac[50]{\micro{m}}{ns},
\ce{H2} always pulls it down (see figures \ref{fig:Ar-CH4-H2-WP} and \ref{fig:Ar-CH4-H2-WP-inv}).
Mixtures with \ce{H2} are thus comparatively slow drift gases. Another difference is the width of the
working points: Adding \ce{H2} to the mixture increases the width much more than the other additives.

\begin{figure}
    \centering
    \includegraphics[width=\plotwidth]{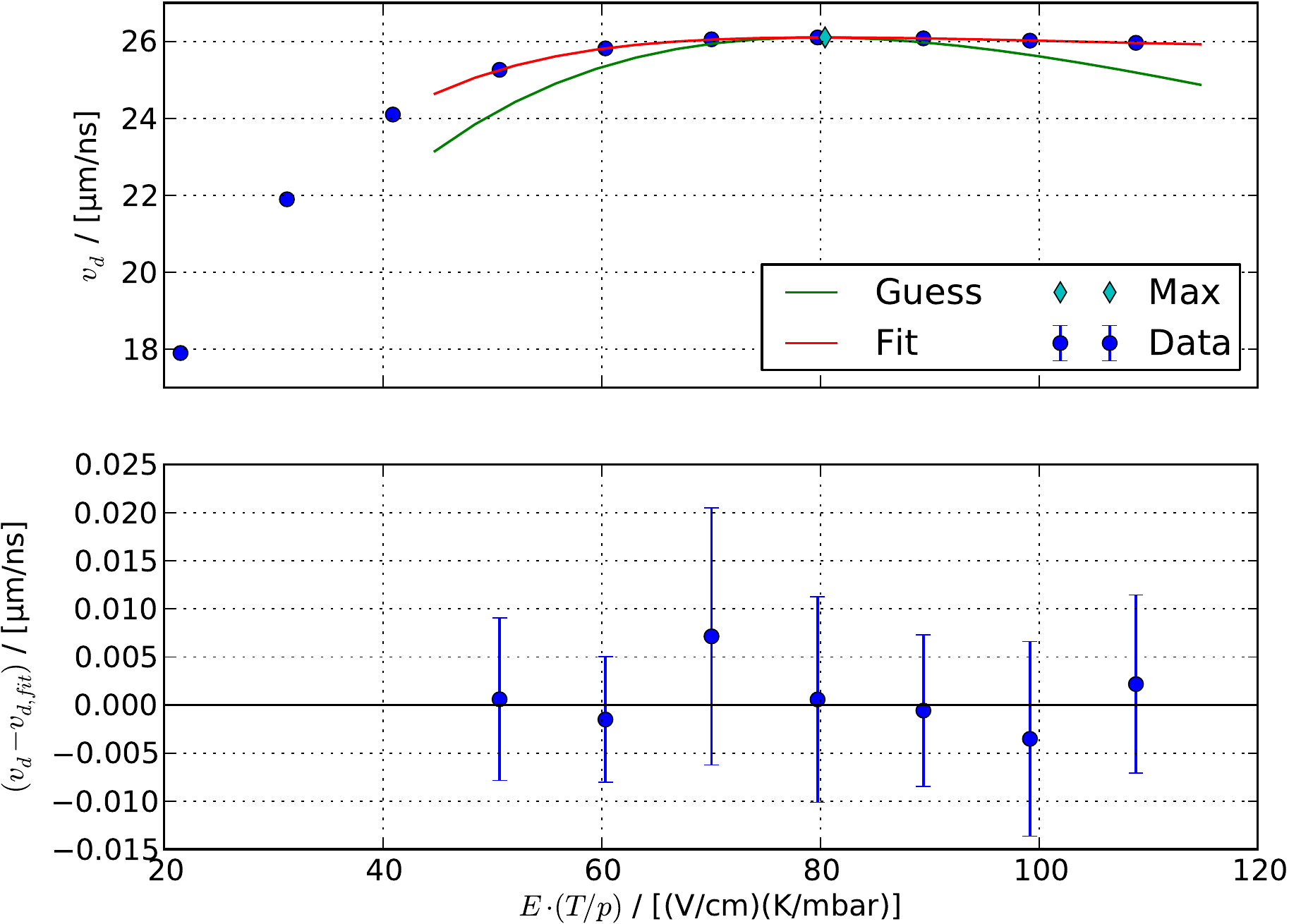}
    \caption[Fit to very wide maximum of \termix{Ar}{89}{CH4}{5}{H2}{6}]
            {Fit to very wide maximum of \termix{Ar}{89}{CH4}{5}{H2}{6}.
             The curve labeled as \enquote{guess} shows the seed parameters for the fit algorithm.}
    \label{fig:wide-max-fit}
\end{figure}

The measured $\ETp|_\text{WP}$ values of mixtures with $\eta_\ce{H2} > \eta_\ce{CH4}$ do not match the simulated values
very well. This is due to the fact that the $v_d$ curves of those mixtures are very flat, which makes the $\ETp$ position of the
maximum very sensitive to changes in the gas composition as well as statistical fluctuations of the measurements.
Despite this, the chosen fit function fits the data points even at very wide maxima, as shown in figure \ref{fig:wide-max-fit}.

\begin{figure}
    \centering
    \includegraphics[width=\plotwidth]{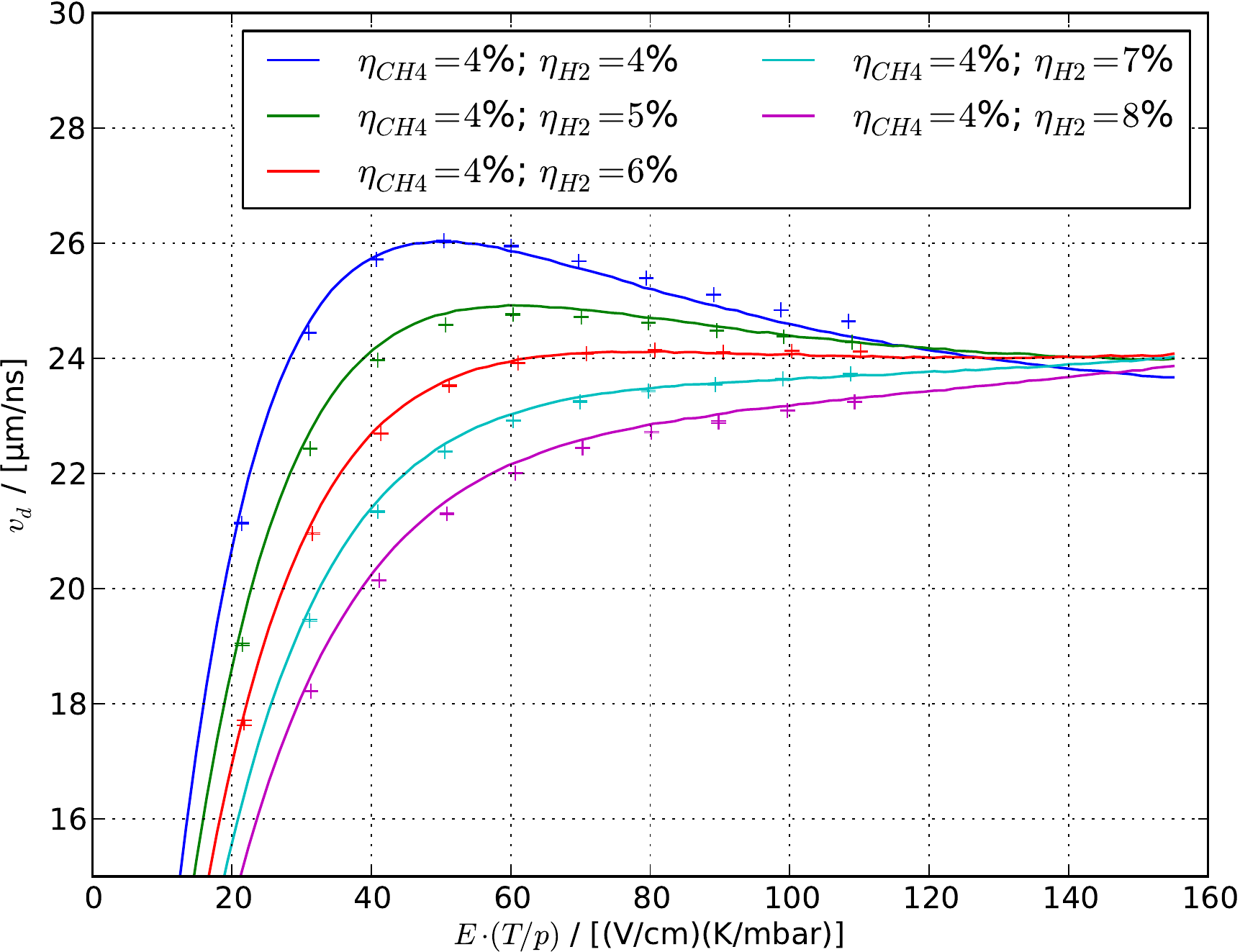}
    \caption{Influence of the \ce{H2} fraction on \ce{Ar}-\ce{CH4}-\ce{H2} mixtures with \unit[4]{\%} \ce{CH4}}
    \label{fig:H2-nomax}
\end{figure}

\begin{figure}
    \centering
    \includegraphics[width=\plotwidth]{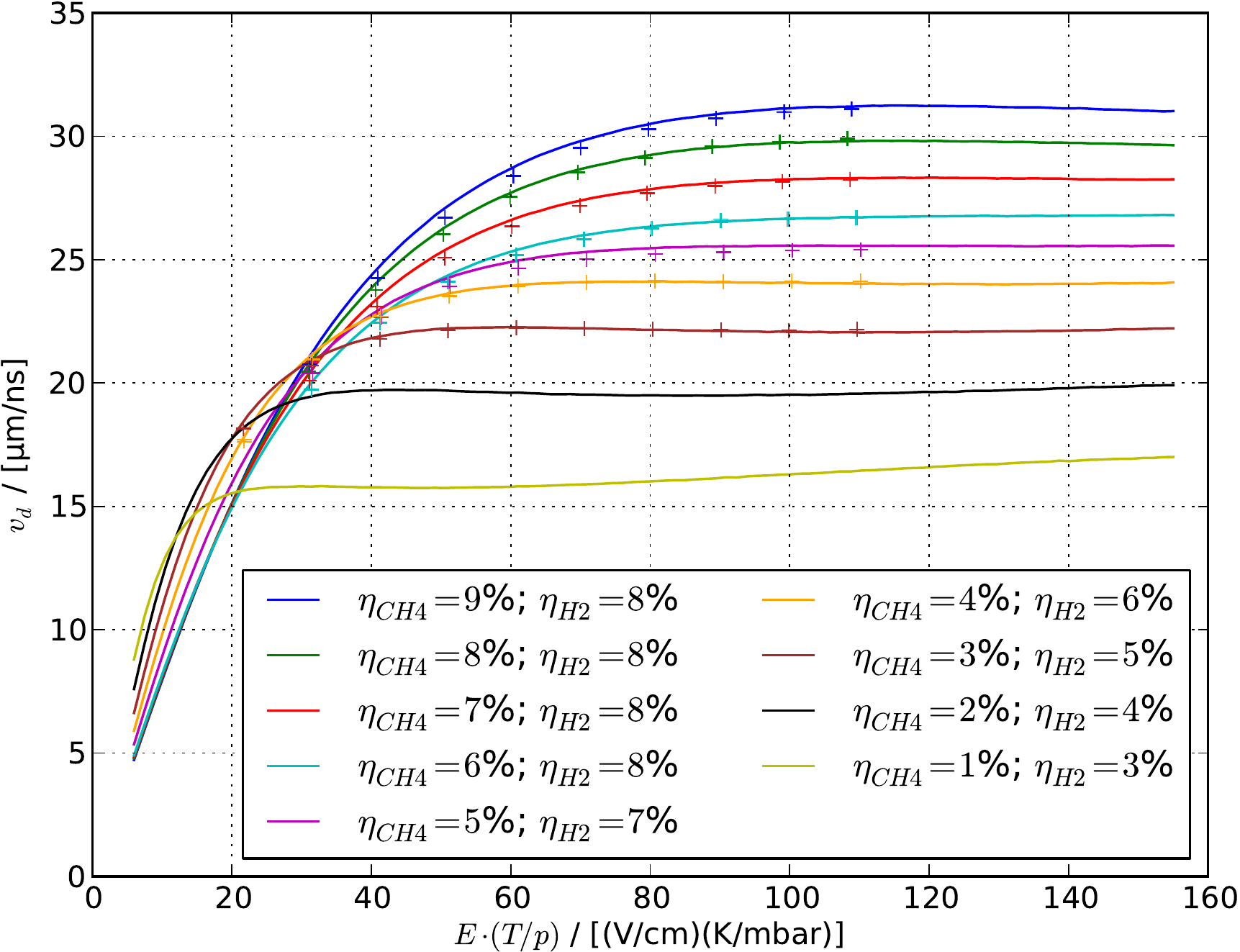}
    \caption{\ce{Ar}-\ce{CH4}-\ce{H2} mixtures that show a wide $v_d$ plateau}
    \label{fig:H2-plateau}
\end{figure}

\begin{figure}
    \centering
    \includegraphics[width=\plotwidth]{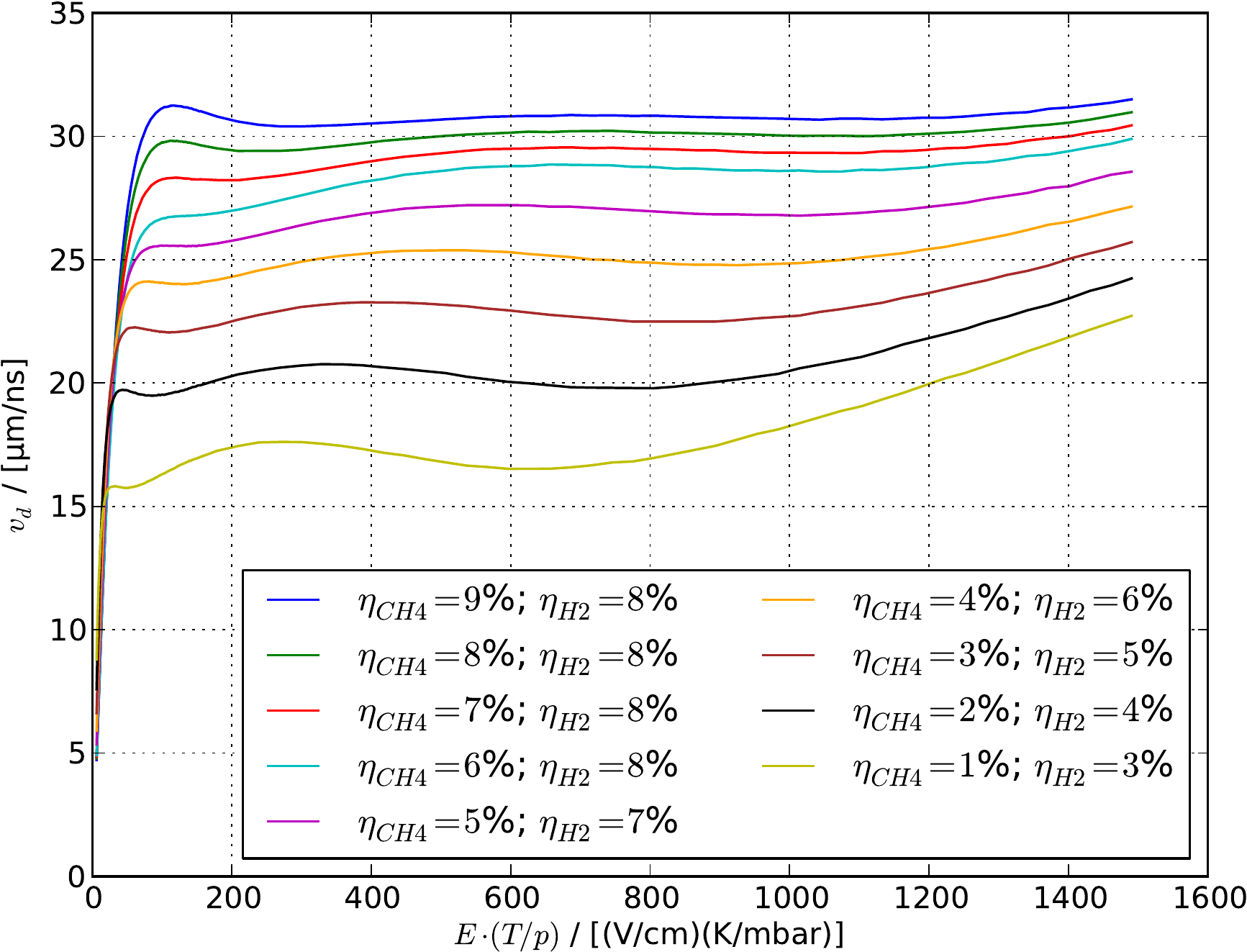}
    \caption[Behaviour of \ce{Ar}-\ce{CH4}-\ce{H2} plateau mixtures up to very high $E$-fields (simulations only)]
            {Behaviour of \ce{Ar}-\ce{CH4}-\ce{H2} plateau mixtures (see fig. \ref{fig:H2-plateau}) up to very high $E$-fields (simulations only)}
    \label{fig:H2-plateau-HV}
\end{figure}

For high \ce{H2} fractions ($\eta_\ce{H2} \gtrsim \eta_\ce{CH4}$) the $v_d$~curves show an interesting behaviour:
Instead of reaching a maximum and then falling off again, they approach a point of minimal slope and
keep rising with higher $\ETp$~values (figure \ref{fig:H2-nomax}). In mixtures with tuned \ce{CH4} and \ce{H2} fractions
the $v_d$~curves show a wide plateau (figure \ref{fig:H2-plateau}), which, for some mixtures, simulations
show to be (more or less) stable up to $\ETp$~values of \unitfrac[\unitfrac[1400]{V}{cm}]{K}{mbar}
(figure \ref{fig:H2-plateau-HV}). This makes these mixtures interesting for applications with very inhomogeneous fields.
Unfortunately, at the moment we cannot measure the drift velocities at such high fields due to limitations in our
experimental setup.

For security reasons we did not use pure \ce{H2}~gas, but pre-mixed \binmix{Ar}{90}{H2}{10} that was mixed
with pure \ce{Ar} and \ce{CH4} to get the desired gas fractions. The gas distributor guarantees a mix accuracy within
\unit[10]{\%} relative to the specified fractions \cite{DIN14175}, so the pre-mixed \binmix{Ar}{90}{H2}{10} has a Hydrogen fraction between
\unit[9]{\%} and \unit[11]{\%}. This uncertainty must be added to the mixing uncertainties from the UGMA. In fact, the pre-mix
error dominates for mixtures with $\eta_{\ce{H2}} > \unit[1]{\%}$.
\[
    \sigma_{\eta_{\ce{H2}}} = \eta_{\ce{H2}} \cdot 0.1
\]

\chapter{Measuring the first Townsend coefficient}
\label{chap:alpha}

\begin{figure}[h!]
    \centering
    \subfigure[Top view]{
        \includegraphics[width=0.47\textwidth]{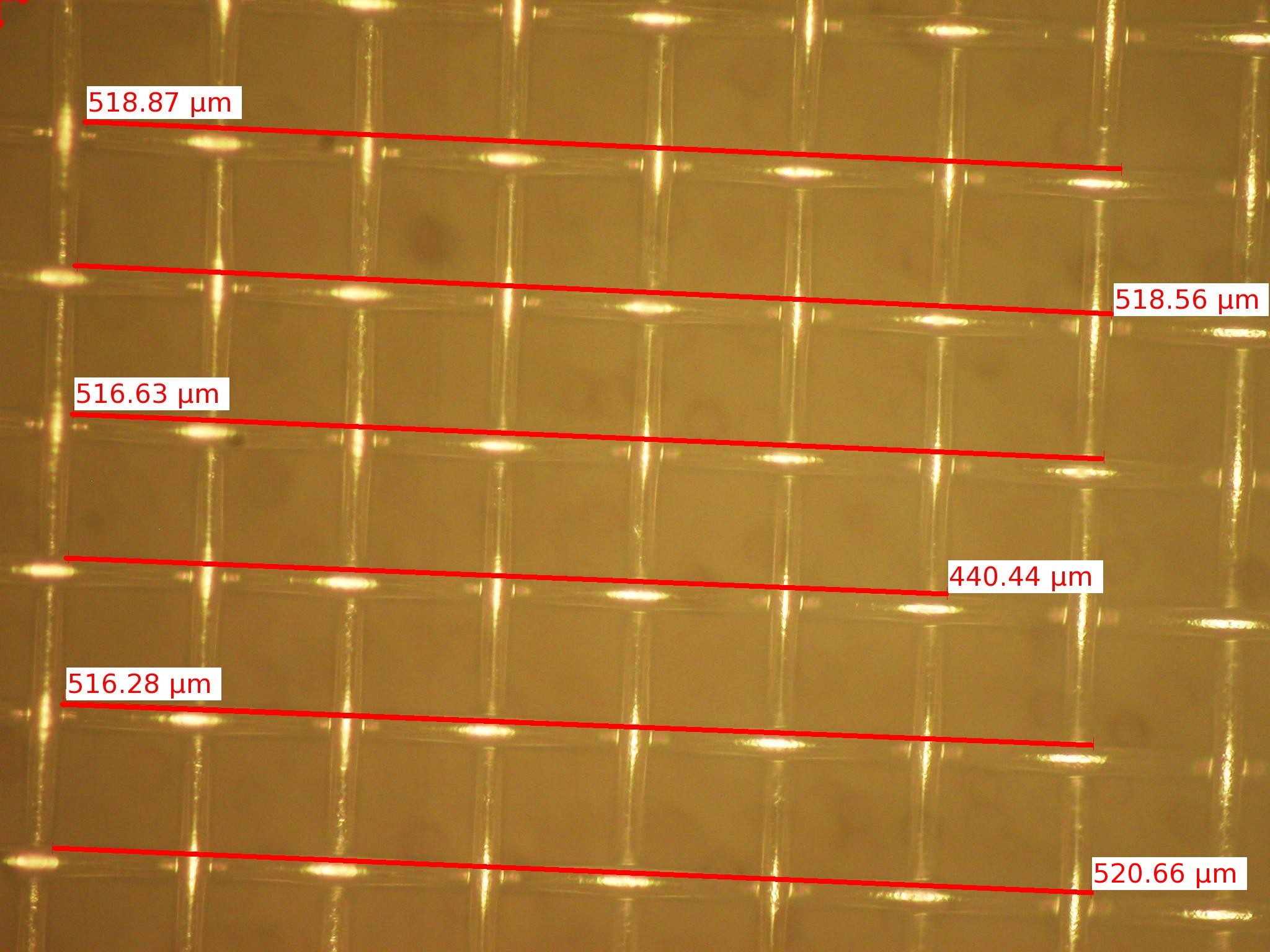}
    }
    \hfill
    \subfigure[Side view]{
        \includegraphics[width=0.47\textwidth]{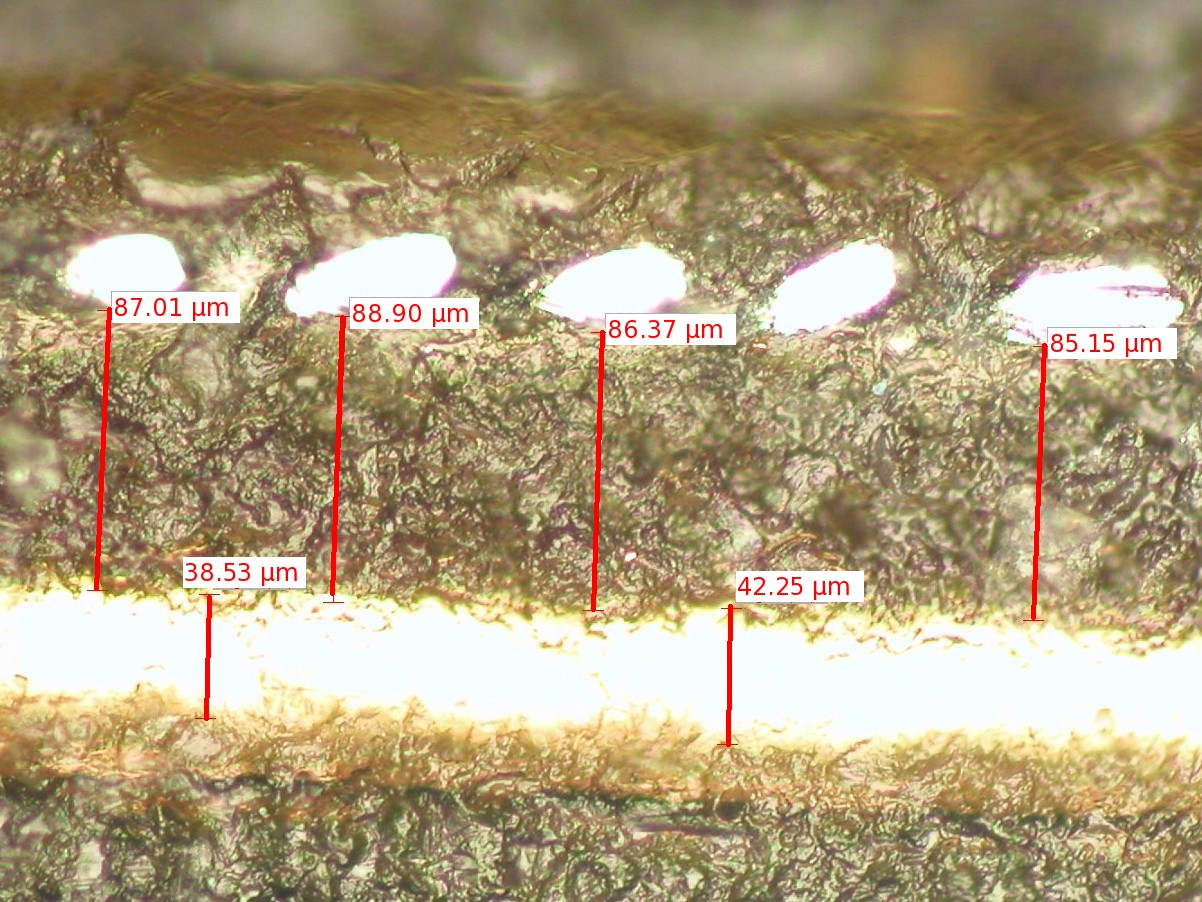}
    }
    \caption[Measurements of the MicroMeGaS geometry]
            {Measurements of the MicroMeGaS geometry}
    \label{fig:MM-geo}
\end{figure}

As it was said in section \ref{sec:amp}, it is not possible to calculate the first Townsend coefficient
$\alpha_T$, because we do not know the probability of Penning and Jesse transfers. Therefore it is necessary
to measure it directly.

\begin{table}
    \centering
    \caption{Comparison of nominal MicroMeGaS geometry with measurements}
    \label{tab:MM-geo}
    \begin{tabular}{rr@{.}lr@{.}l@{\,$\pm$\,}r@{.}l}
        MicroMeGaS parameter & \multicolumn{2}{c}{nominal} & \multicolumn{4}{c}{measured} \\
        \hline
        wire diameter / [µm]        & 18&0  & 14&1 & 0&7 \\
        grating constant / [µm]     & 63&0  & 73&3 & 0&6 \\
        mesh to pad distance / [µm] & 128&0 & 87&7 & 4&8 \\
        \hline
    \end{tabular}
\end{table}

The monitoring chambers include a slot for \isotope[55]{Fe} gamma sources, which can be used to measure the gain
of the MicroMeGaS. Trying to use this to measure $\alpha_T$ proved to be futile, since the
shape of the electric field in the amplification region is non-trivial and very inhomogeneous. Figure~\ref{fig:MM-geo}
shows some geometry measurements that were done with a microscope and table~\ref{tab:MM-geo} shows the measurements in
comparison with the nominal values taken from reference \cite{Abgrall201125}.

The values differ considerably, which is probably
caused by the manufacturing process. The mesh is held in place by two layers of solder resist below, and one layer above it.
The nominal layer width is \unit[64]{µm}, but the mesh is being pressed into the bottom layers during the lamination process
of the top layer. This does \emph{not} hinder the normal operation of the MicroMeGaS, since the exact distance between
mesh and pad is not important as long as it (and thus the gain) is constant across the whole MicroMeGaS.

The reason for the difficulties with the $\alpha_T$~measurement is the fact that the mesh's grating constant is in the same
order of magnitude as the distance between mesh and anode pads. This means one cannot approximate the electric field to
be homogeneous like that of a plate capacitor. This is true for both the nominal and the measured geometries.

\begin{figure}
    \centering
    \includegraphics{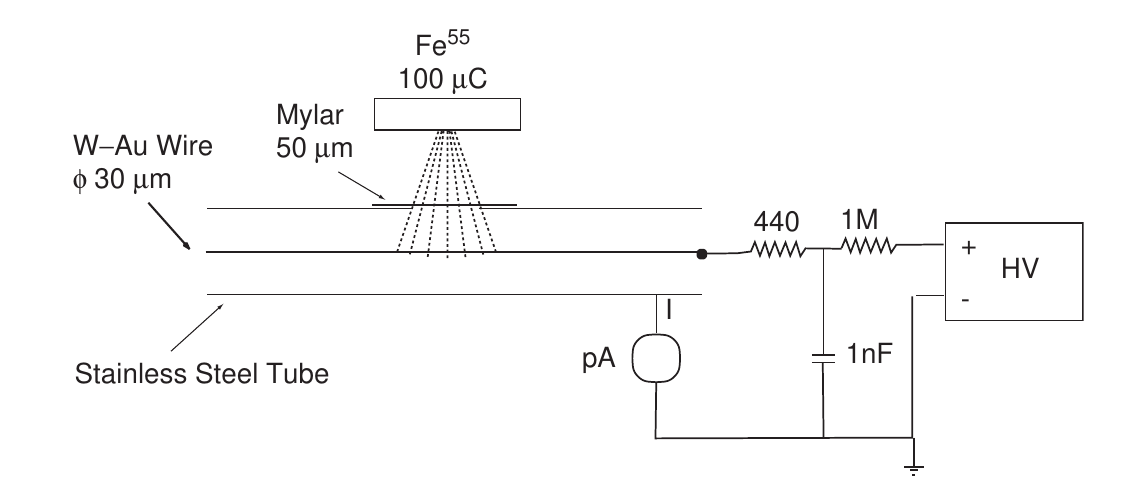}
    \caption[Experimental setup by Auriemma et al.]
            {Experimental setup by Auriemma et al. \cite{AURIEMMA2003}}
    \label{fig:aur-setup}
\end{figure}

Auriemma et al. proposed a method of measuring $\alpha_T$ using a current measurement at a cylindrical wire chamber \cite{AURIEMMA2003}.
Their experimental setup can be seen in figure \ref{fig:aur-setup}. The principle is simple: A radioactive source
continuously ionises the gas in the wire chamber and the electrons drift towards the anode wire, where the
gas amplification takes place. This creates a current which is proportional to the gain $G$ and can be measured.
By varying the anode voltage one can directly measure $\alpha_T(E)$ as follows.

\section{Theory}

The electric field in the gas volume is given by the anode wire radius $r_{\!A}$, the inner radius of the wire chamber
$r_C$ and the voltage $U$ applied between the two.
\begin{equation}
    E(r) = \underbrace{\frac{1}{\ln(\nicefrac{r_C}{r_{\!A}})}}_{:=\,a} \frac{U}{r} \label{eq:E}
\end{equation}
The gain due to gas amplification depends on the integrated Townsend coefficient along the drift path.
\begin{align}
    \ln G &= -\int_{r_0}^{r_{\!A}} \alpha_T\big(E(r)\big) \dd r
          = -\int_{E(r_0)}^{E(r_{\!A})} \alpha_T(E) \underbrace{\diff{r}{E}}_{-a\frac{U}{E^2}} \dd E
          \label{eq:lnG}
\end{align}
The minus sign is due to the fact that $r_0 > r_{\!A}$. If the voltage is varied, the gain varies accordingly.
\begin{align*}
    \diff{\ln G}{U} &= - \diff{}{U} \int_{r_0}^{r_{\!A}} \alpha_T\big(E(r)\big) \dd r
                    = - \int_{r_0}^{r_{\!A}} \diff{\alpha_T}{E}\underbrace{\diff{E}{U}}_{=\frac{a}{r}=\frac{E}{U}} \dd r \\
                    &= - \int_{E(r_0)}^{E(r_{\!A})} \diff{\alpha_T}{E} \frac{E}{U} \underbrace{\diff{r}{E}}_{-a\frac{U}{E^2}} \dd E
                    = \int_{E(r_0)}^{E(r_{\!A})} \diff{\alpha_T}{E} \frac{a}{E} \dd E \\
                    &= \left[ \alpha_T(E) \frac{a}{E} \right]_{E(r_0)}^{E(r_{\!A})}
                        + \underbrace{\int_{E(r_0)}^{E(r_{\!A})} \alpha_T(E) \frac{a}{E^2} \dd E}_{\overset{\eqref{eq:lnG}}{=}{\frac{\ln G}{U}}}
\end{align*}

Since gas amplification starts at very high electric fields, one can assume that the amplification
at the beginning of the drift path is 0 for most electrons.
\begin{gather*}
    \alpha_T\big(E(r_0)\big) = 0 \\
    \diff{\ln G}{U} = \alpha_T\big(E(r_{\!A})\big) \frac{a}{E(r_{\!A})} + \frac{\ln G}{U}
\end{gather*}
One can now solve for $\alpha_T$ and thus measure it directly by varying the voltage.
\[
    \alpha_T\big(E(r_{\!A})\big) = \left( \diff{\ln G}{U} - \frac{\ln G}{U} \right) \frac{E(r_{\!A})}{a}
                     \overset{\eqref{eq:E}}{=} \left( \diff{\ln G}{\ln U} - \ln G \right) \frac{1}{r_{\!A}}
\]

The gain strongly depends on $U$, so it is always $\nicefrac{\dd\ln G}{\dd\ln U} \gg \ln G$ \cite{AURIEMMA2003}. This means that one
can measure $\alpha_T$ even if the value of $G$ is not known, but only a proportional value like
the current $I$.
\begin{align*}
    G &= \frac{I}{I_0} \\
    \alpha_T\big(E(r_{\!A})\big) &= \left( \diff{\ln \nicefrac{I}{I_0}}{\ln U} - \ln \nicefrac{I}{I_0} \right) \frac{1}{r_{\!A}} \\
                             &= \left( \diff{\ln I}{\ln U} - \ln \nicefrac{I}{I_0} \right) \frac{1}{r_{\!A}} \\
                             &\approx \left( \diff{\ln I}{\ln U} \right) \frac{1}{r_{\!A}}
\end{align*}

\section{Modifications}

\begin{figure}
    \centering
    \includegraphics[width=0.80\textwidth]{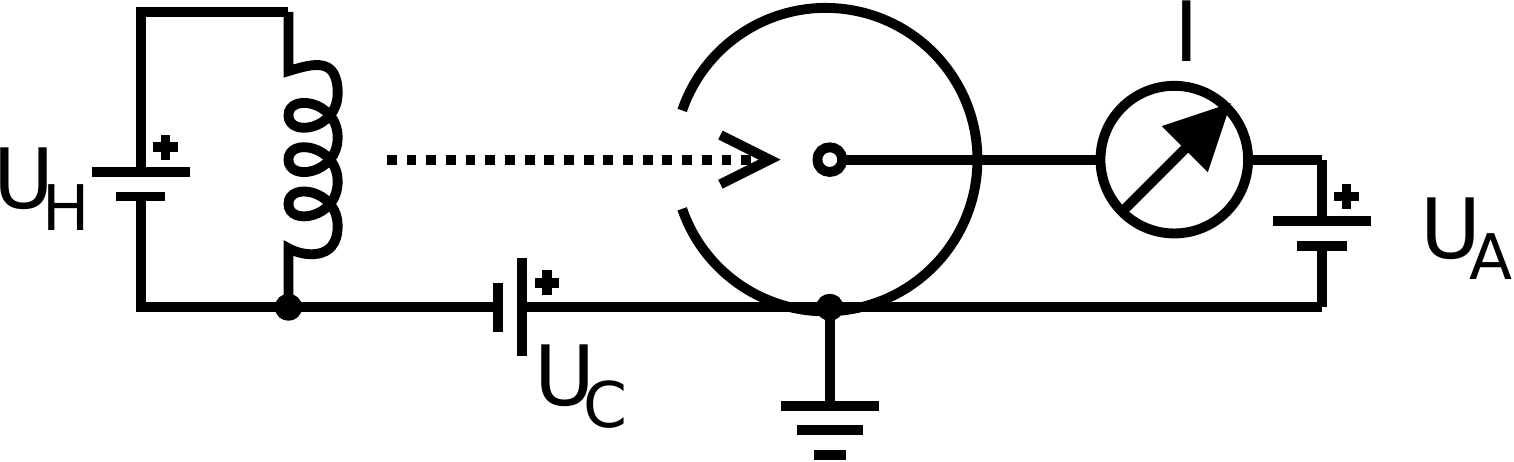}
    \caption[Modified setup for the $\alpha_T$ measurement]
            {Modified setup for the $\alpha_T$ measurement. We use an anode wire with a diameter of \unit[50]{\micro{m}}
             and a cathode cylinder with an inner diameter of about \unit[1]{cm}.}
    \label{fig:alpha-setup}
\end{figure}

The most difficult part with this experimental setup is the current measurement, since the currents
involved are very small ($\sim \unit{pA}$). We therefore modified the setup to increase the primary (unamplified)
current $I_0$ and thus $I$. The new setup can be seen in figure \ref{fig:alpha-setup}. The radioactive source
was replaced with a hot cathode that emits electrons due to their thermal energy. These electrons
are transported to the amplification region by an electric drift field. The currents achievable with this
method are orders of magnitude higher than the ones created by (reasonably dimensioned) radioactive sources.

The wire chamber is not exactly cylindrical, since it needs an opening for the electrons. This deviation
from the ideal cylindrical field will skew the results for $\alpha_T$ somewhat and should be eliminated for
high precision measurements (see section \ref{sec:alpha-upgrade}). Since we are only trying to demonstrate
the feasibility of the measurement, these effects will not be discussed here.

\begin{figure}
    \centering
    \subfigure{
        \includegraphics[width=0.47\textwidth]{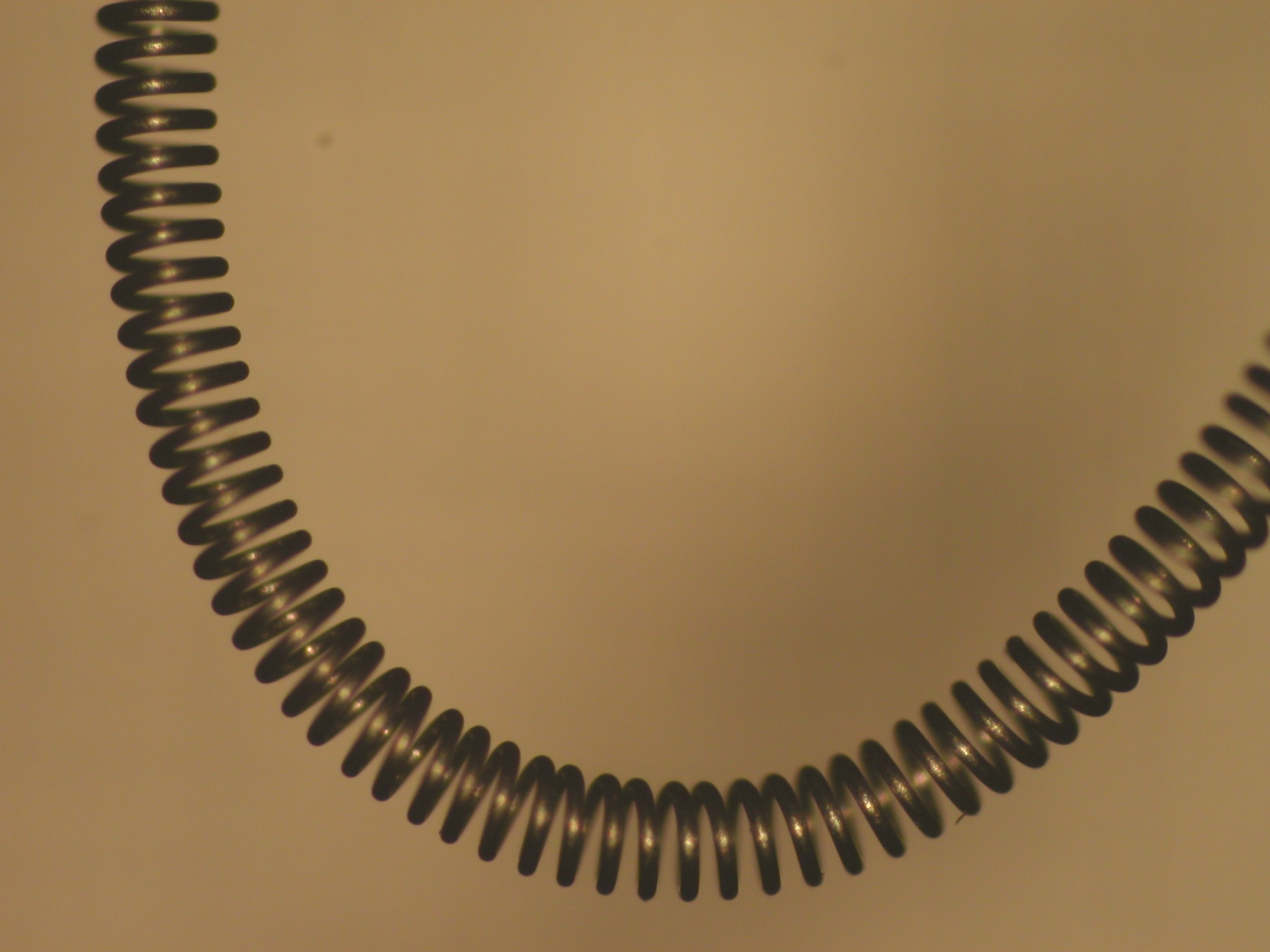}
    }
    \hfill
    \subfigure{
        \includegraphics[width=0.47\textwidth]{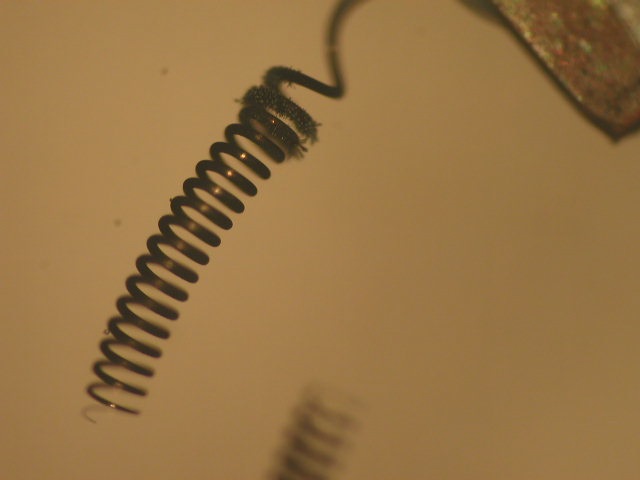}
    }
    \caption[Typical light bulb filament before and after the use as hot cathode in gas]
            {Typical light bulb filament before (left) and after (right) the use as hot cathode in gas}
    \label{fig:bulbs}
\end{figure}

The hot cathode is a simple bicycle light bulb of which the glass body was removed. Unfortunately the filament
degrades as soon as one applies a voltage to it. The life time of the 
lamp depends on the applied voltage as well as the gas mixture in which it is operated. Especially \ce{CO2}
proved to be a hazard for the filament. We assume that the \ce{CO2} in the gas mix dissociates at the hot
cathode and then oxidises the filament. Figure \ref{fig:bulbs} shows a typical filament before and after
its use.

\begin{figure}
    \centering
    \includegraphics[width=\plotwidth]{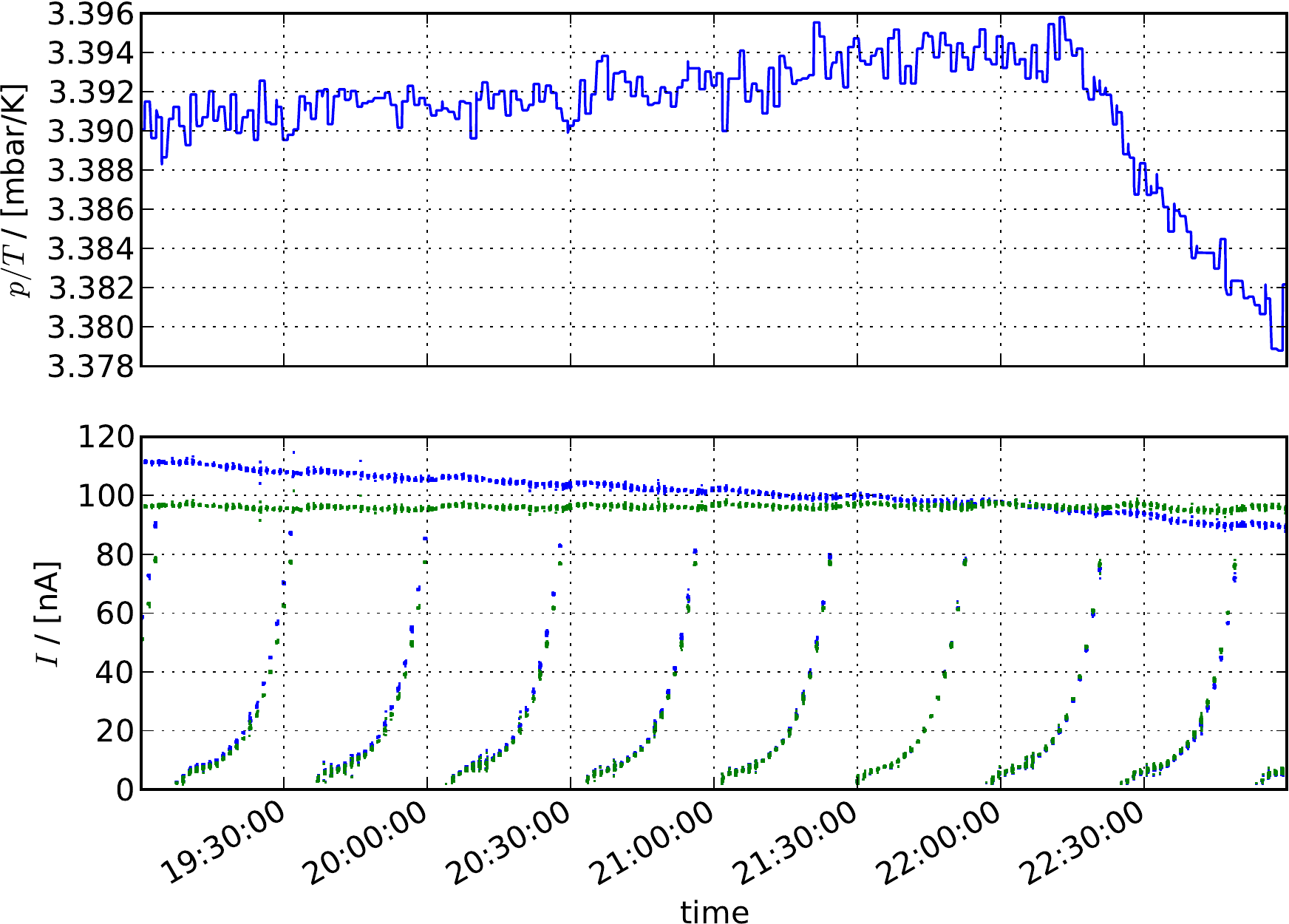}
    \caption[Measured and corrected anode current of P5]
            {Measured (blue) and corrected (green) anode current of P5}
    \label{fig:I-correction-P5}
\end{figure}

\begin{figure}
    \centering
    \includegraphics[width=0.75\textwidth]{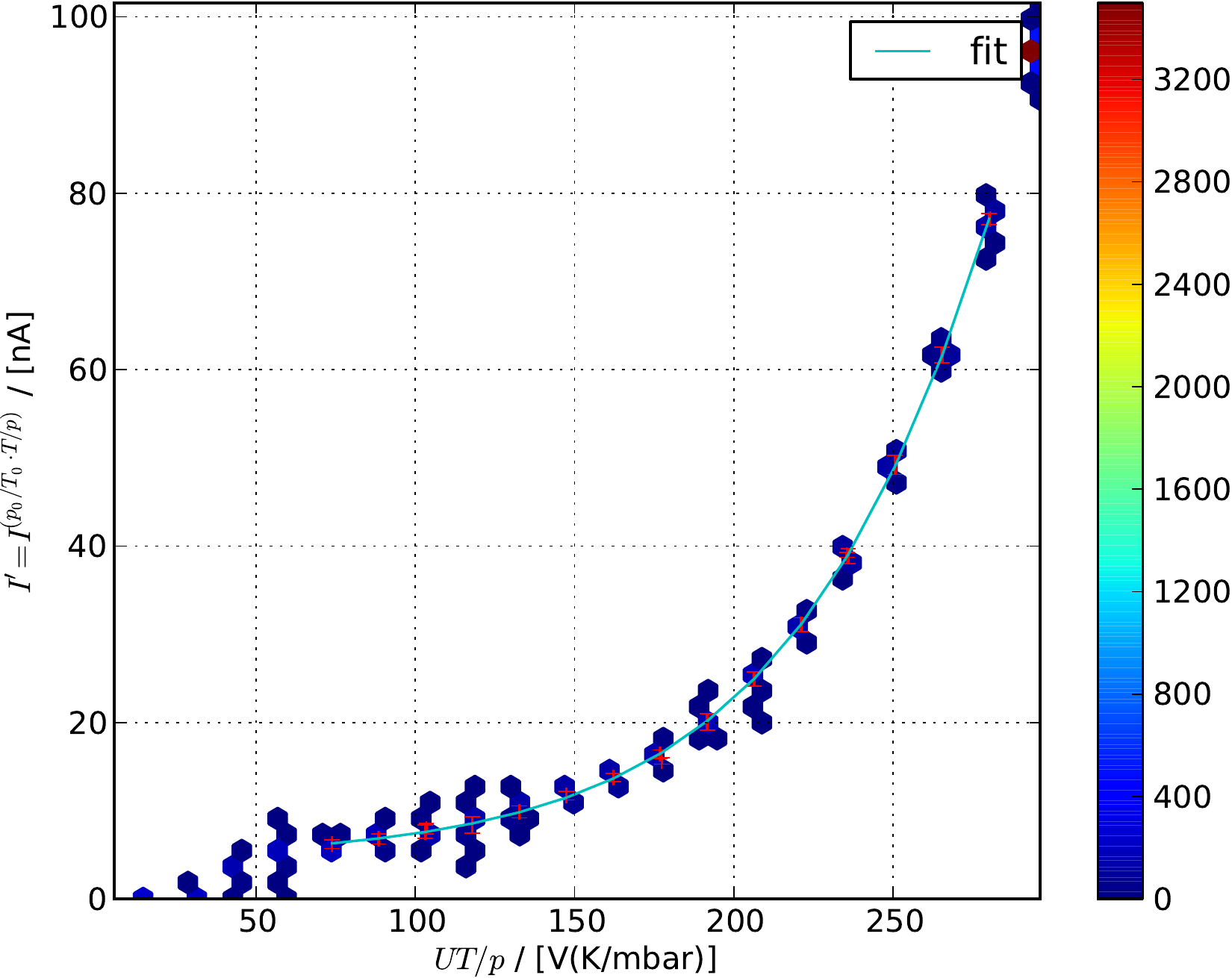}
    \caption[Anode current vs voltage of P5]
            {Anode current vs voltage of P5.
             The fit is a function of the form $c + \exp(a_0 + a_1 x + a_ 2 x^2 + a_3 x^3)$.}
    \label{fig:I-vs-U-P5}
\end{figure}

Since the filament \enquote{suffers} during operation, the primary current $I_0$ will also change
with time. To be able to correct for this, one can choose a $U_A$ profile that inserts a predefined reference
voltage in between the voltage ramp steps. The current can then be normalised to that of the reference
points. One also has to consider the temperature and pressure dependencies and correct for those
influences on the current. The Townsend coefficient can be expressed as a function of $\ETp$ and a
factor $\nicefrac{p}{T}$, so one gets
\begin{gather*}
    \begin{aligned}
        G &= \frac{I}{I_0} = \exp\left( \int \alpha_T\big(\ETp(r), \nicefrac{T}{p}\big) \dd r \right) 
          = \exp\left( \int f\big(\ETp(r)\big) \frac{p}{T} \dd r \right) \\
          &= \Bigg( \exp\left( \int f\big(\ETp(r)\big) \dd r \right) \Bigg)^{\frac{p}{T}} 
          = \Bigg( \underbrace{\exp\left( \int f\big(\ETp(r)\big) \frac{p_0}{T_0} \dd r \right)}
                               _{:=\,G' = \nicefrac{I'}{I_0}} \Bigg)^{\frac{T_0p}{p_0T}} \\
    \end{aligned} \\
    \begin{aligned}
        I &= I'^{\frac{T_0p}{p_0T}} \\
        I' &= I^{\frac{p_0T}{T_0p}}.
    \end{aligned}
\end{gather*}
Figures \ref{fig:I-correction-P5} and \ref{fig:I-vs-U-P5} show the measured and corrected
current for the $\alpha_T$ measurements of P5 (\binmix{Ar}{95}{CH4}{5}).

\begin{figure}
    \centering
    \includegraphics[width=\plotwidth]{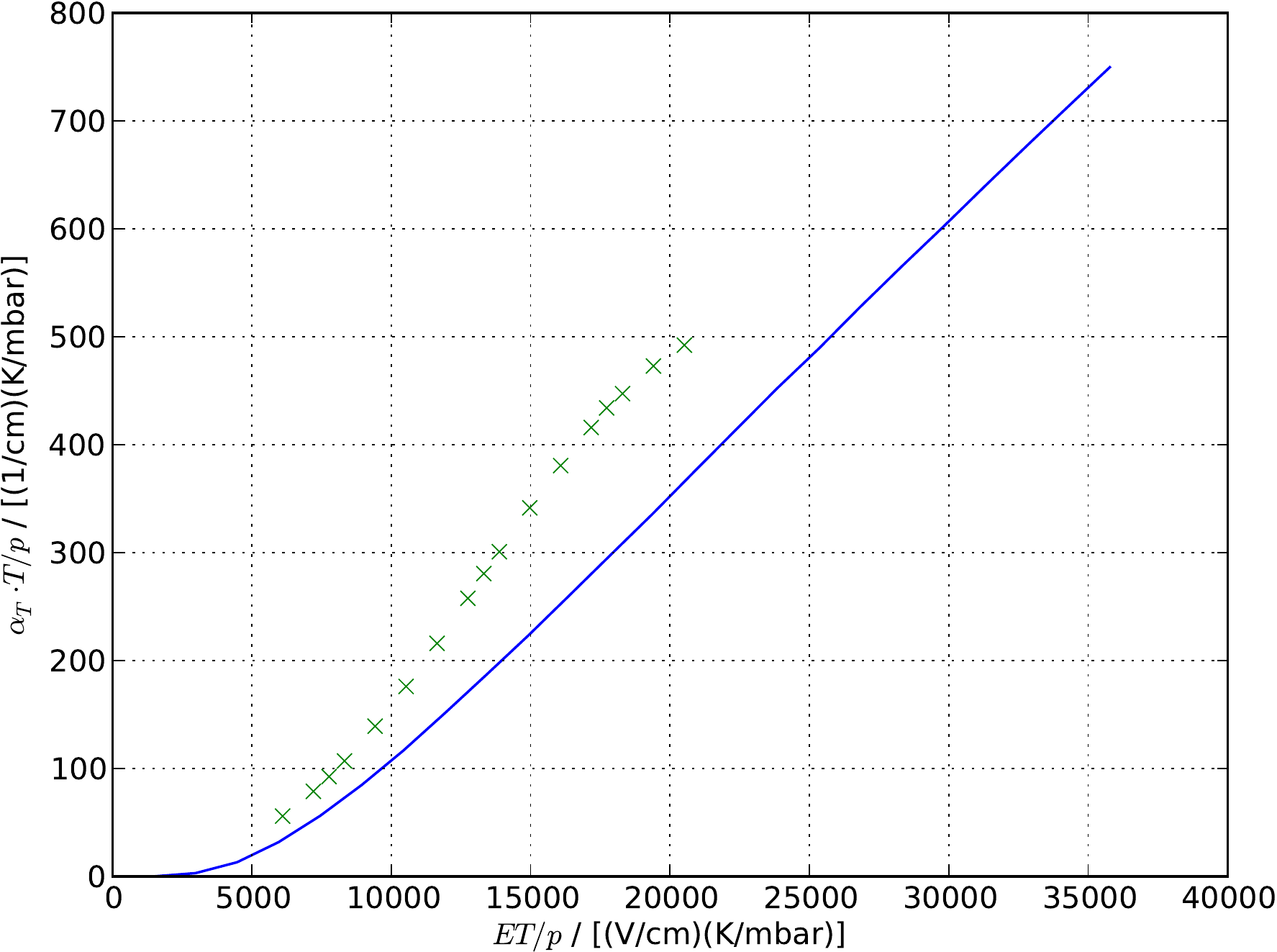}
    \caption{Measured and simulated $\alpha_T$ of P5}
    \label{fig:alpha-P5}
\end{figure}

\begin{figure}
    \centering
    \includegraphics[width=\plotwidth]{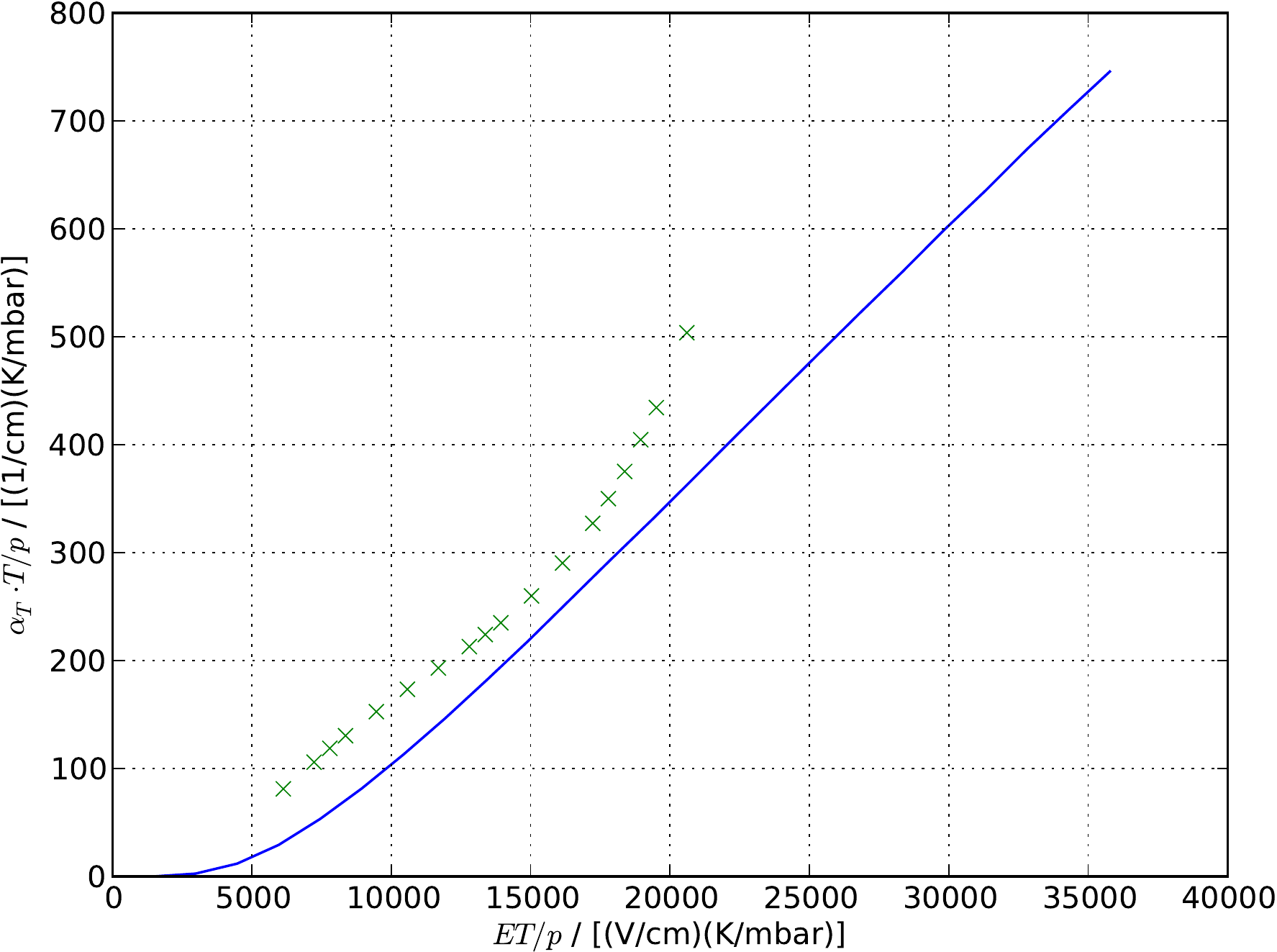}
    \caption{Measured and simulated $\alpha_T$ of P7}
    \label{fig:alpha-P7}
\end{figure}

Figures \ref{fig:alpha-P5} and \ref{fig:alpha-P7} show the simulated and measured Townsend coefficients of
P5 and P7. As one can see, the measurements are somewhat off in comparison with the simulations.
One has to consider, though, that the simulations do not take Penning and Jesse effects into account,
so the measured $\alpha_T$ is expected to be higher than the simulation. Also there has been no
thorough check for systematic influences from the skewed electric field or the possible influence of the voltages
on the number of electrons that make it from the drift field into the cathode tube.

\clearpage

\section{Future improvements}
\label{sec:alpha-upgrade}

\begin{figure}
    \centering
    \includegraphics[width=0.40\textwidth]{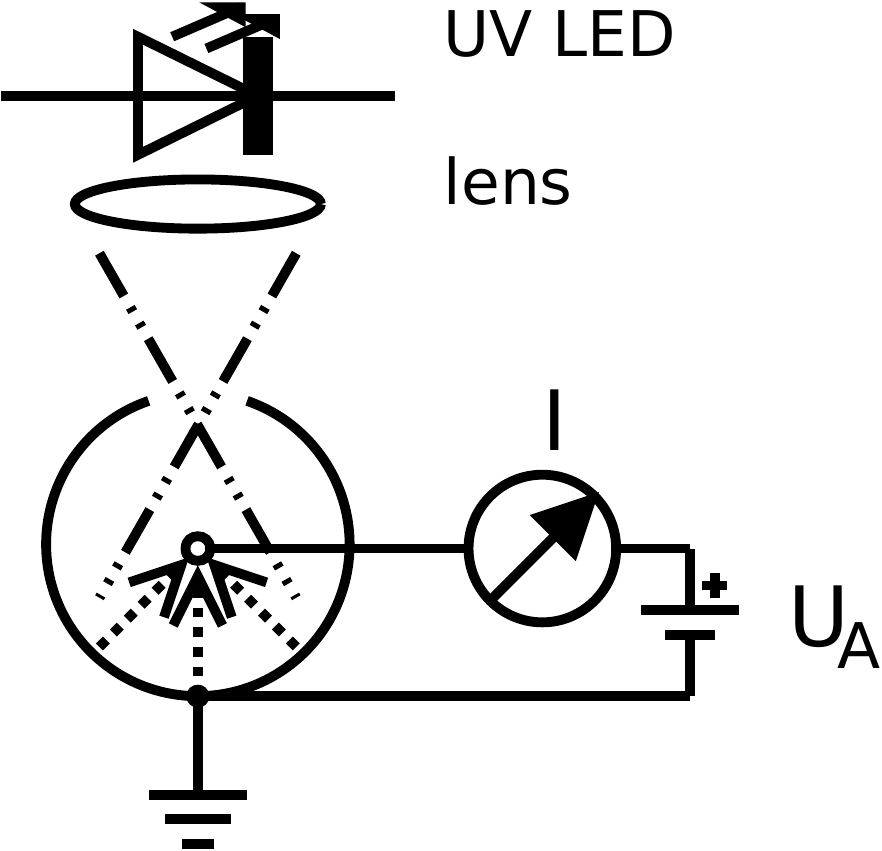}
    \caption{Proposed upgrade of the $\alpha_T$ measurement setup}
    \label{fig:alpha-upgrade}
\end{figure}

The modified method of $\alpha_T$ measurement proved to be feasible and first measurements show promise for
future improvements. The current setup mainly suffers from the degradation of the hot cathode as well as the
field distortions due to the necessary opening in the cathode tube. To improve the setup in those regards,
we suggest an alternative setup as seen in figure~\ref{fig:alpha-upgrade}.

A UV-LED could create free electrons directly inside the cathode tube utilising the photoelectric effect.
Not only would the LED allow for continuous stable operation, but it would also reduce the impact of the
skewed electric field. Since we could make the necessary hole in the cathode tube comparatively small and
the electrons traverse the electric field on the far side of the tube, the effects due to field distortion
can be expected to be minimal.

A possible disadvantage of this setup might be that the UV-LED excites the deployed quenchers in the gas
and thus interferes with the gas electron interaction. This has to be investigated.

\chapter{Conclusion}

The T2K/ND280 monitoring chambers were successfully used for the precision measurement of electron drift
velocities at low electric fields ($< \unitfrac[400]{V}{cm}$). The systematic errors were estimated to
be not larger than \unit[4]{\perthousand}, while the statistical errors are in the order of
\unit[1]{\perthousand} or better. The general shape of the simulations is well reproduced by the measurements,
though deviations that cannot be attributed to the mixing uncertainties remain. It was shown that
one can visualise the influence of different additive fractions by reducing the drift curve to its
working point, the $v_d$~maximum. This is not true for all mixtures as was seen at the example of some
\ce{Ar}-\ce{CH4}-\ce{H2} mixtures, which show a plateau or region of weak $\ETp$-dependence instead of
a maximum.

\begin{figure}[b!]
    \centering
    \includegraphics[width=\plotwidth]{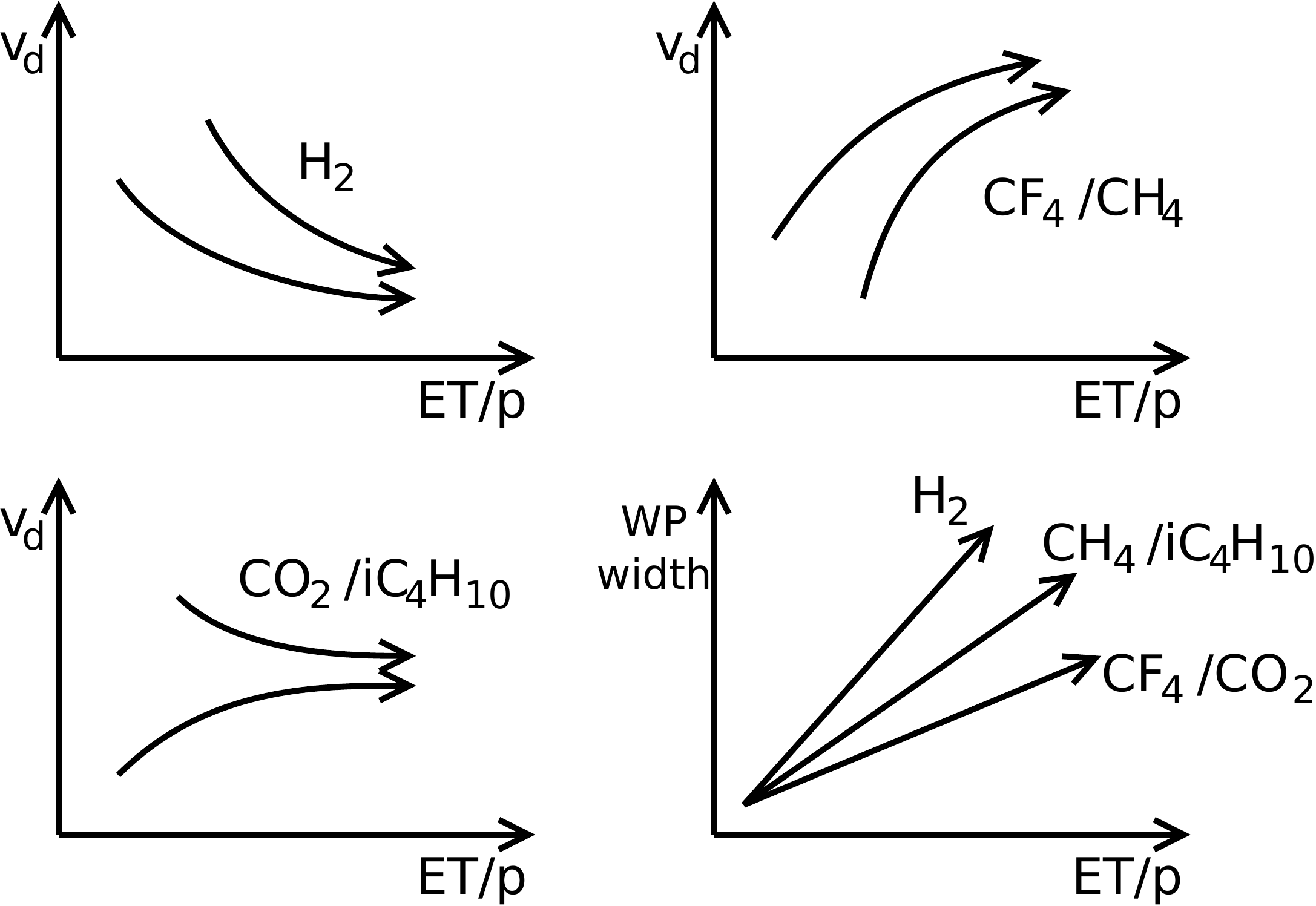}
    \caption[Simplified influences of the additives on the working points]
            {Simplified influences of the additives on the working points}
    \label{fig:simple}
\end{figure}

A simplified depiction of the different influences of the additives on the working points can be seen in
figure \ref{fig:simple}. Both \ce{CF4} and \ce{CH4} generally increase the drift velocity, but one needs
much less \ce{CF4} than \ce{CH4} to achieve the same effect. Conversely, to prevent $v_d$ deviations,
one would have to control the \ce{CF4} fraction to a higher precision than an equivalent \ce{CH4} fraction.
This \enquote{sensitivity} of \ce{CF4} mixtures is amplified by the fact that their working point widths
are smaller than those of \ce{CH4} mixtures.
\ce{CO2} and \ce{iC4H10} pull the working point drift velocity to a value around \unitfrac[50]{\micro{m}}{ns}
and \ce{H2} always pulls the working points down. Compared to the other additives, \ce{H2} causes a much stronger
widening of the drift velocity maxima, which leads to the formation of the aforementioned plateaus.
Mixtures with \ce{H2} are thus less susceptible to changes in $\ETp$.

A modified version of the $\alpha_T$~measurement method introduced by Auriemma et al. \cite{AURIEMMA2003}
was shown to be a feasible alternative for future measurements. Further modifications that might
mitigate the problems with the hot cathode and skewed electric field were suggested and should be
investigated.

\chapter{Acknowledgements}

I want to thank the T2K group under Stefan Roth at III. Physikalisches Institut~B for supporting me during the
work on my thesis. Thanks also go to Johannes Hellmund, Fabian Schneider and Jochen Steinmann for proofreading
this thesis and I especially want to thank Jochen Steinmann for his tireless work at the
UGMA and his help with the monitoring chambers. I also want to thank Stephen Biagi and Heinrich Schindler
for their e-mail correspondence and support in regards to Magboltz and Garfield++ as well as all my colleagues
at Halle Physik for providing an enjoyable working atmosphere and occasionally cake.

\renewcommand{\bibname}{References}
\bibliography{Masterbib,physics_paper}

\appendix

\chapter{Gas properties}
\label{chap:gasprop}

\begin{center}
\begin{tabular}{rr@{.}lr@{.}lr@{.}lr@{.}l}
    Gas & \multicolumn{2}{r}{$\rho_{0} / [\unitfrac{g}{l}]$} & \multicolumn{2}{r}{$M_{mol} / [\unitfrac{g}{mol}]$}
        & \multicolumn{2}{r}{$W_\alpha / [\unit{eV}]$} & \multicolumn{2}{r}{$W_\beta / [\unit{eV}]$} \\
    \hline
    \ce{H2}     & 0&089885 & 2&01588    & 36&4                      & 36&3 \\
    \ce{He}     & 0&178488 & 4&002602   & 46&0                      & 42&3 \\
    \ce{N2}     & 1&250386 & 28&0134    & \multicolumn{2}{c}{--}    & \multicolumn{2}{c}{--} \\
    \ce{O2}     & 1&429033 & 31&9988    & \multicolumn{2}{c}{--}    & \multicolumn{2}{c}{--} \\
    \ce{Ar}     & 1&783956 & 39&948     & 26&4                      & 26&3 \\
    \ce{Xe}     & 5&898003 & 131&293    & 21&7                      & 21&9 \\
    \ce{CO}     & 1&250501 & 28&0101    & \multicolumn{2}{c}{--}    & \multicolumn{2}{c}{--} \\
    \ce{CO2}    & 1&976813 & 44&0095    & 34&3                      & 32&8 \\
    \ce{CH4}    & 0&717459 & 16&0425    & 29&1                      & 27&1 \\
    \ce{CF4}    & 3&946447 & 88&0043    & \multicolumn{2}{c}{--}    & \multicolumn{2}{c}{--} \\
    \ce{C2H4}   & 1&261111 & 28&0532    & \multicolumn{2}{c}{--}    & \multicolumn{2}{c}{--} \\
    \ce{C2H6}   & 1&355125 & 30&069     & 26&6                      & 24&4 \\
    \ce{C3H8}   & 2&010037 & 44&0956    & \multicolumn{2}{c}{--}    & \multicolumn{2}{c}{--} \\
    \ce{iC4H10} & 2&688009 & 58&1222    & \multicolumn{2}{c}{--}    & \multicolumn{2}{c}{--} \\
    \ce{C4H10}  & 2&688697 & 58&1222    & \multicolumn{2}{c}{--}    & \multicolumn{2}{c}{--} \\
    \ce{F6S}    & 6&615833 & 146&055    & \multicolumn{2}{c}{--}    & \multicolumn{2}{c}{--} \\
    \hline
    \multicolumn{9}{r}{$\rho_0$ at $T = \unit[0]{\celsius}$, $p = \unit[1013.25]{mbar}$} \\
    \multicolumn{9}{r}{$\rho_0$ and $M_{mol}$ taken from \cite{NISTCHEM2013}, $W_{\alpha / \beta}$ from \cite{Blum2008}}
\end{tabular}
\end{center}

\chapter{Measurements}
\label{chap:Measurements}

\section{\pdfce{Ar}-\pdfce{CH4}-\pdfce{CO2} mixtures}
\label{sec:Ar-CH4-CO2}

\newcommand{\MeasILD}[3]{
        
    \minisec{\termix{Ar}{#1}{CH4}{#2}{CO2}{#3}}
    \begin{center}
        \includegraphics[width=\plotwidth]{Figures/Ar-CH4-CO2/prof-#2-#3-crop}
    \end{center}

}
\MeasILD{97}{3}{0}
\MeasILD{96}{3}{1}
\MeasILD{95}{3}{2}
\MeasILD{94}{3}{3}
\MeasILD{93}{3}{4}
\MeasILD{96}{4}{0}
\MeasILD{95}{4}{1}
\MeasILD{94}{4}{2}
\MeasILD{93}{4}{3}
\MeasILD{92}{4}{4}
\MeasILD{95}{5}{0}
\MeasILD{94}{5}{1}
\MeasILD{93}{5}{2}
\MeasILD{92}{5}{3}
\MeasILD{91}{5}{4}
\MeasILD{94}{6}{0}
\MeasILD{93}{6}{1}
\MeasILD{92}{6}{2}
\MeasILD{91}{6}{3}
\MeasILD{90}{6}{4}
\MeasILD{93}{7}{0}
\MeasILD{92}{7}{1}
\MeasILD{91}{7}{2}
\MeasILD{90}{7}{3}
\MeasILD{89}{7}{4}
\MeasILD{92}{8}{0}
\MeasILD{91}{8}{1}
\MeasILD{90}{8}{2}

\MeasILD{90}{10}{0}
\MeasILD{89}{10}{1}
\MeasILD{88}{10}{2}
\MeasILD{85}{15}{0}
\MeasILD{84}{15}{1}
\MeasILD{83}{15}{2}
\MeasILD{80}{20}{0}
\MeasILD{79}{20}{1}
\MeasILD{75}{25}{0}

\section{\pdfce{Ar}-\pdfce{CF4}-\pdfce{iC4H10} mixtures}
\label{sec:Ar-CF4-iC4H10}

\newcommand{\MeasTTK}[3]{
        
    \minisec{\termix{Ar}{#1}{CF4}{#2}{iC4H10}{#3}}
    \begin{center}
        \includegraphics[width=\plotwidth]{Figures/Ar-CF4-iC4H10/prof-#2-#3-crop}
    \end{center}

}

\MeasTTK{98}{1}{1}
\MeasTTK{97}{1}{2}
\MeasTTK{96}{1}{3}
\MeasTTK{95}{1}{4}
\MeasTTK{94}{1}{5}
\MeasTTK{93}{1}{6}
\MeasTTK{97}{2}{1}
\MeasTTK{96}{2}{2}
\MeasTTK{95}{2}{3}
\MeasTTK{94}{2}{4}
\MeasTTK{93}{2}{5}
\MeasTTK{96}{3}{1}
\MeasTTK{95}{3}{2}
\MeasTTK{94}{3}{3}
\MeasTTK{93}{3}{4}
\MeasTTK{95}{4}{1}
\MeasTTK{94}{4}{2}
\MeasTTK{93}{4}{3}
\MeasTTK{92}{4}{4}
\MeasTTK{94}{5}{1}
\MeasTTK{93}{5}{2}
\MeasTTK{92}{5}{3}
\MeasTTK{91}{5}{4}

\section{\pdfce{Ar}-\pdfce{CH4}-\pdfce{H2} mixtures}
\label{sec:Ar-CH4-H2}

\newcommand{\MeasAMH}[3]{
        
    \minisec{\termix{Ar}{#1}{CH4}{#2}{H2}{#3}}
    \begin{center}
        \includegraphics[width=\plotwidth]{Figures/Ar-CH4-H2/prof-#2-#3-crop}
    \end{center}

}

\MeasAMH{96}{3}{1}
\MeasAMH{95}{3}{2}
\MeasAMH{94}{3}{3}
\MeasAMH{93}{3}{4}
\MeasAMH{92}{3}{5}
\MeasAMH{95}{4}{1}
\MeasAMH{94}{4}{2}
\MeasAMH{93}{4}{3}
\MeasAMH{92}{4}{4}
\MeasAMH{91}{4}{5}
\MeasAMH{90}{4}{6}
\MeasAMH{89}{4}{7}
\MeasAMH{88}{4}{8}
\MeasAMH{89}{5}{6}
\MeasAMH{88}{5}{7}
\MeasAMH{87}{6}{7}
\MeasAMH{86}{6}{8}
\MeasAMH{91}{7}{2}
\MeasAMH{89}{7}{4}
\MeasAMH{87}{7}{6}
\MeasAMH{86}{7}{7}
\MeasAMH{85}{7}{8}
\MeasAMH{84}{8}{8}
\MeasAMH{83}{9}{8}
\MeasAMH{88}{10}{2}
\MeasAMH{86}{10}{4}
\MeasAMH{84}{10}{6}
\MeasAMH{83}{15}{2}
\MeasAMH{81}{15}{4}
\MeasAMH{78}{20}{2}

\end{document}